\documentclass[reqno,a4paper,dvipdfm,final]{amsart}
\usepackage{amsmath,amsfonts,amssymb,amsthm,amstext,eucal}
\usepackage{amscd,bbold}
\usepackage{array}
\usepackage{epic,eepic}
\usepackage[latin1]{inputenc}
\usepackage[notref,notcite]{showkeys}
\usepackage{hyperref}
\usepackage{cite}
\newcommand{\half}{\tfrac{1}{2}}
\renewcommand{\d}{\partial}
\newcommand{\fg}{\mathfrak{g}}

\newcommand{\fsl}{\mathfrak{sl}}
\newcommand{\fso}{\mathfrak{so}}

\newcommand{\fsu}{\mathfrak{su}}

\newcommand{\SO}{\mathrm{SO}}
\renewcommand{\O}{\mathrm{O}}
\newcommand{\Cl}{\mathrm{C}\ell}
\newcommand{\Spin}{\mathrm{Spin}}

\newcommand{\RR}{\mathbb{R}}

\newcommand{\VV}{\mathbb{V}}
\newcommand{\WW}{\mathbb{W}}
\newcommand{\ZZ}{\mathbb{Z}}

\newcommand{\TT}{\mathbb{T}}

\newcommand{\eM}{\mathcal{M}}
\newcommand{\eN}{\mathcal{N}}

\DeclareMathOperator{\AdS}{AdS}
\DeclareMathOperator{\Sph}{S}

\newcommand{\be}{\boldsymbol{e}}

\newcommand{\bx}{\boldsymbol{x}}

\newcommand{\eG}{\mathcal{G}}

\newcommand{\1}{\mathbb{1}}
\newcommand{\MUNCH}[1]{\relax}
\allowdisplaybreaks[1]
\numberwithin{equation}{section}
\hypersetup{%
  pdftitle = {Quotients of AdS x S: causally well-behaved spaces and
    black holes},
  pdfkeywords = {anti de Sitter, discrete quotients, supergravity,
    classifications, black holes},
  pdfauthor = {José Figueroa-O'Farrill, Owen Madden, Simon F. Ross
    and Joan Sim\'{o}n}
}
\begin{document}
\title[Quotients of $\AdS_{p+1}\times\Sph^q$: causal spaces
  and black holes]{Quotients of
  $\AdS_{p+1}\times\Sph^q$: causally well-behaved spaces
  and black holes}
\author[Figueroa-O'Farrill]{José
  Figueroa-O'Farrill} \address{School of Mathematics, The University
  of Edinburgh, Scotland, United Kingdom}
\email{j.m.figueroa@ed.ac.uk} \author[Madden]{Owen Madden}
\address{Centre for Particle Theory, Department of Mathematical
  Sciences, University of Durham, United Kingdom}
\email{O.F.Madden@durham.ac.uk} \author[Ross]{Simon F. Ross}
\address{Centre for Particle Theory, Department of Mathematical
  Sciences, University of Durham, United Kingdom}
\email{S.F.Ross@durham.ac.uk} \author[Sim\'{o}n]{Joan Sim\'{o}n}
\address{The Weizmann Institute of Physical Sciences, Department of
  Particle Physics, Rehovot, Israel}
\address{Department of Physics
  and Astronomy, David Rittenhouse Laboratories, University of
  Pennsylvania, Philadelphia, United States}
\address{The Kavli Institute of Theoretical Physics, University of
  California, Santa Barbara, United States}
\email{jsimon@bokchoy.hep.upenn.edu} \thanks{EMPG-03-25,
  WIS/02/04-JAN-DPP, UPR-1062-T, DCPT-03/61, NSF-KITP-03-121}
\date{\today}

\begin{abstract}
  Starting from the recent classification of quotients of
  Freund--Rubin backgrounds in string theory of the type
  $\AdS_{p+1}\times \Sph^q$ by one-parameter subgroups of isometries,
  we investigate the physical interpretation of the associated
  quotients by discrete cyclic subgroups.  We establish which
  quotients have well-behaved causal structures, and of those
  containing closed timelike curves, which have interpretations as
  black holes.  We explain the relation to previous investigations of
  quotients of asymptotically flat spacetimes and plane waves, of
  black holes in AdS and of Gödel-type universes.
\end{abstract}

\maketitle
\tableofcontents

\section{Introduction and motivation}
\label{sec:intro}

Taking quotients of smooth (super)gravity backgrounds has long been a
fundamental tool in string theory, both in the context of Kaluza-Klein
reduction, in which one quotients by the action of a continuous group,
and in the orbifold context, in which the group is discrete.
Riemannian singular quotients (orbifolds) provide exact string theory
backgrounds which allow us to understand how string theory resolves
certain types of timelike singularities.  These techniques are also
relevant in the Kaluza--Klein context: an early nontrivial example is
the embedding of the Melvin universe \cite{melvin} in string theory
\cite{GibbWilt,gibmaed}.  This work naturally suggests studying
Lorentzian orbifolds, in the hope of reaching a similar understanding
of certain types of spacelike singularities, in particular those
related to the Big Bang.  Although some progress has been achieved
\cite{KOST, KOSST, BHKN, CorCos1, joan, lms1, lawrence, fabmcg, lms2,
  horpol, CKR, BCKR, EGKR, EGR, PioBer, CorCosrev}, the fate (and
physics) of these singularities remains a very important area of
research in string theory.  Out of this effort we now have a complete
list of smooth quotients of Minkowski spacetime.  This classification
was given in \cite{fofs} recovering previous results on fluxbranes
\cite{DGGH0, DGGH1, DGGH2, GSflux, arkady}\footnote{Related work on
  the physics of fluxbranes can be found in \cite{CGS, CG, Saffin,
    CHC, EmparanFlux, BrecherSaffin, Brechersaffin2, Uranga, EmpGut}.}
and uncovering the existence of an interesting non-static smooth
quotient---the nullbrane---which can be understood as a
desingularisation of the parabolic orbifold \cite{steif}, the
supersymmetric toy model for a Big Crunch-Big Bang transition
singularity, by the introduction of a new scale (modulus) that smoothes
the singularity.

In this paper, we will study discrete cyclic quotients of anti-de
Sitter (AdS) backgrounds in gravity and in string theory.  Because of
its high degree of symmetry, the story for anti-de Sitter space is
particularly interesting, and there is already a rich literature on
physically interesting locally anti-de Sitter spacetimes, with much of
the discussion having focused on the BTZ black hole solutions
\cite{btz1,btz2} and their generalisations \cite{hp, ban1, ban2}.
However, some examples of smooth quotients are also known
\cite{cousshenn, simon, bns} \footnote{Some other work concerning
  orbifolds of AdS can be found in \cite{HorMar, BehLus, GhoMuk,
    Cai,BDHRS,BRSbtz,AST,mcinnes2003,FHLT,mcinnes2004}.}.  Given the 
considerable interest of AdS backgrounds in string theory, the time 
seems ripe for a more systematic investigation of these questions.

In a recent pair of papers \cite{fofs2,mr} we have classified
quotients of AdS by one-parameter subgroups of isometries.  The
emphasis in \cite{fofs2} being on AdS backgrounds in string theory, it
was necessary to classify quotients of geometries of the form
$\AdS_{p+1} \times \Sph^q$ by one-parameter subgroups of isometries.
As such backgrounds are maximally supersymmetric, it was also natural
to study the question of how much supersymmetry was preserved by the
quotient and in \cite{fofs2} there is a detailed analysis of this
question and the related issue of the existence of a spin structure on
the quotient.

Our purpose in the present paper is to study the geometry of the
discrete cyclic quotients associated to such one-parameter subgroups,
paying close attention to their causal structure, and to develop a
formalism to discuss the geometry and physical interpretation of all
smooth quotients.

Many of the quotients classified in \cite{fofs2,mr} contain closed
timelike curves and while there may be some interest in studying
such quotients, we shall nevertheless concentrate our attention on
those quotients for which there is a well-founded expectation that
they will provide good backgrounds for string propagation.  We will
therefore focus on and discuss in detail two kinds of quotients that
can be given a simple physical interpretation: smooth quotients with a
well-behaved causal structure, and those which can be given a black
hole interpretation following \cite{btz1,btz2}.  At the end of this
work, we shall briefly comment on the relation between some of our 
spacetimes having closed causal curves and
Gödel-type universes recently discussed in the literature
\cite{godel,GGHPR,BGHV,HarTak}.  The connection arises because
certain quotients commute with the Penrose limit
\cite{penrose,Gueven,BFOHP,BFOP}. Thus, one can identify which
discrete quotients of $\AdS_{p+1}\times S^q$ backgrounds give rise 
to compactified pp-waves having closed timelike curves after taking
the Penrose limit, the latter being T-dual to G\"{o}del-type universes.

We find that there are two types of quotients with well-behaved causal
structures.  First, there are quotients where an action on AdS alone
is well behaved.  These are generalisations of the two cases studied
previously:
\begin{enumerate}\renewcommand{\theenumi}{\roman{enumi}}
\item self-dual orbifolds of $\AdS_3$ \cite{cousshenn,bns} and
  their higher-dimensional generalisations, having no analogue in
  asymptotically flat configurations; and
\item the AdS analogue of the flat nullbrane construction
  \cite{simon}, consisting of a double null rotation action on
  $\SO(2,p)$ $p\geq 4$.  This is the near horizon geometry of a stack
  of D3-branes in the nullbrane vacuum for $p=4$ and a stack of
  M5-branes in the same vacuum for $p=6$.
\end{enumerate}
We give a comprehensive discussion of the structure of these
quotients, extending previous results. For the double null rotation,
we construct a new symmetry-adapted coordinate system, and find
interesting relations to compactified plane waves. We comment on
related issues in the nullbranes in an appendix. 

Secondly, there are quotients where the norm of the AdS isometry is
non-negative, but not always positive, so the pure AdS action would
have singularities or closed null curves.  These can be removed by a
suitable action on the transverse sphere if the latter is
odd-dimensional.  This second type is qualitatively new.  These
non-trivial actions on AdS can be divided into three categories:
\begin{enumerate}\renewcommand{\theenumi}{\roman{enumi}}
\item discrete quotients by rotations in AdS, the
  higher-dimensional analogues of the $\AdS_3$ conical defects;
\item discrete quotients by a null rotation, whose description in the
  Poincaré patch corresponds to a spacelike translation (in pure
  $\AdS_3$, these would give rise to the massless BTZ black hole
  \cite{btz2}) and whose sphere deformations are the near horizon
  limit of brane configurations in fluxbrane vacua classified in
  \cite{FigSimBranes, FigSimGrav}; and
\item discrete quotients defined by an everywhere null vector
  field in $\AdS_p$ $(p\geq 3)$, whose description in the Poincaré
  patch corresponds to a `translation' along a lightlike direction.
  Once more, when deformed by a non-trivial action on a transverse
  sphere, this corresponds to the near horizon counterpart of the
  corresponding quotients classified in \cite{FigSimBranes,
    FigSimGrav}.
\end{enumerate}

It is important to stress that any of the string theory backgrounds
discussed in this paper are related to many others through U-duality
and by Kaluza--Klein reductions from or liftings to M-theory.  We
shall not pursue this possibility in this paper, even though it is
natural to wonder about the dual incarnations of our backgrounds.

In studying quotients with a black hole interpretation, we confirm and
elucidate the conclusion of \cite{hp}, that for $p>2$, the only
locally AdS$_{p+1}$ black hole solution is the higher-dimensional
generalisation of the non-rotating BTZ black hole, discussed
previously in \cite{ban1,ban2}.  We explain the origin of this
restriction in general.  We discuss the relation to other recent work
and comment on the proper interpretation of another solution presented
in \cite{ban2}.

We begin in Section~\ref{sec:review} by reviewing the classification
of quotients, setting up the notation that will be used in the
remainder of the paper, discussing Killing vectors on the sphere and
determining the conditions under which a discrete cyclic quotient of
$\AdS \times \Sph$ will admit a spin structure.
Section~\ref{sec:cause} explains the relation between the
classification of Killing vectors in $\AdS$ and the existence of
closed timelike curves in the resulting discrete quotients.  In
Section~\ref{sec:nonsingular} we discuss causally well-behaved
quotients, and Section~\ref{sec:bh} demonstrates that the only black
hole solution is the generalisation of the non-rotating BTZ black
hole.  We finish with a small digression on Penrose limits of discrete
quotients, and the relation between Gödel-type universes and some
quotients of $\AdS$ having closed timelike curves.  Some technical
details are relegated to the appendices.

\section{Conventions and background material}
\label{sec:review}

In this section we will briefly review the geometrical set up and the
results of \cite{fofs2,mr} in an attempt to make the present paper
self-contained.

\subsection{Anti-de~Sitter isometries}
\label{sec:adsisom}

The \emph{dramatis personae} of this paper are quotients of AdS
backgrounds, either of anti-de~Sitter space $\AdS_{p+1}$ itself in the
context of pure gravity, or of Freund--Rubin backgrounds of the form
$\AdS_{p+1} \times \Sph^q$ in supergravity and string theory.

As usual in Physics, throughout this paper $\AdS_{p+1}$ ($p\geq 2$)
shall denote the \emph{simply-connected} anti-de~Sitter space.  In
other words, $\AdS_{p+1}$ (with radius of curvature $R$) is the
universal cover of the quadric traced by the equation
\begin{equation}
  \label{eq:ads}
  -(x^1)^2 - (x^2)^2 + \sum_{i=3}^{p+2} (x^i)^2 = -R^2
\end{equation}
in the pseudo-euclidean space $\RR^{2,p}$ with coordinates
$(x^1,x^2,\dots,x^{p+2})$.  The isometry group of the quadric is
$\O(2,p)$, which acts linearly on $\RR^{2,p}$ and preserves the
quadric.  This is analogous to the case of the sphere $\Sph^q$ (of
radius of curvature $R$), which can be identified with the
corresponding quadric in the euclidean space $\RR^{q+1}$ and whose
group of isometries is $\O(q+1)$ acting linearly in $\RR^{q+1}$ and
preserving the quadric.  However, whereas the sphere (for $q>1$) is
simply-connected, the quadric \eqref{eq:ads} is not.  Indeed, its
fundamental group is $\ZZ$ if $p>2$ and $\ZZ \oplus \ZZ$ if $p=2$.
This means that although the isometry group of the quadric
\eqref{eq:ads} is $\O(2,p)$, that of $\AdS_{p+1}$ is a non-trivial
central extension by $\ZZ$ or $\ZZ\oplus\ZZ$.

In string theory, Freund--Rubin backgrounds of the form $\AdS_{p+1}
\times \Sph^q$ are not fully specified by the geometry alone, but
require in addition specifying fluxes, which in these backgrounds
coincide with the volume forms of the relevant factors.  In other
words, both factors come with orientation.  This means that the
symmetries of a Freund--Rubin background are the
orientation-preserving isometries of the underlying geometries.  For
$\Sph^q$ this is the Lie group $\SO(q+1)$, whereas for $\AdS_{p+1}$ it
is the infinite cover of $\SO(2,p)$ obtained by centrally extending
this group by the fundamental group of the quadric, as explained in
\cite[Section~5.1.2]{fofs2}.  We will denote this group by
$\widetilde{\SO(2,p)}$.  Annoyingly, it cannot be embedded in a matrix
group; that is, it does not admit any finite-dimensional faithful
linear representations.  Crucially, however, $\widetilde{\SO(2,p)}$
has two features in common with its quotient $\SO(2,p)$.  First of
all, they share the same Lie algebra $\fso(2,p)$ and furthermore,
since conjugation by central elements is trivial, the adjoint action
of $\widetilde{\SO(2,p)}$ on $\fso(2,p)$ factors through $\SO(2,p)$.
Similarly, the action of the spin cover $\widetilde{\Spin(2,p)}$ of
$\widetilde{\SO(2,p)}$ on the spinor representations factors through
$\Spin(2,p)$.  These happy facts allow a complete analysis of
one-parameter subgroups and also the determination of the
supersymmetry preserved by a quotient.

\subsection{One-parameter subgroups of isometries of $\AdS_{p+1}$}
\label{sec:so2n}

By definition, a one-parameter subgroup $\Gamma$ of a Lie group $G$ is
the image under the exponential map of a one-dimensional subspace of
its Lie algebra $\fg$.  In other words, $\Gamma$ consists of group
elements of the form $\exp(t X)$, where $t \in \RR$ and $X \in \fg$.
The topology of $\Gamma$ is either $\RR$ or $\Sph^1$, depending on
whether or not $\exp(t X)$ is the identity element in $G$ for some
nonzero $t$.  If $2\pi T>0$ is the smallest such $t$, then the
exponential map defines a diffeomorphism of the circle $\RR/2\pi T\ZZ$
with $\Gamma$, otherwise it defines a diffeomorphism of $\RR$ with
$\Gamma$.

Every one-parameter subgroup $\Gamma \subset G$ gives rise to an
infinite family (indexed by the subgroup itself) of discrete cyclic
subgroups $\Gamma_\gamma$ generated by an element $\gamma \in \Gamma$.
If $\gamma$ has infinite order, then $\Gamma_\gamma \cong \ZZ$,
whereas if the order is $N$, then $\Gamma_\gamma \cong \ZZ_N$.  All
infinite cyclic subgroups of $G$ \emph{in the image of the exponential map} are
obtained in this way.  In the cases when $\Gamma \cong \Sph^1$, we
will restrict our attention to elements $\gamma$ of finite order.
Quotienting a manifold $M$ on which $G$ acts by the action of
$\Gamma_\gamma$ consists in identifying points of $M$ which are
related by the action of $\gamma$.  Since $\gamma = \exp(\ell X)$ for
some $X \in \fg$ and some $\ell > 0$, quotienting by $\Gamma_\gamma$
consists in identifying points in $M$ which are related by flowing
along the integral curve of the Killing vector $\xi_X$ corresponding
to $X$ for a time $\ell$.

As explained, for example, in \cite{fofs}, if $\Gamma$ and $\Gamma'$
are conjugate subgroups of isometries of a space $M$, then their
quotients $M/\Gamma$ and $M/\Gamma'$ are isometric, the isometry being
induced from the isometry of $M$ which conjugates $\Gamma$ into
$\Gamma'$.  Therefore to classify such quotients $M/\Gamma$, it is
enough to classify subgroups up to conjugation.  For one-parameter
subgroups this corresponds to classifying adjoint orbits in the Lie
algebra $\fg$.  Furthermore, by reparametrising the subgroup if
needed, one can further projectivise the Lie algebra and declare
collinear elements as equivalent.

Therefore to classify conjugacy classes of one-parameter subgroups of
isometries of $\AdS_{p+1}$ for $p\geq 2$ it is equivalent to classify
equivalence classes of elements $X\in\fso(2,p)$ under
\begin{equation}
  \label{eq:equiv}
  X\sim t g X g^{-1} \quad \text{where $t\in\RR^\times$ and
    $g\in\SO(2,p)$.}
\end{equation}
Such a classification was established in \cite{fofs2,mr} and we review
it now.

Every $B \in \fso(2,p)$ defines a skew-symmetric endomorphism of
$\RR^{2,p}$, which we also denote by $B$.  Associated to each such
endomorphism there is an orthogonal decomposition
\begin{equation*}
  \RR^{2,p} = \VV_1 \oplus \dots \oplus \VV_k
\end{equation*}
into indecomposable nondegenerate subspaces stabilised by $B$; that
is, for each $i$, $B(\VV_i) \subset \VV_i$, the inner product
restricts non-degenerately to each $\VV_i$, and the restriction $B_i$
of $B$ to $\VV_i$ does not decompose further into nondegenerate
blocks.  Conversely out of such \emph{elementary blocks} $B_i$ one can
build the original endomorphism $B$.  In this way, the original
problem is essentially mapped into the classification of normal forms
of skew-symmetric endomorphisms of $\RR^{m,n}$ with $m\leq 2$ and
$n\leq p$ up to conjugation by isometries.  The latter are listed in
Table~\ref{tab:blocksSO}, where we found it convenient to identify the
endomorphism with the corresponding bilinear form, and to write these
in terms of the usual basis $\be_{ij} = \be_i \wedge \be_j$ for
$\Lambda^2 \RR^{2,p}$ consisting of wedge products of the elements of
the ordered frame $(\be_i)$, where $\be_1$, $\be_2$ denote the two
timelike directions, the remaining ones being spacelike. The superscript
$(m,n)$ on the elementary blocks specifies the subspace $\RR^{m,n}$
that they act on.
The Killing
vector in $\RR^{2,p}$ associated to the two-form
\begin{equation*}
  X = \half \sum_{i,j} B^{ij} \be_{ij} \in \Lambda^2\RR^{2,p} \cong
  \fso(2,p)
\end{equation*}
is given by
\begin{equation*}
  \xi_X = \half \sum_{i,j} B^{ij} (x_i \d_j - x_j \d_i) = \sum_{i,j}x^i
  B_i{}^j \d_j~.
\end{equation*}
It is clearly tangent to the quadric and it lifts to a Killing vector
field on $\AdS_{p+1}$ which we also denote $\xi_X$.

\begin{table}[h!]
  \centering
  \setlength{\extrarowheight}{3pt}
  \renewcommand{\arraystretch}{1.3}
    \begin{tabular}{|>{$}l<{$}|>{$}l<{$}|}\hline
      \multicolumn{1}{|c|}{Block} & \multicolumn{ 1}{c|}{Two-Form} \\
      \hline\hline
      B^{(0,2)}(\varphi) & \varphi\, \be_{34} \\
      B^{(1,1)}(\beta) & \beta \,\be_{13} \\
      B^{(2,0)}(\varphi) & \varphi \,\be_{12} \\
      B^{(1,2)} & \be_{13} - \be_{34} \\
      B^{(2,1)} & \be_{12} - \be_{23} \\
      B^{(2,2)}_\pm & \pm \be_{12} + \be_{13} \mp \be_{24} - \be_{34} \\
      B^{(2,2)}_\pm(\beta) &  \pm \be_{12} + \be_{13} \mp \be_{24} - \be_{34}
      + \beta (\be_{14} \mp \be_{23}) \\
      B^{(2,2)}_\pm(\varphi) & \pm \be_{12} + \be_{13} \mp \be_{24} - \be_{34}
      + \varphi (\pm \be_{12} + \be_{34}) \\
      B^{(2,2)}_\pm(\beta,\varphi) & \varphi (\pm \be_{12} - \be_{34})
      + \beta (\be_{14} \mp \be_{23}) \\
      B^{(2,3)} & \be_{12} - \be_{24} + \be_{13} - \be_{34} + \be_{15} - \be_{45} \\
      B^{(2,4)}_\pm (\varphi) & \be_{15} - \be_{35} \pm \be_{26} - \be_{46}
      + \varphi (\mp \be_{12} + \be_{34} + \be_{56}) \\
      \hline
    \end{tabular}
  \vspace{8pt}
  \caption{The elementary blocks as two-forms.}
  \label{tab:blocksSO}
\end{table}

Let us briefly discuss the interpretation of each of these elementary
blocks to help the reader get used to our notation.  We shall denote
boost parameters by $\beta$ and rotation parameters by $\varphi$.
There are three inequivalent two-dimensional elementary blocks: a
spacelike rotation $B^{(0,2)}(\varphi)$, a boost $B^{(1,1)}(\beta)$
and a timelike rotation $B^{(2,0)}(\varphi)$.  In three dimensions,
normal forms either reduce to the previous ones or preserve null
directions.  Since we work in non-Lorentzian signature, we must
distinguish among two different null rotations: a null rotation
$B^{(1,2)}$ involving two spacelike directions and a null rotation
$B^{(2,1)}$ involving two timelike directions.  There are four types
of non-trivial four-dimensional elementary blocks: a linear
combination $B^{(2,2)}_\pm$ of timelike and spacelike null rotations,
a deformation $B^{(2,2)}_\pm(\beta)$ of the latter by the addition of
a linear combination of boosts, a different deformation
$B^{(2,2)}_\pm(\varphi)$ involving the addition of a timelike rotation
and a spacelike rotation, and finally a linear combination
$B^{(2,2)}_\pm(\beta,\varphi)$ of two actions involving a timelike and
spacelike rotation with parameter $\varphi$ (up to signs) on one side
and a linear combination of boosts on the other side.  There is only
one five-dimensional elementary block, $B^{(2,3)}$, which can be
interpreted as the linear combination of a timelike null rotation and
two spacelike null rotations sharing the time direction and one of the
spacelike directions.  The last elementary block,
$B^{(2,4)}_\pm(\varphi)$, appears in six dimensions, and it consists
of a double spacelike null rotation acting on orthogonal subspaces,
deformed by a simultaneous rotation in the plane formed by the two
timelike directions and two orthogonal spacelike planes.

Let us remark the appearance of pairs of elementary blocks
$B^{(m,n)}_\pm$, with or without parameter, in the classification
in table \ref{tab:blocksSO}.  It can be checked that one element of the
pair is always mapped into the other by an orientation-reversing
transformation.  Therefore, no classification based on the isometry
group $\O(2,p)$ can distinguish between these objects.  Analogously,
orientation-reversing transformations act nontrivially in the
parameters $(\beta,\varphi)$ in those elementary blocks which do not
come in pairs, allowing us to restrict their range.  In this section,
we shall follow the $\SO(2,p)$ classification (unless otherwise
stated), but in the rest of the paper, when discussing the geometrical
interpretation of the different discrete quotients, we shall omit
these distinctions.  This is because the metric in
$\AdS_{p+1}$ is invariant under orientation reversing transformations,
therefore the geometry itself will not change among the members of the
pair.  The distinction will arise in the sign of the fluxes that
stabilise the classical configurations: the members of a pair will
have opposite sign fluxes.  This fact can certainly have consequences
concerning the supersymmetry preserved by the members of the pair.

The small number of elementary blocks notwithstanding, the taxonomy of
inequivalent discrete quotients increases quickly with dimension due
to the possibility of combining the action of different blocks acting
in orthogonal subspaces of $\RR^{2,p}$.  Lack of spacetime prevents us
from discussing all possible quotients in detail.  There are several
criteria which we could employ to narrow our choice of quotients.  For
example, we could focus on supersymmetric quotients,
everywhere-spacelike and non-singular quotients, etc.  Our primary
criterion will be that a quotient should have a well-behaved causal
structure: our subsequent discussion will focus on those discrete
quotients that are either free of closed timelike curves, or in which
the closed timelike curves are `expungeable', in the sense that a
spacetime free of closed timelike curves can be obtained by
quotienting only part of AdS, and that the boundary so introduced lies
behind a horizon.  In the latter case, the resulting causally
well-behaved singular spacetime is interpreted as an analogue of a
black hole, following \cite{btz1,btz2}.

The causal properties of the quotient are determined primarily by the
norm of the Killing vector field generating it.  It is therefore
important to study the norm of the Killing vectors associated with the
two-forms listed in Table~\ref{tab:blocksSO}.  These are given in
Table~\ref{tab:normsSO}, where the following notation is used.  We
write explicitly the coordinates $x_i$ of the subspace $\WW \subset
\RR^{2,p}$ on which the elementary blocks act non-trivially and write
$\bx_\perp$ for the coordinates of the perpendicular subspace
$\WW^\perp$.  The norm is defined on the quadric \eqref{eq:ads}, but
can be pulled back to functions on AdS which are invariant under the
deck transformations generated by the fundamental group of the
quadric.

\begin{table}[h!]
  \centering
  \setlength{\extrarowheight}{3pt}
  \renewcommand{\arraystretch}{1.3}
    \begin{tabular}{|>{$}l<{$}|>{$}l<{$}|}\hline
      \multicolumn{1}{|c|}{Block} & \multicolumn{1}{c|}{Norm} \\
      \hline\hline
      B^{(0,2)}(\varphi) & \varphi^2\, (x_3^2 + x_4^2) \\
      B^{(1,1)}(\beta) & \beta^2 \,(R^2 + \|\bx_\perp\|^2 -x_2^2) \\
      B^{(1,2)} & (x_1 + x_4)^2 \\
      B^{(2,0)}(\varphi) & -\varphi^2 \,(R^2 + \|\bx_\perp\|^2) \\
      B^{(2,1)} & -(x_1 + x_3)^2 \\
      B^{(2,2)}_\pm & 0 \\
      B^{(2,2)}_\pm(\beta) &  \beta^2\,(R^2 + \|\bx_\perp\|^2)
      + 4\beta\,(x_1+x_4)(x_3\pm x_2) \\
      B^{(2,2)}_\pm(\varphi) & -\varphi^2\,(R^2 + \|\bx_\perp\|^2)
      + 2\varphi\,\left(\left(x_1+x_4\right)^2 + \left(x_3 \pm x_2\right)^2
      \right) \\
      B^{(2,2)}_\pm(\beta,\varphi) & (\beta^2 - \varphi^2)(R^2 +
      \|\bx_\perp\|^2) - 4\beta\varphi\,\left(x_1x_3 \pm x_2x_4\right) \\
      B^{(2,3)} & (x_4 - x_1)^2 -4\left(x_2 + x_3\right) x_5 \\
      B^{(2,4)}_\pm (\varphi) & -\varphi^2\,(R^2 + \|\bx_\perp\|^2) +
      (x_1 - x_3)^2 + (x_4 \mp x_2)^2 \\
      & \hfill -4\varphi\left(\left(x_4 \mp x_2\right)x_5
      + \left(x_1 - x_3\right) x_6\right) \\
      \hline
    \end{tabular}
  \vspace{8pt}
  \caption{The elementary blocks and their norms.}
  \label{tab:normsSO}
\end{table}

We can see from Table~\ref{tab:normsSO} that some Killing vectors are
timelike in some regions of AdS, leading to closed timelike curves in
the associated discrete quotients.  Indeed, we see that for
$B^{(2,0)}(\varphi)$, $B^{(1,1)}(\beta)$, $B^{(2,1)}$, $B^{(2,2)}_\pm
(\beta)$, $B^{(2,2)}_\pm (\beta,\varphi)$, $B^{(2,3)}$, $B^{(2,4)}_\pm
(\varphi)$ and $B^{(2,2)}_\pm (\varphi < 0)$, the norm is not bounded
below.  For $B^{(2,2)}_\pm (\varphi > 0)$, the norm can be negative,
but is bounded from below; whereas for $B^{(0,2)}(\varphi)$ and
$B^{(1,2)}$ and $B^{(2,2)}_\pm$ the norm is always non-negative.

The Killing vector $\xi$ which generates the quotient will be the sum
of such elementary blocks and its norm on AdS will influence the
causal structure of the quotient.  We therefore consider the possible
endomorphisms in signature $(2,p)$ that can be constructed from
elementary blocks acting in orthogonal subspaces.  In
Tables~\ref{tab:positive}, \ref{tab:bounded} and \ref{tab:unbounded}
we classify them in terms of the norms of the associated Killing
vectors in AdS.  It should be stressed that even though we used the
notation adapted to an $\SO(2,p)$ classification, we have not
constrained the range of the different parameters appearing in these
endomorphisms.  For a complete discussion concerning these
constraints, we refer the reader to \cite{fofs2}.

\begin{table}[h!]
  \centering
  \setlength{\extrarowheight}{3pt}
  \renewcommand{\arraystretch}{1.3}
    \begin{tabular}{|>{$}l<{$}|}\hline
      \multicolumn{1}{|c|}{Endomorphism} \\
      \hline\hline
      \oplus_i B^{(0,2)}(\varphi_i) \\
      B^{(1,1)}(\beta_1) \oplus B^{(1,1)}(\beta_2) \oplus_i
      B^{(0,2)}(\varphi_i) \quad \text{\emph{if}
        $|\beta_1|=|\beta_2| > 0$} \\
      B^{(1,2)} \oplus_i B^{(0,2)}(\varphi_i) \\
      B^{(1,2)} \oplus B^{(1,2)} \oplus_i B^{(0,2)}(\varphi_i) \\
      B^{(2,2)}_\pm \oplus_i B^{(0,2)}(\varphi_i) \\
      \hline
    \end{tabular}
  \vspace{8pt}
  \caption{Killing vectors with everywhere non-negative norm}
  \label{tab:positive}
\end{table}

\begin{table}[h!]
  \centering
  \setlength{\extrarowheight}{3pt}
  \renewcommand{\arraystretch}{1.3}
    \begin{tabular}{|>{$}l<{$}|}\hline
      \multicolumn{1}{|c|}{Endomorphism} \\
      \hline\hline
      B^{(2,0)}(\varphi) \oplus_i B^{(0,2)}(\varphi_i) \quad
      \text{\emph{if} $p$ is even \emph{and} $|\varphi_i|
        \geq \varphi > 0$ \emph{for all} $i$} \\
      B^{(2,2)}_\pm(\varphi)  \oplus_i B^{(0,2)}(\varphi_i) \quad
      \text{\emph{if} $|\varphi_i| \geq |\varphi| \geq 0$
        \emph{for all} $i$} \\
      \hline
    \end{tabular}
  \vspace{8pt}
  \caption{Killing vectors allowing negative norm but bounded below}
  \label{tab:bounded}
\end{table}

\begin{table}[h!]
  \centering
  \setlength{\extrarowheight}{3pt}
  \renewcommand{\arraystretch}{1.3}
    \begin{tabular}{|>{$}l<{$}|}\hline
      \multicolumn{1}{|c|}{Endomorphism} \\
      \hline\hline
      B^{(1,1)}(\beta_1) \oplus B^{(1,1)}(\beta_2) \oplus_i
      B^{(0,2)}(\varphi_i) \quad \text{\emph{unless} $|\beta_1| = |\beta_2|
        > 0$} \\
      B^{(1,2)} \oplus B^{(1,1)}(\beta) \oplus_i B^{(0,2)}(\varphi_i) \\
      B^{(2,0)}(\varphi) \oplus_i B^{(0,2)}(\varphi_i) \quad
      \text{\emph{unless} $p$ is even \emph{and}
        $|\varphi_i| \geq |\varphi|$ \emph{for all} $i$}\\
      B^{(2,1)} \oplus_i B^{(0,2)}(\varphi_i) \\
      B^{(2,2)}_\pm(\beta) \oplus_i B^{(0,2)}(\varphi_i) \\
      B^{(2,2)}_\pm(\varphi)  \oplus_i B^{(0,2)}(\varphi_i) \quad
      \text{\emph{unless} $|\varphi_i| \geq \varphi > 0$
        \emph{for all} $i$} \\
      B^{(2,2)}_\pm(\beta,\varphi)  \oplus_i B^{(0,2)}(\varphi_i)  \\
      B^{(2,3)} \oplus_i B^{(0,2)}(\varphi_i)  \\ 
      B^{(2,4)}_\pm (\varphi) \oplus_i B^{(0,2)}(\varphi_i) \\
      \hline
    \end{tabular}
  \vspace{8pt}
  \caption{Killing vectors with norm unbounded below}
  \label{tab:unbounded}
\end{table}

The quotients generated by the Killing vectors in
Table~\ref{tab:unbounded} clearly contain closed timelike curves
corresponding to the very orbits of the Killing vector in regions
where it is timelike.  Furthermore, even when we consider quotients of
$\AdS_{p+1} \times \Sph^q$ by adding a nontrivial action on the
sphere, the resulting Killing vector will still be timelike somewhere,
so the quotients will still have closed timelike curves.  Therefore
the only way in which these quotients will enter into our discussion
is in asking whether any of them lead to `black hole' spacetimes.  We
shall discuss this issue in Section~\ref{sec:bh}.

The quotients generated by the Killing vectors in
Table~\ref{tab:bounded} also clearly contain closed timelike curves.
This time however the Killing vector can be made everywhere spacelike
by adding a suitable action on an odd-dimensional sphere.  However, we
will show in the next section that this is not sufficient to ensure
the absence of closed timelike curves.  Therefore the quotients of
$\AdS_{p+1} \times \Sph^q$ associated to the Killing vectors in this
table will not lead to causally regular quotients either.  In summary,
the only quotients we will consider in Section~\ref{sec:nonsingular},
where we discuss causally non-singular quotients, are those in
Table~\ref{tab:positive}.

\subsection{Infinitesimal isometries of spheres}
\label{sec:sphere}

Here we set up the notation to describe the Killing vectors on
spheres.  For this purpose, we find convenient to identify the
$q$-sphere of radius $R$ with the quadric traced by
\begin{equation}
  \sum_{i=1}^{q+1} x_i^2 = R^2
 \label{eq:sphere}
\end{equation}
in $\RR^{q+1}$.  This has the virtue that the isometry group of the
quadric, $\O(q+1)$ acts linearly in the ambient euclidean space.  As
we did for $\AdS_{p+1}$, we shall restrict this group to the subgroup
$\SO(q+1)$ which preserves the orientation.

The conjugacy theorem for Cartan subalgebras of $\fso(q+1)$ allows us
to bring any Killing vector $\xi_S$ on $\Sph^q$ to the form
\begin{equation}
   \label{eq:ksphere}
   \xi_S = \sum_{i=1}^r \theta_i\,R_{2i-1,2i}~,
 \end{equation}
 where $r=\lfloor\frac{q+1}2\rfloor$, $R_{ij}$ stands for a rotation
in the $ij$-plane and the $\theta_i$ are real parameters specifying
the rotation angles.  This still leaves the freedom to conjugate by
the Weyl group, which we can fix by arranging the parameters in such a
way that
\begin{equation*}
  \theta_1 \geq \theta_2 \geq \dots \geq |\theta_r|~.
\end{equation*}
For odd-dimensional spheres, Killing vectors with all $\theta_i \neq
0$ are everywhere nonvanishing, whereas in even-dimensional spheres
every vector field, Killing or not, has a zero.

It will be convenient in what follows to construct a coordinate system
for $\Sph^q$ adapted to a given Killing vector $\xi_S$; that is, one
in which $\xi_S=\partial_\psi$.  Let us describe in detail the case of
even-dimensional spheres.  First, rewrite \eqref{eq:sphere} as
\begin{equation}
  \label{eq:esphere}
  \sum_{i=1}^{r} |z_i|^2 + (x_{2r+1})^2 = R^2~,
\end{equation}
in which we introduce $r$ complex coordinates for the two-planes where
the action of \eqref{eq:ksphere} may be non-trivial.  A natural way to
solve \eqref{eq:esphere} is by
\begin{equation}
 \label{eq:espherecoor}
  \begin{aligned}[m]
    x_{2r+1} & = R\,\cos\theta \\
    z_i & = R\,\sin\theta\,\rho_i\,e^{i\varphi_i} \quad \text{where}
    \quad \sum_{i=1}^{r} \rho_i^2 = 1~.
  \end{aligned}
\end{equation}
It is clear that in coordinates $\{\theta,\rho_i,\varphi_i\}$,
\begin{equation*}
  \xi_S = \sum_{i=1}^{r}\theta_i\,\partial_{\varphi_i}~,
\end{equation*}
whence by a linear transformation in the space $\{\varphi_i\}$ we can
rewrite $\xi_S$ as $\partial_\psi$.  Indeed, assume $\theta_1\neq 0$,
and consider
\begin{equation}
\begin{aligned}[m]
  \psi &= \theta_1^{-1}\,\varphi_1~, \\
  \tilde{\varphi}_i & = \varphi_i - \theta_i\,\theta_1^{-1}\,\varphi_1
  \quad i=2,\dots ,r ~.
\end{aligned}
\end{equation}
By construction, $\xi_{S}$ becomes $\partial_\psi$.

The case of odd-dimensional spheres follows formally from the above by
setting $\theta=\pi/2$ in the above expressions.

\subsection{Spin structures and supersymmetry}
\label{sec:spin}

A supergravity background must admit a spin structure, since the
fermionic fields, although set to zero in a classical background, and
the supersymmetry parameters are sections of (possibly twisted) spinor
bundles.  This is not necessarily the case in string/M-theory as the
phenomenon of `supersymmetry without supersymmetry' illustrates
\cite{DLP,DuffLuPopeUnt,PSS}.  This has been recently discussed in
\cite{HullHolonomy} and in the present context of quotients in
\cite{fofs2}.  We will add nothing to this discussion here.  Indeed,
as in \cite{fofs2}, we will adopt a conservative point of view and
require the underlying spacetime of a supergravity background to be
spin and will only consider supersymmetries which are realised
geometrically as Killing spinors.

A natural question in this context is then the following. Let 
$(M,g,\dots)$ be a supergravity background with $(M,g)$ a Lorentzian
spin manifold and $\Gamma$ a discrete (cyclic) group of
orientation-preserving isometries acting freely and properly
discontinuously on $M$ (so that the quotient $M/\Gamma$ is smooth).
When will $M/\Gamma$ be spin?  Furthermore, if $(M,g,\dots)$ is a
supersymmetric background, how much supersymmetry (if any at all) will
the quotient preserve?  These questions were answered in
\cite{fofs2} for the case of $\Gamma$ a one-parameter group: in
principle for an arbitrary background, and explicitly for
Freund--Rubin backgrounds of the form $\AdS_{p+1} \times \Sph^q$.

If $\Gamma$ is a one-parameter group of isometries (hence
automatically orientation-preserving) acting freely on a spin manifold
$M$ with smooth quotient $M/\Gamma$, then $M/\Gamma$ is spin if and
only if the action of $\Gamma$ on the bundle $P_{\SO}(M)$ of oriented
orthonormal frames lifts to an action on the spin bundle
$P_{\Spin}(M)$ in such a way that the natural surjection
\begin{equation*}
  \theta: P_{\Spin}(M) \to P_{\SO}(M)
\end{equation*}
is $\Gamma$-equivariant.  In this case, the spin bundle
$P_{\Spin}(M/\Gamma)$ on the quotient is given by
\begin{equation*}
  P_{\Spin}(M/\Gamma) := P_{\Spin}(M)/\Gamma~.
\end{equation*}
Indeed, equivariance guarantees that this bundle covers
\begin{equation*}
  P_{\SO}(M/\Gamma) := P_{\SO}(M)/\Gamma
\end{equation*}
twice and agrees fibrewise with the spin cover of the special
orthogonal group.

The same is true for $\Gamma$ a discrete group acting freely and
properly discontinuously on a spin manifold $M$.  For a general spin
manifold $M$, it is not easy to determine when the action of $\Gamma$
on $P_{\SO}(M)$ lifts equivariantly to the spin bundle; however, as
explained in \cite{fofs2}, for backgrounds of the form
\begin{equation*}
  M = \AdS_{p+1} \times \Sph^q
\end{equation*}
we fare much better.  Indeed, for this geometry the criterion for the
existence of a spin structure in $M/\Gamma$ translates into a simple
calculation in a Clifford algebra.

For simplicity we will consider discrete cyclic groups generated by an
element $\gamma$ in the image of the exponential map $\exp: \fg \to
G$ between the the Lie algebra and Lie group of
(orientation-preserving) isometries of $M$; that is,
\begin{equation*}
  \gamma = \exp(\ell X)
\end{equation*}
for some $X\in\fg$ and some $\ell > 0$.  Then $\Gamma$ acts on the
(unique) spin bundle on $\AdS_{p+1} \times \Sph^q$ if and only if
$\Gamma$ embeds isomorphically in $\Spin(2,p) \times_{\ZZ_2}
\Spin(q+1) \subset \Cl(2,p+q+1)$.  Since $\Gamma$ is generated by
$\gamma$, this is a simple criterion:  does there exist
\begin{equation*}
  \widetilde\gamma \in \Spin(2,p) \times_{\ZZ_2} \Spin(q+1) \subset
  \Cl(2,p+q+1)
\end{equation*}
which lifts $\gamma$ and which has the same order?

The element $\gamma$ has two possible lifts $\pm \widetilde\gamma$.
If $\gamma$ has infinite order, so that $\Gamma \cong \ZZ$, then so
does $\widetilde\gamma$ and thus it also generates a group
$\widetilde\Gamma \cong \ZZ$ which therefore covers $\Gamma$
isomorphically.  Therefore, if $\Gamma \cong \ZZ$, the quotient
\begin{equation*}
  \left( \AdS_{p+1} \times \Sph^q \right)/\Gamma
\end{equation*}
is spin.

Now suppose that $\gamma$ has finite order $N$.  Then all we know is
that $(\pm \widetilde\gamma)^N$ covers the identity, whence
\begin{equation*}
  (\pm \widetilde\gamma)^N  = \pm \1~,
\end{equation*}
and the question is whether there exists a choice of lift such that $
(\pm \widetilde\gamma)^N  = \1$.

Clearly if $N$ is odd, then either $(\widetilde\gamma)^N = \1$ or $(-
\widetilde\gamma)^N = \1$, whence if $\Gamma \cong \ZZ_N$, $N$ odd,
the quotient is spin.

The only possible obstruction arises when $N$ is even.  In this case
the choice of lift is immaterial and either $\widetilde\gamma^N = \1$
or $\widetilde\gamma^N = -\1$ and one needs to do a calculation to
settle this issue.

This obstruction arises only if $\gamma = \exp(\ell X)$ for $\ell > 0$
and
\begin{equation*}
  X = \varphi_1 \be_{34} + \cdots + \varphi_r \be_{2r+1,2r+2} +
  \theta_1 R_{12} + \cdots + \theta_s R_{2s-1,2s}~,
\end{equation*}
where $r = \lfloor\frac{p-1}{2}\rfloor$ and $s =
\lfloor\frac{q+1}{2}\rfloor$.  Let $\gamma = \exp(\ell X)$.  Then
$\gamma$ has order $N$ if and only if
\begin{equation*}
  \ell \varphi_i = \frac{2\pi n_i}{N} \qquad \text{and} \qquad
  \ell \theta_j = \frac{2\pi m_j}{N}~,
\end{equation*}
where $n_i$, $m_j$ are integers with
\begin{equation*}
  \gcd(n_1,\dots,n_r,m_1,\dots,m_s) = 1~.
\end{equation*}
This last condition ensures that the order of $\gamma$ is precisely
$N$ and not a smaller divisor.  Let $\gamma_i$ and $\Gamma_i$ be the
gamma matrices for $\Cl(2,p)$ and $\Cl(q+1)$, respectively, embedded
in
\begin{equation*}
  \Cl(2,p+q+1) \cong \Cl(2,p) \widehat\otimes \Cl(q+1)~,
\end{equation*}
where $\widehat\otimes$ denotes the $\ZZ_2$-graded tensor product.
Then the two lifts of $\gamma$ in
\begin{equation*}
  \Spin(2,p) \times_{\ZZ_2} \Spin(q+1) \subset \Cl(2,p+q+1)
\end{equation*}
are given by $\pm \widetilde\gamma$, where
\begin{multline*}
  \widetilde\gamma = \left(\1 \cos\frac{\ell\varphi_1}2 +
    \gamma_{34} \sin \frac{\ell\varphi_1}2\right) \cdots 
  \left(\1 \cos\frac{\ell\varphi_r}2 + \gamma_{2r+1,2r+2} \sin
    \frac{\ell\varphi_r}2\right) \\
  \times \left(\1 \cos\frac{\ell\theta_1}2 +
    \Gamma_{12} \sin \frac{\ell\theta_1}2\right) \cdots 
  \left(\1 \cos\frac{\ell\theta_s}2 + \Gamma_{2s-1,2s} \sin
    \frac{\ell\theta_s}2\right)~,
\end{multline*}
whence
\begin{multline*}
  \widetilde\gamma^N = \left(\1 \cos\frac{N\ell\varphi_1}2 +
    \gamma_{34} \sin \frac{N\ell\varphi_1}2\right) \cdots 
  \left(\1 \cos\frac{N\ell\varphi_r}2 + \gamma_{2r+1,2r+2} \sin
    \frac{N\ell\varphi_r}2\right) \\
  \times \left(\1 \cos\frac{N\ell\theta_1}2 +
    \Gamma_{12} \sin \frac{N\ell\theta_1}2\right) \cdots 
  \left(\1 \cos\frac{N\ell\theta_s}2 + \Gamma_{2s-1,2s} \sin
    \frac{N\ell\theta_s}2\right)~.
\end{multline*}
Using now that $N\ell\varphi_i = 2 \pi n_i$ and $N\ell\theta_j = 2\pi
m_j$, this evaluates to
\begin{equation*}
  \widetilde\gamma^N = (-1)^{n_1 + \dots + n_r + m_1 + \dots + m_s} \1~.
\end{equation*}
Therefore we conclude that when $\Gamma \cong \ZZ_N$, $N$ even, the
quotient is spin if and only if
\begin{equation*}
  \sum_{i=1}^r n_i + \sum_{j=1}^s m_j \quad \text{is even.}
\end{equation*}

\section{Causal properties of $\AdS_{p+1}$ quotients and their
  deformations}
\label{sec:cause}

In Section~\ref{sec:so2n} we reviewed the classification of
one-parameter subgroups of isometries of $\AdS_{p+1}$.  We divided
these into three different subsets according to whether
\begin{itemize}
\item the norm of the associated Killing vector field is non-negative
  (Table~\ref{tab:positive});
\item the norm can take negative values, but is bounded below 
  (Table~\ref{tab:bounded}); and
\item the norm can take arbitrarily negative values
  (Table~\ref{tab:unbounded}).
\end{itemize}
As explained above, this distinction is important in the context of
Freund--Rubin backgrounds of the form $\AdS_{p+1} \times \Sph^q$,
since the spherical component of the Killing vector can in some cases
render its norm positive everywhere.  Indeed, odd-dimensional spheres
admit Killing vectors whose norm is pinched away from zero, whence the
total Killing vector
\begin{equation}
  \label{eq:defkilling}
  \xi = \xi_{\AdS} + \xi_{\Sph}
\end{equation}
may be spacelike even if $\xi_{\AdS}$ is not.  This can only happen if
the norm of $\xi_{\AdS}$ is bounded below, since the norm of
$\xi_{\Sph}$ is bounded above by compactness of $\Sph^q$.

In this section, we will explain in detail the connection between this
classification and the appearance of closed timelike curves in
quotients involving these Killing vectors.

If we were just considering quotients of AdS, of course, the
connection would be immediate.  Indeed, the quotient consists in
identifying points which are obtained by flowing along the integral
curves of $\xi_{\AdS}$ for some time $\ell > 0$.  Let $\xi_{\AdS}$ be
timelike in a nonempty region $D \subset \AdS_{p+1}$ and let $x \in
D$.  Since the norm of $\xi_{\AdS}$ is constant along its integral
curves, the integral curve passing through $x$ is timelike and hence
lies in $D$.  Therefore the point $\gamma \cdot x$ is also in $D$ and
the segment of the integral curve from $x$ to $\gamma\cdot x$ becomes,
in the quotient, a closed timelike curve.  A similar argument shows
that the quotient has closed null curves in the region of
$\AdS/\Gamma$ where $\xi_{\AdS}$ is null.

The situation for quotients of $\AdS_{p+1} \times \Sph^q$ is similar.
Indeed, the same argument as for quotients of $\AdS$ shows that if
$\xi = \xi_{\AdS} + \xi_{\Sph}$ is not everywhere spacelike, then any
associated discrete cyclic quotient will have closed causal curves.

How about if $\xi$ is everywhere spacelike?  The property of being
spacelike everywhere is a necessary condition for the absence of
closed causal curves, but it is certainly not sufficient (see
\cite{causal} for another example where it fails to be sufficient and
a statement of a sufficient condition, and \cite{MaozSimon} for a
discussion on this topic and its relation with U-duality).  Indeed, we
will show presently that even when $\xi$ is everywhere spacelike, if
$\xi_{\AdS}$ is timelike in some region $D \subset \AdS_{p+1}$, then
any discrete cyclic quotient associated to $\xi = \xi_{\AdS} +
\xi_{\Sph}$ will have closed timelike curves in the region $(D \times
\Sph^q)/\Gamma$ of the quotient.  The key point in the argument is to
exploit the fact that the sphere has bounded diameter in order to
construct a timelike curve between two points identified by the action
of $\Gamma$ which, as in \cite{causal}, is different from the integral
curve of $\xi$.

\begin{figure}[h!]
  \begin{center}
    \begin{picture}(175,70)(15,15)
      \thinlines
      \put(20,25){\vector(1,0){105}}
      \put(25,20){\vector(0,1){55}}
      \put(27,75){$\theta$}
      \put(120,20){$\tau$}
      \put(20,64){$2\pi$}
      \drawline[-100](25,65)(120,65)
      \put(35,38.3333){\circle*{1}}
      \put(30,37.3333){$x$}
      \put(85,51.6666){\circle*{1}}
      \put(87,50.6666){$\gamma^N \cdot x$}
      \drawline(30,25)(45,65)
      \drawline(45,25)(60,65)
      \drawline(60,25)(75,65)
      \drawline(75,25)(90,65)
      \drawline(90,25)(105,65)
      \dottedline(27,30.3333)(43,46.3333)
      \dottedline(27,46.3333)(43,30.3333)
      \dottedline(78,43.6666)(93,59.6666)
      \dottedline(78,59.6666)(93,43.6666)
      \thicklines
      \drawline(35,38.3333)(85,51.6666)
    \end{picture}
    \caption{Closed timelike curve in a discrete quotient of the Lorentzian
      cylinder.  The dotted lines represent the ``lightcones'' at $x$
      and at $\gamma^N \cdot x$.  Notice that although the orbit of
      $\xi$ is spacelike, the straight line between $x$ and $\gamma^N
      \cdot x$ is timelike.}
    \label{fig:cylinder}
  \end{center}
\end{figure}
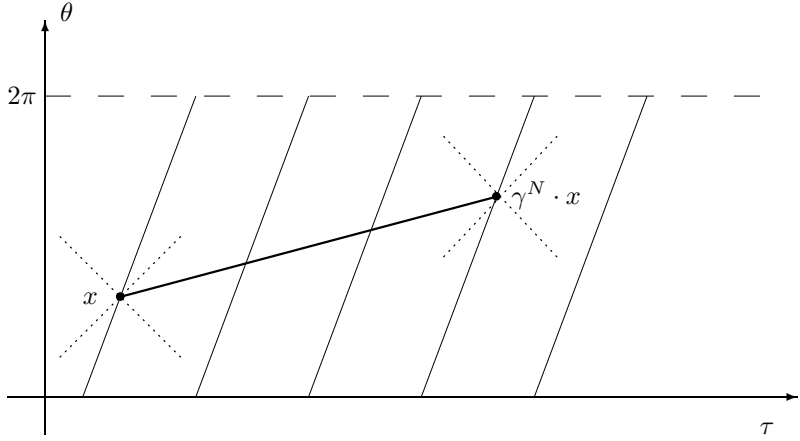

Let us first illustrate this construction with a simple example, which
is depicted in Figure~\ref{fig:cylinder}.

Let $C = (\RR/2\pi\ZZ) \times \RR$ denote a Lorentzian cylinder
coordinatised by $(\theta,\tau)$ and flat metric $d\theta^2 -
d\tau^2$.  Let $\xi = \d_\theta + \alpha \d_\tau$ be a spacelike
Killing vector, so that $\alpha^2 < 1$.  The integral curve of $\xi$
through a point $(\theta_0,\tau_0)$ is the curve
\begin{equation*}
  t \mapsto (\theta_0 + t, \tau_0 + \alpha t)~.
\end{equation*}
Let us define an action of $\ZZ$ on $C$, generated by the operation of
flowing along the integral curves of $\xi$ for a time $\ell > 0$:
\begin{equation*}
  (\theta,\tau) \mapsto (\theta + \ell, \tau + \alpha\ell)~.
\end{equation*}
Consider the two points $(\theta,\tau)$ and $(\theta + N\ell, \tau +
\alpha N \ell)$, which are identified in the quotient $C/\ZZ$.
The geodesic joining this point to $(\theta, \tau)$ is the straight line
\begin{equation*}
  t \mapsto ([\theta + t N \ell], \tau + \alpha N \ell)~,
\end{equation*}
where $[-]$ denotes the residue modulo $2\pi$.  The norm of the
velocity of this curve is therefore
\begin{equation*}
  [N\ell]^2 - N^2 \alpha^2 \ell^2 \leq 4\pi^2 - N^2 \alpha^2 \ell^2~,
\end{equation*}
which is clearly negative for $N$ large enough.  This curve is
therefore a closed timelike curve in the quotient $C/\ZZ$.

Now let us go back to the general case.  Let $\gamma = \exp(\ell X)$
for some $X \in \fg$ and $\ell > 0$, and let $\xi = \xi_{\AdS} +
\xi_{\Sph}$ be the Killing vector corresponding to $X$, with
$\xi_{\AdS}$ timelike in some nonempty region $D \subset \AdS_{p+1}$.
Let $x \in D \times \Sph^q$.  Since the norms of each component
$\xi_{\AdS}$ and $\xi_{\Sph}$ are separately conserved along the
integral curves of $\xi$, these belong to $D \times \Sph^q$, and hence
so does $\gamma \cdot x$.  For those Killing vectors with AdS
component in Table~\ref{tab:bounded}, the associated discrete cyclic
groups $\Gamma$ have infinite order, so we can consider points $x$ and
$\gamma^N \cdot x$ for $N$ arbitrarily large, which will give rise to
the same point in the quotient.  We will construct a curve
\begin{equation*}
  c : [0,N\ell] \to \AdS_{p+1} \times \Sph^q
\end{equation*}
between $c(0) = x = (x_{\AdS},x_{\Sph})$ and $c(N\ell) = \gamma^N
\cdot x = ((\gamma^N\cdot x)_{\AdS},(\gamma^N\cdot x)_{\Sph})$ which
will be timelike for $N$ sufficiently large and hence becomes a closed
timelike curve in the quotient.

The curve $c$ is uniquely specified by its two components: $c_{\AdS}$
on $\AdS_{p+1}$ and $c_{\Sph}$ on $\Sph^q$.  We will take $c_{\AdS}$
to be the integral curve of $\xi_{\AdS}$, and $c_{\Sph}$ to be a
minimum-length geodesic between $x_{\Sph}$ and $(\gamma^N \cdot
x)_{\Sph}$.  Let $L$ denote the diameter of the sphere; that is, the
supremum of the geodesic distances between any two points.  Then the
arc-length along $c_{\Sph}$ satisfies
\begin{equation*}
  \int_0^{N\ell} \|\dot c_{\Sph}\| dt = N \ell \|\dot c_{\Sph}\| \leq L~,
\end{equation*}
where the equality is because $\|\dot c_{\Sph}\|$ is constant along
$c_{\Sph}$ and the inequality is because $c_{\Sph}$ is
length-minimising.  Therefore,
\begin{equation*}
  \|\dot c\|^2 = \|\dot c_{\AdS}\|^2 + \|\dot c_{\Sph}\|^2 \leq
  \|\xi_{\AdS}\|^2  +  \frac{L^2}{N^2\ell^2}~,
\end{equation*}
which is negative in $D \times \Sph^q$ for $N$ large enough.

Let us remark that this argument applies to any Freund--Rubin
background of the form $\AdS \times N$, or more generally $M \times
N$, with $M$ Lorentzian admitting such isometries, at least when $N$
is complete.  Indeed, the supergravity equations of motion force $N$
to be Einstein with positive scalar curvature.  By the Bonnet--Myers's
theorem (see, e.g., \cite[Section~9.3]{doCarmo}), if $N$ is complete,
then it has bounded diameter.

This leaves the cases in Table~\ref{tab:positive}, where the AdS
Killing vector is nowhere timelike.  It is clear that the above
argument for closed timelike curves fails in this case.  One should
note that this still does not directly imply the absence of closed
timelike curves; however, we will see in the next section that there
are in fact no closed timelike curves in any of these cases.

We should also note that in the cases where the Killing vector is null
somewhere, namely $\oplus_i B^{(0,2)}(\varphi_i)$, $B^{(1,2)} \oplus_i
B^{(0,2)}(\varphi_i)$ and $B^{(2,2)}_\pm \oplus_i
B^{(0,2)}(\varphi_i)$, we can use a similar argument to see that
\emph{some} quotients of $\AdS_{p+1} \times \Sph^q$ still produce
closed causal curves.  The point is that if we choose $\ell$ such that
$\exp(\ell X_{\Sph}) \in \SO(q+1)$ has order $N$, then $x$ and $x' =
\gamma^N \cdot x$ can be null separated, as $x'_{\Sph} = x_{\Sph}$,
and the separation in the AdS factor is null if $\|\xi_{\AdS}\| = 0$
at $x$.  Physically, this corresponds to deforming by a rotation with
rational angles on $\Sph^q$.

Clearly, however, deformations for which $\gamma_{\Sph}$ does not have
finite order do exist, and will not lead to closed causal curves by
any of our arguments above.  Hence, we should discuss all the cases
listed in Table~\ref{tab:positive} in the next section, as they can
all give rise to causally non-singular quotients.

\section{Causally non-singular quotients}
\label{sec:nonsingular}

In this section, we shall discuss in detail the geometry of the
discrete quotients that are free of closed causal curves.  These are
based on the two-forms listed in table \ref{tab:positive},
conveniently deformed when necessary by some non-trivial action on an
odd sphere leaving no invariant directions, so that the full Killing
vector field \eqref{eq:defkilling} is spacelike everywhere.

Before initiating such a task, we would like to comment on the general
philosophy that we shall apply in each of the particular geometries to
be discussed.  Just by inspection of table \ref{tab:positive}, we know
that given any two-form in that list, we can study the geometry of the
corresponding discrete quotient in different dimensional AdS 
spacetimes, starting with the minimal $(n,m)$ signature in the
embedding space $\RR^{(n,m)}$ that allows the action of the
corresponding decomposable block.  Besides that, we can also study
further deformations on the sphere sector of the discrete quotient.
It is therefore natural to start our analysis in the lowest
dimensional $\AdS_{p+1}\times\Sph^q$ spacetime allowing our causally
non-singular quotients, and afterwards, extend such an analysis to
higher dimensions.

This latter extension is entirely straightforward.  Indeed, given some
adapted coordinate system describing the action of $\xi_{\AdS}$ in
$\AdS_{n+1}$, it is very simple to construct an adapted coordinate
system describing the action of the same Killing vector field in
$\AdS_{p+1}$ with $p>n$.  This is just obtained by considering
the standard $\AdS_{n+1}$ foliation of $\AdS_{p+1}$ given in terms
of the embedding coordinates by\footnote{In the following, we shall
set the radius of curvature R to one.}
\begin{equation}
  \begin{aligned}[m]
    x^i &= \,\cosh\chi\,\hat{x}^i \quad i=1,\dots ,n+2 \\
    x^m &= \,\sinh\chi\,\hat{x}^m \quad m=1,\dots p-n
  \end{aligned}
 \label{eq:adsslicing}
\end{equation}
where $\chi$ is non-compact and $\{\hat{x}^{i}\}$ satisfy the quadric
defining relation giving rise to $\AdS_{p+1}$, whereas $\{\hat{x}^m\}$
parametrise an $\Sph^{p-n-1}$ sphere of unit radius.  For $p=n+1$, 
the range of $\chi$ is given by $-\infty <
\chi < +\infty$, whereas for $p-n\geq 2$, it is simply given by
$\chi\geq 0$.  The metric description of $\AdS_{p+1}$ in the
$\AdS_{n+1}$ foliation defined in \eqref{eq:adsslicing} is
\begin{equation}
  g_{\AdS_{p+1}} = (\cosh\chi)^2\,g_{\AdS_{n+1}}+ (d\chi)^2
  + (\sinh\chi)^2\,g_{\Sph^{p-n-1}}~.
 \label{eq:adsk}
\end{equation}

The foliation given by (\ref{eq:adsk}) also gives us an interesting
description of the asymptotic boundary.  If we assume $p-n\geq2$,
taking the limit $\chi \to \infty$ and conformally rescaling by a
factor of $e^{-2 \chi}$, we can describe the asymptotic boundary in
terms of an AdS$_{n+1} \times \Sph^{p-n-1}$ metric,\footnote{For
  $p-n=1$, we would have $-\infty < \chi < \infty$, and conformally
  rescaling by a factor of $e^{-2 |\chi|}$ as we take the limit
  $|\chi| \to \infty$, we would get a description of the boundary in
  terms of two AdS$_{p}$ patches, each covering one of the hemispheres
  of the $\Sph^{p-1}$ in the usual Einstein static universe $\RR \times
  \Sph^{p-1}$ description of the boundary of AdS$_{p+1}$.}
\begin{equation}
  g_{\partial} = g_{\rm AdS_{n+1}} + g_{\Sph^{p-n-1}}~.
 \label{eq:adsbdy}
\end{equation}
To see the relation of this coordinate system to the usual Einstein
static universe description of the conformal boundary, let us write
the AdS$_{n+1}$ metric in global coordinates,
\begin{equation}
  g_{\rm AdS_{n+1}} = -\cosh^2 \rho\, dt^2 + d\rho^2 + \sinh^2
  \rho\,g_{\Sph^{n-1}}~.
\end{equation}
Then defining $\cos \theta = 1/\cosh \rho$, we can rewrite
\eqref{eq:adsbdy} as
\begin{equation}
  g_{\partial} = \frac{1}{\cos^2 \theta} (-dt^2 + d\theta^2 +
  \sin^2 \theta g_{\Sph^{n-1}} + \cos^2 \theta g_{\Sph^{p-n-1}}).
\end{equation}
This shows that the metric in \eqref{eq:adsbdy} is indeed conformal to
the Einstein static universe metric on $\RR \times \Sph^{p-1}$, where
we are writing the $\Sph^{p-1}$ as an $\Sph^{p-n-1}$ fibred over an
$\Sph^n$.  The coordinates of (\ref{eq:adsbdy}) cover all of the
Einstein static universe apart from the $\RR \times \Sph^{n-1}$
submanifold where $\cos \theta = 0$, which is conformally rescaled to
become the boundary of the AdS$_{n+1}$ factor in \eqref{eq:adsbdy}.

If there is a global adapted coordinate system for the action of
$\xi_{\AdS}$ on AdS$_{n+1}$, we can use the above foliation to
construct an adapted coordinate system for the action on AdS$_{p+1}$.
If we deform the action by $B^{(0,2)}(\varphi_i)$ blocks, these will
act as rotations of the $\Sph^{p-n-1}$ factor in the above foliation.

When we consider the deformation of our AdS quotient by some
non-trivial action on the transverse sphere, we have two approaches to
the construction of an overall adapted coordinate such that the total
Killing vector $\xi = \partial_\varphi$ for some coordinate $\varphi$.
In most of the cases we consider\footnote{The only exceptions are
  where the AdS Killing vector has fixed points.}, there is a globally
well-defined adapted coordinate on AdS$_{p+1}$ such that $\xi_{\AdS} =
\partial_\phi$.  As noted in Section~\ref{sec:sphere}, there is always
a global adapted coordinate system for the Killing vectors in the
sphere, in which $\xi_{\Sph}$ acts by a simple ``translation'', i.e.
$\xi_{\Sph} = \partial_\psi$.  Consequently, the full generator of the
discrete quotient is
\begin{equation}
  \xi = \partial_\phi + \gamma \partial_\psi,
\end{equation}
By a linear transformation, $\varphi = \phi, \psi^\prime= \psi -
\gamma \phi$, we are able to write $\xi = \partial_{\varphi}$.  This
coordinate system is very convenient for studying the causal structure
and asymptotic structure of the resulting quotient, so this is the
technique we shall mostly employ.

Unfortunately, there are examples where there is no such global
adapted coordinate system on AdS.  The example of this type we shall be
concerned with is the quotient by a Killing vector with a single
$B^{(1,2)}$ block.  In this case, we need to use a different technique,
exploiting the existence of adapted coordinates on the sphere.  The
full Killing vector field \eqref{eq:defkilling} can always be written
as
\begin{equation}
  \xi = \partial_\psi + \xi_{\AdS}~.
\end{equation}
We can therefore write $\xi$ as a dressed version of its
``translation'' component according to
\begin{equation}
  \xi = U\,\partial_\psi\,U^{-1} \qquad \text{where}\quad U =
  \exp\left(-\psi\,\xi_{\AdS}\right)~.
 \label{eq:dressing}
\end{equation}
Consequently, if the original coordinate system was given by
$\{\psi,z^l\}$, where $z^l$ stand for all the remaining coordinates
describing the manifold $\AdS_{p+1}\times \Sph^{q}$, it is natural to
change coordinates to an adapted coordinate system defined by
\begin{equation}
  y = U\,z~,
 \label{eq:adress}
\end{equation}
which indeed satisfies the property $\xi\, y = 0$, so that $\{y^l\}$
are good coordinates for the space of orbits.  Equivalently, $\xi =
\partial_\psi$ in the coordinates \eqref{eq:adress}.  Thus, we obtain
an adapted coordinate system on the full quotient for any AdS Killing
vector.  For the case at hand, we split the coordinates $\{z^l\}$
appearing in the above discussion into
$\{z^l\}=\{\tilde{\varphi}_i,\vec{x}\}$, where $\{\vec{x}\}$ stand
for the embedding coordinates of $\AdS_{p+1}$ in $\RR^{2,p}$.  Since
$\xi_{\AdS}$ is a Lorentz transformation in $\RR^{2,p}$, its action on
$\vec{x}$ can be defined by
\begin{equation}
  \xi_{\AdS}\,\vec{x} = B\,\vec{x}~,
 \label{eq:Bmatrix}
\end{equation}
where $B$ is a $(p+2)\times (p+2)$ constant matrix.  Thus, $\vec{y}
(\psi,\vec{x})= e^{-\psi\,B}\,\vec{x}$, so that
\begin{equation}
  d\vec{x} = e^{\psi\,B}\left(d\vec{y} + B\,\vec{y}\,d\psi\right)~.
\end{equation}
One can now compute the metric in adapted coordinates
$\{\psi,\tilde{\varphi}_i,\vec{y}\}$.  This can be written as
\begin{equation}
  g = \|\xi_{\Sph}\|^2\,\left(d\psi + B_1\right)^2 + \tilde{g} +
  g_{\AdS_{p+1}} +
  2\,d\psi\cdot\hat{\xi}_{\AdS} + \|\xi_{\AdS}\|^2\,d\psi^2~,
 \label{eq:asmetric}
\end{equation}
where the first two terms are just describing the metric on $\Sph^q$
in the adapted coordinate system $\{\psi,\tilde{\varphi}_i\}$
introduced in Section~\ref{sec:sphere}, and $\hat{\xi}_{\AdS}$
stands for the one-form associated with the Killing vector
$\xi_{\AdS}$, that is,
\begin{equation}
  \hat{\xi}_{\AdS} = \eta_{ij}\xi_{\AdS}^j\,dy^{i} =
  \eta_{ij}\,\left(B\cdot y\right)^j\,dy^{i}~.
\end{equation}

After these general considerations, we shall now proceed to discuss
the different geometries that appear in these discrete quotients of
$\AdS_{p+1}\times \Sph^{q}$.

\subsection{Non-everywhere spacelike $\xi_{\AdS}$}
\label{sec:notsl}

Let us first discuss the three cases in which $\xi_{\AdS}$ is not
always spacelike.  The first of these is where the two-form is
$\oplus_i B^{(0,2)}(\varphi_i)$, corresponding to the quotient of
$\AdS_{p+1}$ by some combination of rotations in orthogonal two-planes
$\RR^2$ in the embedding space.  These quotients produce special cases
of the conical defects, which were discussed extensively in, for
example \cite{bbkr}.  An interesting discussion of the properties of
the supersymmetric orbifolds in string theory is also given
in \cite{mart:orb1,mart:orb2}.  We will not discuss this case further
here, except to note that it is for these quotients where the
existence of a spin structure is not guaranteed.  The condition for
the existence of a spin structure was stated at the end of
Section~\ref{sec:spin}.

To consider the other two cases in Table~\ref{tab:positive} which are
not always spacelike, $B^{(1,2)} \oplus_i B^{(0,2)} (\varphi_i)$ and
$B^{(2,2)}_\pm \oplus_i B^{(0,2)}(\varphi_i)$, we follow our general
strategy, and start by describing the action of $B^{(1,2)}$ or
$B^{(2,2)}_\pm$ in $\AdS_3$.  The action of a more general Killing
vector of this form on $\AdS_{p+1}$ can then be built up by considering
the $\AdS_3$ action deformed by the rotations $B^{(0,2)}(\varphi_i)$ on
the $\Sph^{p-3}$ in the $\AdS_3 \times \Sph^{p-3}$ foliation of
\eqref{eq:adsk}.  We will then add in the deformation on a transverse
sphere $\Sph^q$ to obtain an everywhere spacelike quotient.

For the quotient of $\AdS_3$ by $B^{(1,2)}$, the relevant Killing
vector is
\begin{equation}
  \xi_{\AdS} = \be_{13} - \be_{34}.
\end{equation}
This Killing vector is spacelike almost everywhere, $||\xi_{\AdS}||^2
= (x_1 + x_4)^2$.  There is a single other Killing vector in
$\fso(2,2)$ which commutes with this one, $\xi_1 = \be_{12} -
\be_{24}$.  It has norm $||\xi_1 ||^2 = -(x_1 + x_4)^2$.
The most convenient coordinate system for studying this quotient is
Poincaré coordinates.  The form of the Killing vectors in
Poincaré coordinates is reviewed in appendix~\ref{sec:patch}.  It
is easy to see from those expressions that in the case of $B^{(1,2)}$,
we can orient the coordinates so that $\xi_{\AdS} = \partial_x$ and
$\xi_1 = \partial_t$, where the $\AdS_3$ metric in Poincaré
coordinates is
\begin{equation} \label{poinmet}
  g_{\AdS_3} = \frac{1}{z^2} \left(-dt^2 + dz^2 + dx^2\right)~.
\end{equation}
We see that the effect of the quotient is simply to make the
coordinate $x$ periodic.  The Killing vector $\xi_{\AdS}$ becomes null
on the Poincaré horizon $z=\infty$ where this coordinate system
breaks down.  In terms of the embedding coordinates, this is the
surface $x_1 + x_4 = 0$, where $\xi_{\AdS} = x_3 ( \partial_1 -
\partial_4)$.  We note that this symmetry has a null line of fixed
points at $x_1 + x_4 = x_3 = 0$ (parametrised by $x_1 - x_4$).  Away
from the fixed points, the identification along $\xi_{\AdS}$ will
generate closed null curves in the Poincaré horizon.  These can be
eliminated by deforming this quotient by a suitable action on an
odd-dimensional sphere.  Since we do not have a good global coordinate
system on this quotient, the best way to describe the causally regular
deformed quotient will be to use the coordinates adapted to the action
on the transverse sphere, as described at the end of the last
subsection.  We will not give the details of the application of this
general technique for this particular case; we just remark that for
this case, the matrix $B$ defined in \eqref{eq:Bmatrix} is
\begin{equation}
B =
 \begin{pmatrix}
   0 & 0 & -1 & 0\\
   0 & 0 & 0 & 0 \\
    -1 & 0 & 0 & 1 \\
    0 & 0 & -1 & 0
  \end{pmatrix}~.
\end{equation}
Following the supersymmetry analysis in \cite{fofs2}, it
is easy to conclude that for a suitable choice of sphere deformation,
the above quotient preserves $\nu=\frac{1}{4}$ of the vacuum
supersymmetry, that is, it has four supercharges.

For the case where we introduce a deformation on a transverse
$\Sph^3$, we can interpret the quotient as the near horizon geometry
of a D1-D5 system that has been quotiented by the action generated by
\begin{equation*}
  \xi = \partial_x + \theta_1\,R_{12}+\theta_2\,R_{34}~,
\end{equation*}
in which $x$ stands for the common direction shared by the D1-D5
system, and $R_{ij}$ stand for rotations transverse to the D1-D5's.  In
the language developed in \cite{FigSimBranes,FigSimGrav}, this
asymptotically flat spacetime would correspond to a D1-D5 system in a
generic intersection of flux 7-branes vacuum.  Whenever
$\theta_1=\pm\theta_2$, it would be interpreted as a D1-D5 system in
the flux 5-brane vacuum, which also 
has four supercharges.  Note that the standard supersymmetry enhancement due to
the near horizon limit is lost in this quotient, as the generator $\partial_x$,
which does not break any supersymmetry in the asymptotically flat spacetime
construction, becomes a null rotation generator from the AdS
perspective, which breaks one half of the supersymmetry.

We would also like to understand the boundary of this quotient.  In the
Poincaré coordinates (\ref{poinmet}), the global AdS boundary is
written in terms of an infinite series of flat space patches,
\begin{equation}
g_{\partial} = -dt^2 + dx^2.
\end{equation}
The action of the Killing vector on the AdS boundary compactifies the
spatial coordinate $x$; it might therefore seem that the quotient will
have an infinite sequence of boundaries.  However, the Killing vector
only has isolated fixed points on the boundary, at the points where
the line of fixed points $x_1 + x_4 = x_3 = 0$ meets the boundary.  In
Poincaré coordinates, these correspond to the points at past and
future timelike infinity and at spacelike infinity.  The different
boundary patches are therefore connected.  We can extend the
Poincaré coordinates to cover more of the boundary by defining
\begin{equation}
v = t - x, \tan T = t.
\end{equation}
The boundary metric then becomes
\begin{equation}
\label{nrbdy}
 g_{\partial} = \frac{1}{\cos^2 T} ( - 2 dv\, dT + \cos^2 T dv^2),
\end{equation}
and the Killing vector we quotient along is $\xi_{\AdS} =
\partial_v$.  Since we only have a conformal structure on the boundary,
we can ignore the overall factor in this metric.  In the resulting
metric, we see that the direction we quotient along is spacelike
except when $T = (n+1/2) \pi$, where it becomes null.  These points
correspond to one half of future and past null infinity in the
original Poincaré coordinates.  This coordinate system covers the
whole of the conformal boundary with the exception of a null line
corresponding to one half of past and future null infinity in each
Poincaré patch.  We could construct a similar coordinate system by
defining $u=t+x$---it would then cover that half but not the one where
$t-x$ remains finite.  We can think of the field theory dual to the
quotient along a null rotation as living on the cylindrical space
described in \eqref{nrbdy}, which has closed null curves at $T =
(n+1/2) \pi$.\footnote{There are some obvious similarities between
this construction and the Milne coordinate system on the orbifold of
flat space by a boost.}  Since the deformation by an action on a
transverse sphere does not alter the action on the boundary, it cannot
remove these closed lightlike curves in the dual theory.

A more interesting example of a not everywhere spacelike quotient is
$B^{(2,2)}_\pm$, where the Killing vector we quotient along is
\begin{equation}
  \xi^\pm_{\AdS} = \pm (\be_{12} - \be_{24}) + (\be_{13} - \be_{34})
\end{equation}
respectively. Both are null everywhere, $\|\xi^\pm_{\AdS}\|^2 = 0$.
From now on, we shall focus on $\xi^+_{\AdS}$; there is an analogous
discussion and structure for $\xi^-_{\AdS}$.  There are three other
Killing vectors in $\fso(2,2)$ commuting with $\xi^+_{\AdS}$,
\begin{equation}
 \label{slkv}
  \xi_1 = \be_{24} + \be_{13}, \quad \xi_2 = \be_{12} + \be_{34}, \quad
  \xi_3 = \be_{14} - \be_{23}~.
\end{equation}
These satisfy
\begin{equation}
  [\xi_i, \xi_j] = 2\epsilon_{ijk} \xi_k,
\end{equation}
so they define an $\fsl(2,\mathbb{R})$ symmetry which commutes with
$\xi^+_{\AdS}$.  This $\fsl(2,\mathbb{R})$ structure appears because when
we write $\fso(2,2) = \fsl(2,\mathbb{R}) \oplus \fsl(2,\mathbb{R})$,
the $B^{(2,2)}_\pm$ Killing vector lies entirely in one of the
$\fsl(2,\mathbb{R})$ factors.  A similar structure will reappear for
the same reason in our discussion of the self-dual orbifold in section
\ref{sec:sdorbdef}; it was first identified in that context
in \cite{cousshenn}.

We would like to adopt a coordinate system adapted to this
symmetry.  Since the $\xi_i$ do not commute, we can only adapt our
coordinates to one of them.  We note that $\|\xi_1\|^2 = \|\xi_3\|^2 =
1$, $\|\xi_2\|^2 = -1$.  Since our interest is in causal structure, it
seems natural to adapt the coordinates to the timelike vector
$\xi_2$.  We therefore want to construct a coordinate system
$(t,v,\rho)$ on $\AdS_3$ such that $\xi^+_{\AdS} = \partial_v$ and
$\xi_2 = \partial_t$.  This requires
\begin{equation}
  \begin{aligned}[m]
    \frac{\partial (x^4-x^1)}{\partial v} &= 0, \quad \frac{\partial (x^4+x^1)}
    {\partial v} = -2(x^3-x^2), \\
    \frac{\partial (x^3-x^2)}{\partial v} &= 0, \quad \frac{\partial (x^3+x^2)}
    {\partial v} = 2(x^4- x^1)\,;  \\
    \frac{\partial (x^4-x^1)}{\partial t} &= x^3 - x^2, \quad
    \frac{\partial (x^4+x^1)}{\partial t} = x^3+ x^2, \\
    \frac{\partial (x^3-x^2)}{\partial t} &= - (x^4-x^1), \quad
    \frac{\partial (x^3+x^2)}{\partial t} = - (x^4+x^1)~.
  \end{aligned}
\end{equation}
A combination which is thus independent of $t,v$ is $(x^4-x^1)^2 +
(x^3-x^2)^2$.  We will choose the $\rho$ coordinate so that this
combination is $e^{2 \rho}$.  A suitable coordinate system satisfying
these criteria and the condition $-x_1^2 - x_2^2 + x_3^2 + x_4^2 = -1$
defining the $\AdS_3$ embedding is
\begin{equation}
  \begin{aligned}[m]
    x^4 - x^1 &= e^\rho \sin t, \\
    x^4 + x^1 &= -e^{-\rho} \sin t - 2v\, e^\rho \cos t, \\
    x^3 - x^2 &= e^\rho \cos t, \\
    x^3 + x^2 &= -e^{-\rho} \cos t + 2v\, e^\rho \sin t ~.
  \end{aligned}
 \label{eq:null3}
\end{equation}
The inverse coordinate transformation is given by
\begin{equation}
  \begin{aligned}[m]
    e^{2 \rho} &= (x^4 -x^1)^2 + (x^3 -x^2)^2, \\
    \tan t &= \frac{x^4- x^1}{x^3 - x^2}, \\
    v &= e^{-2\rho}\left\{ \left[ (x^3 + x^2) + e^{-2\rho} (x^3-x^2)
    \right]^2 + \left[ (x^4+x^1) + e^{-2\rho} (x^4-x^1) \right]^2
    \right\}~.
  \end{aligned}
\end{equation}
Since these give finite values of $t,v,\rho$ for all points in
$\AdS_3$, this coordinate system covers the whole spacetime.  In terms
of these coordinates, the metric is
\begin{equation}
 \label{global22}
  g_{\AdS_3} = -dt^2 + d\rho^2 - 2 e^{2\rho} dv \,dt~.
\end{equation}
In this coordinate
system, the other two Killing vectors are
\begin{equation}
  \begin{aligned}[m]
    \xi_1 &= \sin 2t\, \partial_\rho + \cos 2t\, ( \partial_t - e^{-2\rho}
    \partial_v), \\
    \xi_3 &= -\cos 2t \,\partial_\rho + \sin 2t \,( \partial_t - e^{-2\rho}
   \partial_v)~.
  \end{aligned}
\end{equation}
We see that making identifications along the Killing vector
$\partial_v$ will produce closed null curves.  To eliminate these
closed null curves, we should introduce a deformation by a rotation on
the transverse sphere.  To simplify the discussion, we shall work it
out explicitly for a transverse $\Sph^3$, having in mind the standard
way of embedding $\AdS_3$ in type IIB string theory, as the near
horizon geometry of the D1-D5 system, giving rise to $\AdS_3\times
\,\Sph^3\times \TT^4$.  As discussed in Section~\ref{sec:sphere},
there are several inequivalent quotients that one can take of
$\Sph^3$.  We will focus on a particular quotient which preserves
supersymmetry, namely the quotient where $\xi_S = \partial_\psi$ when
we write the $\Sph^3$ metric as
\begin{equation}
g_{\Sph^3} = d\theta^2 + d\psi^2 + d\varphi^2 + 2\cos 2\theta\,
  d\psi\cdot d\varphi .
 \label{eq:metrics3}
\end{equation}
Thus, we consider the quotient along a total Killing vector $\xi =
\xi_{\AdS} + \gamma \xi_S = \partial_v + \gamma
\partial_\psi$.  Since we have a global adapted coordinate system
(\ref{global22}) on the AdS part of the quotient, it is convenient to
construct the global coordinate system on the full $\AdS_3 \times \Sph^3$
quotient by defining $\psi' = \psi - \gamma v$.  The six-dimensional
metric is then
\begin{equation} \label{fullg22}
g =  -dt^2 + d\rho^2 - 2 e^{2\rho} dv dt + d\theta^2 + (d\psi' +
  \gamma dv)^2 + d\varphi^2 + 2\cos 2\theta\,
  (d\psi'+ \gamma dv) \cdot d\varphi.
\end{equation}
The quotient is now along $\xi = \partial_v$.  We can see that this is
an everywhere spacelike direction; $\|\xi\|^2 = \gamma^2$.  This is a
necessary but not a sufficient condition for the absence of closed
causal curves, but it is easy to check explicitly that there are no
closed causal curves in the bulk of the quotient manifold in this
case.  As shown in \cite{fofs2}, the corresponding type IIB
configuration preserves $\nu=\frac{1}{8}$ of the vacuum
supersymmetry, that is, it has four supercharges. It is interesting to
point out that if we would have considered the action on the three
sphere \eqref{eq:metrics3} generated by $\xi_S=\partial_\varphi$, the
corresponding quotient $\xi=\xi_{\AdS} + \gamma\,\xi_S$ would have
preserved $\nu=\frac{1}{4}$ of the full type IIB supersymmetry.

It is interesting to note that, like the null rotation, the
$B^{(2,2)}_\pm$ Killing vector also has a simple action in
Poincaré coordinates.  We can orient the coordinates so that
$\xi_{\AdS} = \partial_t + \partial_x$ in the metric
(\ref{poinmet}).  The additional symmetry $\partial_t - \partial_x$
that is manifest in these coordinates can be written in terms of the
$\fsl(2,\mathbb{R})$ Killing vectors (\ref{slkv}) as the combination
$\xi_2 - \xi_1$.  Although the Poincaré coordinates are not a
global coordinate system for the quotient, they allow us to relate
these quotients and quotients of branes in asymptotically flat
spacetimes: the $B^{(2,2)}_\pm$ quotients can be understood as the near
horizon geometries of a D1-D5 system quotiented by the discrete action
generated by
\begin{equation}
  \xi = \pm\partial_t + \partial_x + \theta_1\,R_{12}+\theta_2\,R_{34}~.
\end{equation}
The physical interpretation of these quotients is unclear.  They can be
supersymmetric, and they are free from closed causal curves.  It might
be possible to give them some interpretation using a limiting
procedure in which one finally identifies bulk points along a ``null
translation'', by infinitely boosting a spacelike translation.  In
this case, there is still a supersymmetry enhancement since the
asymptotically flat quotient has four supercharges.

To discuss the conformal boundary of this quotient, we will use a
technique that will be used again in section~\ref{sec:dnrdef}, and
relate the spacetime to a plane wave. If we set $r = e^{-\rho}$, the
metric (\ref{fullg22}) becomes
\begin{multline} \label{cg22}
g = \frac{1}{r^2}[- 2 dv dt - r^2 dt^2 + dr^2 + r^2 (d\theta^2 + (d\psi' +
  \gamma dv)^2 + d\varphi^2 \\ + 2\cos 2\theta\,
  (d\psi'+ \gamma dv) \cdot d\varphi)].
\end{multline}
The conformally related metric in square brackets is a symmetric
six-dimensional plane wave, written in a polar coordinate system
deformed so that $\partial_v$ is a mixture of the null translation
symmetry of the plane wave and a rotation in the four transverse
spacelike coordinates. 

The conformal mapping between an $\AdS_3 \times S^3$ space and a plane
wave is implicit in previous work~\cite{bn} which showed that such
plane waves can be conformally mapped onto the Einstein static
universe. That is, since both spaces are conformally flat, we would
expect them to be conformally related. It is interesting to note the
relative simplicity of the relation: $\AdS_3 \times S^3$ corresponds
to the plane wave with the axis $r=0$ excluded, rescaled by a factor
of $1/r^2$.

More important for our present purpose is that the Killing vector we
wish to quotient along, $\partial_v$, annihilates the conformal factor
(as does $\xi_2 = \partial_t$), so we can use this conformal map to
study the boundary of the quotient spacetime, and not just to study
global $\AdS_3 \times S^3$. Note that unlike the double null rotation
in section \ref{sec:dnrdef}, the other Killing symmetries $\xi_1$ and
$\xi_2$ of this quotient do not also commute with the conformal
rescaling. They will hence appear as conformal isometries in the
boundary theory.

The conformal boundary of the quotient~\eqref{cg22} lies at $r=0$, and
has the metric (up to conformal transformations)
\begin{equation}
 \label{bdy22}
  g_{\partial} = - 2 dv\, dt. 
\end{equation}
Since $v$ is periodically identified in the quotient, there is a
compact null direction through every point in the boundary.  As in the
null rotation case, these closed null curves in the conformal boundary
cannot be removed by a sphere deformation.  This fact can explicitly
by checked in \eqref{fullg22}. It is interesting to note that we get
the same metric on the conformal boundary here as on either of the two
boundaries in the self-dual orbifold discussed in the next subsection.

If we regard \eqref{fullg22} simply as a coordinate system on $\AdS_3
\times S^3$, we can relate this description of the conformal boundary
to the usual two-dimensional $\mathbb{R} \times S^1$ Einstein static
universe boundary of global $\AdS_3 \times S^3$. In global
coordinates, the Killing vector field is given by
\begin{equation}
  \xi = (1+ \cos (\tau-\varphi)) (\partial_\tau - \partial_\varphi),
\end{equation}
where we are using the global coordinates introduced in
appendix~\ref{sec:patch}, and further writing $\hat x_3 = \cos
\varphi$, $\hat x_4 = \sin \varphi$, so that the metric on the
boundary reads
\begin{equation}
  g_\partial = -d\tau^2 + d\varphi^2~.
\end{equation}
We see that the quotient is along a null direction, and has a single
null line of fixed points at $\tau - \varphi = \pi$ (mod $2\pi$).
While the coordinate system \eqref{fullg22} covers all of global
$\AdS_3 \times S^3$, it does not cover all of its conformal boundary,
as these symmetry-adapted coordinates break down on the fixed points
of $\xi_{\AdS}$. The coordinates of \eqref{fullg22} cover all of the
boundary apart from this null line. They are related to the global
description above in the same way that a symmetric plane wave is
related to the Einstein static universe in higher-dimensional
cases~\cite{bn} (in two dimensions, there is no non-trivial plane
wave). Thus we see that \eqref{bdy22} provides a natural description
of the asymptotic boundary of the quotient, corresponding to excluding
these fixed points in discussing the quotient.

While it is clear that the deformed quotient \eqref{fullg22} is free
of closed causal curves, we can show that this quotient does not
preserve the stable causality of the original $\AdS_3 \times S^3$
space. If we write \eqref{fullg22} in the form appropriate for
Kaluza-Klein reduction along $v$,
\begin{multline} 
g =  -(1+ \gamma^{-2} e^{4\rho}) dt^2 + d\rho^2 + d\theta^2 +
\sin^2 2\theta d\varphi^2 + 2\gamma^{-1} e^{2\rho} dt (d\psi' + \cos
2\theta d\varphi)\\ + (\gamma dv + d\psi' + \cos 2\theta d\varphi -
\gamma^{-1} e^{2\rho} dt)^2, 
\end{multline}
we see that the lower-dimensional metric obtained by Kaluza-Klein
reduction along $v$ will have closed null curves, since the compact
circle parametrised by $\psi'$ is null. This implies that there can be
no time function $\tau$ on $\AdS_3 \times S^3$ such that
$\mathcal{L}_{\xi} \tau = 0$, for if there was, the
Kaluza-Klein reduced metric would be stably causal, which is
inconsistent with the appearance of closed null curves in the latter.
Thus, the discrete quotient cannot satisfy the condition
of~\cite{causal}, and does not preserve stable causality.

Following the discussion around \eqref{eq:adsk}, it is straightforward
to describe the quotient generated by $\xi^+_{\AdS}$ in higher
dimensional $\AdS_{p+1}$ spaces.  By construction, the global
symmetries of such a higher dimensional quotient will be the ones
discussed before times $\SO(p-2)$, corresponding to the rotational
symmetry transverse to the subspace where $\xi^+_{\AdS}$ acts.  Notice
that in this case, the metric on the boundary is conformally
equivalent to a plane wave metric,
\begin{equation}
  g_\partial = -2\,dv\,dt - r^2 dt^2 + dr^2 +
  r^2 g_{\Sph^{p-3}} ~. 
\end{equation}
In higher dimensions, there exists the possibility to deform the
quotient by rotations, i.e. $\oplus_i B^{(0,2)}(\varphi_i)$.  Let us
focus on $\AdS_5$, for algebraic simplicity.  The metric for $\AdS_5$
in the $\AdS_3$ foliation adapted to the action of $\xi^+_{\AdS}$ is
given by
\begin{equation}
  g_{\AdS_5} = \cosh^2\chi\,\left(-dt^2 + d\rho^2 -2e^{2\rho}\,dv\,dt\right)
  + d\chi^2 + \sinh^2\chi\,d\theta^2~.
\end{equation}
The deformation consists in acting on the angular direction $\theta$
through the generator $\xi = \varphi\,\partial_\theta$.  Thus, it is
convenient to introduce the new coordinate $\theta^\prime = \theta -
\varphi\,v$, so that $\xi^+_{\AdS}+ \xi = \partial_v$.  The metric on
the deformed quotient is
\begin{equation}
  g_{\AdS_5/\Gamma} = \cosh^2\chi\,\left(-dt^2 + d\rho^2
    -2e^{2\rho}\,dv\,dt\right) + d\chi^2 +
  \sinh^2\chi\,\left(d\theta^\prime + \varphi\,dv\right)^2~,
\end{equation}
where, once again, $v\sim v + 2\pi$.  As expected, the periodic
coordinate $v$ becomes everywhere spacelike except at the fixed point
of the deformed action.  This is just a consequence of the fact that
the norm of the deformed Killing vector is $\|\xi^+_{\AdS} + \xi\|^2 =
\varphi^2[(x^5)^2 + (x^6)^2] = \varphi^2\,\sinh^2\chi$, which
certainly vanishes at the origin of the 56-plane, where the fixed
point of $\xi$ lies.

This particular deformation $(\varphi\neq 0)$ breaks all the
supersymmetry and it can be interpreted as the near horizon geometry
of a bunch of parallel and coincident D3-branes quotiented by the
action of a null translation plus a rotation.  It is certainly possible
to turn on supersymmetric deformations in higher dimensional AdS
spacetimes.  In particular, it is possible to consider families of two
parameter deformations corresponding to $B^{(0,2)}(\varphi_1)\oplus
B^{(0,2)}(\varphi_2)$ in $\AdS_7$.  Whenever $\varphi_1=\pm \varphi_2$,
the quotient will preserve supersymmetry.  The corresponding
asymptotically flat interpretation would be in terms of parallel and
coincident M5-branes quotiented by the action of a null translation
plus a certain rotation in $\RR^4$.  The supersymmetric deformation
would correspond to the action having an $\fsu(2)$ holonomy.

\subsection{Self-dual orbifolds and their deformations}
\label{sec:sdorbdef}

The fifth two-form appearing in table \ref{tab:positive},
$B^{(1,1)}(\beta_1) \oplus B^{(1,1)}(\beta_2) \oplus_i
B^{(0,2)}(\varphi_i)$ with $|\beta_1|=|\beta_2|$, can be interpreted
as the deformation of the self-dual orbifolds of $\AdS_3$, first
introduced in \cite{cousshenn}, and recently discussed in \cite{bns}.
The norm of $\xi_{\AdS}$ is spacelike everywhere.  Therefore, one can
study these geometries with or without any further non-trivial action
on transverse spheres.

As already indicated above, the minimal dimension where this discrete
quotient exists is for $p=2$, i.e.  $\AdS_3$.  The addition of any
rotation parameter $\varphi_i$ would increase this dimension by two.
Since the elementary indecomposable block acting on $\AdS_3$ is a
linear combination of boosts in $\RR^{2,2}$, this discrete quotient
does not have an analogue in an asymptotically flat spacetime, in the
sense that there is no quotient whose near horizon limit gives rise to
these self-dual orbifolds.

The anti-de Sitter action, including the deformation parameters
$\{\varphi_1\}$, integrates to the following $\RR$-action on
$\RR^{2,p}$:
\begin{equation}
  \begin{pmatrix}
    x^1\\x^2\\x^3\\x^4\\x^{2i+5}\\x^{2i+6}\\
  \end{pmatrix}
  \mapsto
  \begin{pmatrix}
    x^1\cosh \beta t \pm x^3 \sinh \beta t\\
    x^2\cosh \beta t + \pm x^4\sinh \beta t\\
    x^3\cosh \beta t \pm x^1\sinh \beta t\\
    x^4\cosh\beta t \pm x^2\sinh\beta t \\
    x^{2i+5}\cos\varphi_i t - x^{2i+6}\sin\varphi_i t\\
    x^{2i+6}\cos\varphi_i t + x^{2i+5}\sin\varphi_i t
  \end{pmatrix}~, \quad \forall\,\text{i}
\end{equation}
where we set $\beta_1=\beta$ and $\beta_2=\pm \beta$.  Notice that the above
action is manifestly free of fixed points for
any value of the boost and rotation parameters $\{\beta,\varphi_i\}$.

In the following, we shall review the main features of the self-dual
orbifolds of $\AdS_3$, extending the discussion to uncover their
embeddings in higher dimensional anti-de Sitter spacetimes and their
deformations both by rotations in anti-de Sitter and non-trivial
actions on transverse spheres, afterwards.

\subsubsection{Pure AdS}
\label{sec:sdorbads}

Let us start our discussion by focusing on $\AdS_3$, so that there are
no $B^{(0,2)}(\varphi_i)$ blocks.  In this case, as first described
in \cite{cousshenn}, the quotient preserves an $\mathbb{R} \times
\fsl(2,\mathbb{R})$ subalgebra of the original $\fso(2,2) =
\fsl(2,\mathbb{R}) \oplus \fsl(2,\mathbb{R})$ isometry algebra.  A
suitable system of global coordinates adapted to the quotient and the
timelike vector in $\fsl(2,\mathbb{R})$ is \cite{cousshenn}
\begin{equation}
  \begin{aligned}[m]
    x^1 &= \cosh z\cosh\beta\phi\cos t -
    \sinh z\sinh\beta\phi\sin t, \\
    x^2 &= \cosh z\cosh\beta\phi\sin t
    + \sinh z\sinh\beta\phi\cos t, \\
    x^3 &= -\cosh z\sinh\beta\phi\cos t
    + \sinh z\cosh\beta\phi\sin t, \\
    x^4 &= \pm \left(\cosh z\sinh\beta\phi\sin t
    - \sinh z\cosh\beta\phi\cos t\right)~.
  \end{aligned}
 \label{eq:sdadapted}
\end{equation}
The sign ambiguity in the last line of (\ref{eq:sdadapted}) corresponds
to the two distinct cases $\beta_2 = \pm \beta_1$ in the $\SO(2,n)$
classification reviewed in section~\ref{sec:so2n}.  This illustrates
explicitly that these two cases are related by an
orientation-reversing symmetry of AdS$_3$, namely the reflection $x_4
\to -x_4$.  It is important to stress that, at this point, the
coordinates $\{t,\phi,z\}$ are just some particular global
description for $\AdS_3$.  All of them are defined in the range
$-\infty < t,\phi,z < +\infty$.  It is only when we identify
points in $\AdS_3$ along some discrete step generated by
$\xi_{\AdS}=\partial_\phi$ that our discrete quotients will differ from
$\AdS_3$ globally, by making the adapted coordinate $\phi$ a compact
variable with period $2\pi$ in some normalisation, i.e. $\phi \sim
\phi + 2\pi$.

As first proved in \cite{cousshenn} for $\AdS_3$, corroborated in \cite{bns}
and extended to any higher dimensional AdS spacetime in \cite{fofs2}, the
supersymmetry preserved by these self-dual orbifolds is one--half
of the original one.

The metric in adapted coordinates \eqref{eq:sdadapted} looks like
\begin{equation}
  g_{sd} = -dt^2 + \beta^2\,d\phi^2 + dz^2
  -2\beta\sinh 2z\,dt\,d\phi~.
 \label{eq:sdplus}
\end{equation}
Thus, it describes a non-static but stationary spacetime.  One
interesting feature which has not previously been noted is that $t$ is
a global time function, since $\nabla_\mu t \nabla^\mu t = -1/\cosh^2
2z$, so the self-dual orbifolds are stably causal, and hence do not
contain closed timelike curves.  This metric can be
interpreted as an $\Sph^1$ fibration over $\AdS_2$, as the following
rewriting indicates
\begin{equation}
  g_{sd} = -\cosh^2 2z dt^2 + dz^2
  + \left(\beta\,d\phi - \sinh 2z\,dt\right)^2~.
 \label{eq:sdplua}
\end{equation}
This quotient was recently analysed in detail in \cite{bns}, where its
isometries, geodesics, asymptotic structure and holography in this
background were extensively studied.

An important point to note from that analysis is the structure of the
conformal boundaries.  It was shown in \cite{bns} that the quotient has
two disconnected conformal boundaries.  If we consider the coordinate
transformation
\begin{equation*}
  \sinh z = \tan\theta \quad
  \theta\in\left(-\frac{\pi}{2},\frac{\pi}{2}\right)~,
\end{equation*}
the metric \eqref{eq:sdplus} becomes
\begin{equation}
  g_{sd} = \frac{1}{\cos^2\theta}\left(\cos^2\theta\,(-dt^2 +
  \beta^2\,d\phi^2)
  + d\theta^2 - 4\beta\sin\theta\,dt\,d\phi
  \right)~,
\end{equation}
from which we learn that the metric on both conformal boundaries,
located at $\theta\to\pm\frac{\pi}{2}$ is given by
\begin{equation}
  g_\partial = \pm dt\,d\phi~.
 \label{sdplusboundary}
\end{equation}
Thus, there are closed lightlike curves on the conformal boundary.
The appearance of two disconnected boundaries can be further
understood by noting that in the adapted coordinates
\eqref{eq:sdadapted}, the original $\AdS_3$ conformal boundary is
covered by four connected patches located at $z\to \pm\infty$ and
$\phi\to\pm\infty$.  After the discrete identification, two of these
patches no longer belong to our space, leaving as a consequence, the
existence of two boundaries at $z\to\pm\infty$, being disconnected.
These boundaries are causally connected through the bulk, as was shown
in \cite{bns} by analysing the geodesics in this space.

Unlike the previous cases, this quotient has no natural interpretation
as arising from a quotient of an asymptotically flat spacetime.  This
is related to the fact that the quotient does not take a simple form
in Poincaré coordinates.  However, Strominger \cite{strom:ads2}
showed that these self-dual orbifolds emerge as the local
description of a very--near horizon geometry when focusing on the
vicinity of the horizon of an extremal BTZ black hole.

Thus, even though this quotient does not emerge directly from the
D1-D5 perspective, it is nevertheless possible to set-up an
asymptotically flat spacetime which reproduces the self-dual orbifolds
in two steps \cite{bns}.  This is achieved by adding some momentum
along the common direction shared by the D1's and D5's, and taking the
standard near horizon limit, keeping the momentum density fixed.  One
then focuses on the vicinity of the horizon resulting from the
previous limit.  This procedure generalises the construction in
\cite{cvetic} to the D1-D5 system, and it provides an independent way
of understanding the DLCQ holography proposed in \cite{bns}.

Following our general discussion presented at the beginning of section
\ref{sec:nonsingular}, it is straightforward to extend the analysis to
higher dimensional $\AdS_{p+1}$ spaces, for $p\geq 3$.  Indeed, we can
use the foliation in \eqref{eq:adsslicing} and replace the
$\{\hat{x}_i\}$ appearing there with the $R=1$ version of
\eqref{eq:sdadapted}.  The resulting metric is
\begin{equation}
  g_{sd_{p+1}} = (\cosh\chi)^2\,g_{sd}+ (d\chi)^2
  + (\sinh\chi)^2 \,g_{\Sph^{p-3}}~.
\end{equation}
where $g_{sd}$ is the metric given in \eqref{eq:sdplus}.

This allows us to see that in these higher dimensional cases, the
boundary of the quotient will be connected.  The point is that the
boundary of the quotient in higher dimensions is given in these
coordinates by $\chi \to \infty$, as discussed earlier.  Thus, the
boundary of the higher-dimensional quotients naturally contains a copy
of the bulk of the AdS$_3$ quotient.  Since the AdS$_3$ quotient is
connected, this implies that the boundary of the quotient is connected
in higher dimensions.  It also shows us that unlike the AdS$_3$ case,
in higher dimensions there is a natural non-degenerate metric on the
boundary of the quotient.

\subsubsection{Deformation by $B^{(0,2)}$}

Even though we could discuss the turning on of the deformation
parameters $\varphi_i$ in the general case, we shall just briefly
mention their main new features in the string theory embeddings
described above.  This means that we shall concentrate on $\AdS_5$ and
$\AdS_7$, since these deformations are not available for $\AdS_4$.

This programme is particularly simple to carry on already in the
foliation defined by \eqref{eq:adsslicing}.  As previously mentioned,
$B^{(0,2)}(\varphi_i)$ blocks correspond to rotations in $\RR^2$ planes
in the embedding
space, and in the coordinates of \eqref{eq:adsk}, these motions can be
globally described as a single ``translation'' along one of the angular
variables of the $\Sph^{n-1}$ factor.  The definition of the adapted
coordinate system in which $\oplus_i B^{(0,2)}(\varphi_i)$ takes the
form of a single ``translation'' is precisely parallel to the
discussion for the transverse $\Sph^q$ given in Section~\ref{sec:sphere}.

As an example, consider $\AdS_5$.  In this case, we can only turn on
one parameter, $\varphi_1=\varphi$.  It is clear that rotations in
$\RR^2$ correspond to motions along the $\Sph^1$ transverse to the
$\AdS_3$ foliation of $\AdS_5$ in \eqref{eq:adsk}, for $p-n=2$.  If we
parameterise this circle by $\theta$, the Killing vector field
$\xi_{\AdS}$ generating the full action of the deformed discrete
quotient is given by
\begin{equation}
  \xi_{\AdS} = \partial_\phi + \varphi\,\partial_\theta~,
\end{equation}
in the adapted coordinates defined by \eqref{eq:adsslicing} and
\eqref{eq:sdadapted}.

It is now just a matter of applying a linear transformation in the
$\{\phi,\theta\}$ plane, which will generate an extra fibration, to
rewrite the metric in a globally defined coordinate system adapted to
the deformed Killing vector field $\xi_{\AdS}$.  This metric is given
by
\begin{equation}
  g = \cosh^2\chi\,g_{sd} + d\chi^2 +
  \sinh^2\chi\left(d\theta + \varphi\,d\phi\right)^2~.
 \label{eq:sdconical}
\end{equation}
By construction, this deformation will break all the spacetime
supersymmetry.

The techniques for $\AdS_7$ are exactly the same, but there is a
richer structure of possibilities since we have an $\Sph^3$
transverse to the $\AdS_3$ action, which allows to turn on two
inequivalent parameters $\{\varphi_1, \varphi_2\}$
\begin{equation*}
  \varphi_1\,R_{12} + \varphi_2\,R_{34}~,
\end{equation*}
where $R_{ij}$ stands for a rotation generator in the ij-plane
belonging to $\RR^4$, where the 3-sphere is embedded as a quadric.
Let us describe this 3-sphere in terms of standard complex
coordinates
\begin{equation}
  \begin{aligned}[m]
    z_1 & = x^1 + i\,x^2 = \cos\theta\,e^{i(\psi + \varphi)}\,,\\
    z_2 & = x^3 + i\,x^3 = \sin\theta\,e^{i(\psi - \varphi)}\,.
  \end{aligned}
\end{equation}
A supersymmetric quotient \cite{fofs2} is given by the choice
$\varphi_1=-\varphi_2=\theta_1$. The metric describing the global
quotient is given by    
\begin{multline}
  g_{\AdS_7/\Gamma} = \cosh^2\chi\,g_{sd} + d\chi^2 + \sinh^2\chi\,
  \left(d\theta^2 + \left(d\varphi + \theta_1\phi\right)^2 \right.\\
  \left. + d\psi^2 + 2\cos 2\theta\,\left(d\varphi + \theta_1\,d\phi\right)
  \cdot d\psi\right)~.
\end{multline}
Adding a transverse four-sphere and a constant flux on it, the
above configuration is supersymmetric.  It actually preserves
$\nu=\frac{1}{2}$ of the supersymmetries preserved by the original vacuum.
Thus, it has sixteen supercharges. It is worthwhile mentioning that
the deformation described by $\varphi_1=-\varphi_2$ does not break any
further supersymmetry. It is a further action that we can consider in
our spacetime for free, supersymmetry wise. Contrary to what intuition
may suggest, as explained in more detail in \cite{fofs2}, the
deformation $\varphi_1=\varphi_2$ breaks all the supersymmetry.


\subsubsection{Sphere deformations}
\label{sec:sdorbsphere}

Let us start our discussion on sphere deformations of self-dual
orbifolds on the embedding of $\AdS_3\times S^3$ in type IIB.
The most general action that we can write down on $S^3$ is given in
terms of two real parameters
\begin{equation}
  \xi_S = \theta_1\,R_{12} + \theta_2\,R_{34}~.
 \label{eq:sphere3}
\end{equation}
Because of the freedom that we have to quotient by the action of the
Weyl group, we can always choose to work on the fundamental region
defined by $\theta_1\geq |\theta_2|$.

Among all these quotients, only a subset preserve supersymmetry.
In particular, if we consider the action generated by $\be_{13} \pm
\be_{24}$ on $\AdS_3$, the only supersymmetric deformations are given
by $\theta_1=\pm\theta_2$, the signs being correlated. Interestingly,
such deformations still preserve the same amount of supersymmetry as
the self-dual orbifolds themselves. Thus, these supersymmetric
deformations are for free, as pointed out in \cite{fofs2}, where the
reader can also find the explanation for this phenomenon.

The discussion proceeds in an analogous way for higher dimensional AdS
spacetimes. If we consider the eleven dimensional configuration
$\AdS_4 \times S^7$, their deformations are characterised by four real
numbers
\begin{equation}
  \xi_S = \theta_1\,R_{12} + \theta_2\,R_{34} + \theta_3\,R_{56} +
  \theta_4\,R_{78}~.
 \label{eq:sphere7}
\end{equation}
Due to the Weyl group action, we can restrict ourselves to the region
defined by $\theta_1\geq\theta_2\geq \theta_3\geq |\theta_4|$. As
discussed in \cite{fofs2}, there are several loci in this parameter
space where supersymmetry is allowed. If $\theta_1=\theta_2$ and
$\theta_3=-\theta_4$ the quotient preserves $\nu=\frac{1}{4}$.
Whenever one of the relations
\begin{eqnarray*}
  \theta_1 - \theta_2 + \theta_3 + \theta_4 & = & 0~,\\
  \theta_1 + \theta_2 - \theta_3 + \theta_4 & = & 0~,\\
  \theta_1 - \theta_2 - \theta_3 - \theta_4 & = & 0~,
\end{eqnarray*}
is satisfied, the supersymmetry will be $\nu=\frac{1}{8}$. Finally,
there is enhancement whenever $\theta_1=\theta_2=\theta_3=-\theta_4$,
giving rise to $\nu=\frac{3}{8}$.

The discussion for $\AdS_5\times S^5$ is fairly simple. The action on
the 5-sphere is given in terms of three real parameters
\begin{equation}
  \xi_S = \theta_1\,R_{12} + \theta_2\,R_{34} + \theta_3\,R_{56}~.
 \label{eq:sphere5}
\end{equation}
The deformation preserves $\nu=\frac{1}{4}$ for $\theta_1=\theta_2$
and $\theta_3=0$. It preserves $\nu=\frac{1}{8}$ if
$\theta_1\pm\theta_2\pm\theta_3=0$, with uncorrelated signs. See
\cite{fofs2} for more details.

The only supersymmetric deformation for $\AdS_7\times S^4$ out of the
two parameter family 
\begin{equation}
  \xi_S = \theta_1\,R_{12} + \theta_2\,R_{34}~,
 \label{eq:sphere4}
\end{equation}
is given by $\theta_1=\theta_2$, also preserving $\nu=\frac{1}{4}$.

As an explicit example of a supersymmetric deformation of the self-dual
orbifold, we shall present one particular example of the above
discussion, one embedded in $\AdS_5\times S^5$. More precisely,
we shall focus on $\theta_1=2$, $\theta_2=\theta_3=1$.  A simple description of
this quotient can be obtained by parametrising the 5-sphere in terms
of the coordinates
\begin{equation}
  \begin{aligned}[m]
    z_1 &= x^1 + ix^2 = \cos\theta_1\,e^{i(\varphi_1 + 2\psi)}\, \\
    z_2 &= x^3 + ix^4 = \sin\theta_1\,\cos\theta_2\,e^{i(\psi + \varphi)}\, \\
    z_3 &= x^5 + ix^6 = \sin\theta_1\,\sin\theta_2\,e^{i(\psi - \varphi)}~.
  \end{aligned}
 \label{eq:f3coord}
\end{equation}
One can check that $\xi_S=\partial_\psi$.  This is an example in which
both $\xi_{\AdS}$ and $\xi_S$ are described in terms of adapted
coordinates.  Thus, by a simple linear transformation, we can easily
write the fully adapted ten dimensional metric as
\begin{multline}
  g = \cosh^2\chi\,g_{sd} + d\chi^2 + \sinh^2\chi\,d\theta^2 + d\theta_1^2
  + \sin^2\theta_1\,d\theta_2^2 + \cos^2\theta_1\,\left(d\varphi_1 +
  2(d\psi + d\phi)\right)^2 \\
  + \sin^2\theta_1\,\left((d\psi+d\phi)^2 + d\varphi^2 + 2\cos 2\theta_2\,
  (d\psi+d\phi)\cdot \varphi\right)~.
\end{multline}
As can be checked from the review of the results in \cite{fofs2}
presented at the beginning of this subsection, this particular example
preserves $\nu=\frac{1}{8}$ of the vacuum supersymmetry. Thus, it has 
four supercharges.

Of course, there is no conceptual difficulty in dealing with
deformations that contain both two forms $\oplus_i
B^{(0,2)}(\varphi_i)$ on $\AdS$ and non-trivial sphere actions. The
supersymmetric quotients can also be found in \cite{fofs2}.

\subsection{Double null rotation and its deformations}
\label{sec:dnrdef}

The third two-form appearing in Table~\ref{tab:positive}, $B^{(1,2)}
\oplus B^{(1,2)} \oplus_i B^{(0,2)}(\varphi_i)$, can be interpreted as
a deformation, with deformation parameters $\{\varphi_i\}$, of the
double null rotation discrete quotient considered in \cite{simon}.
Indeed, it consists of the simultaneous action of two spacelike null
rotations in transverse $\RR^{1,2}$ subspaces, and a set of rotations
with parameters $\varphi_i$ in different transverse $\RR^2$ planes.
Since the norm of $\xi_{\AdS}$ is positive everywhere, even for
$\varphi_i=0$ $\forall$ $i$, there is no need to deform the previous
action by a non-trivial one on a transverse sphere to get an
everywhere spacelike Killing vector field $\xi$ in
\eqref{eq:defkilling}.

The minimal dimension where such an object exists is for $p=4$, i.e.
$\AdS_{5}$, in which case there are no $B^{(0,2)}(\varphi_i)$
blocks.  The pure double null rotation discrete quotient has a very
natural interpretation in the Poincaré patch: it consists of the
combined action of a null rotation plus a spacelike
translation.  Consequently, it has a very straightforward origin in
terms of the geometry of a bunch of parallel D3-branes: the pure
double null rotation discrete quotient in $\AdS_5$ is the near horizon
geometry corresponding to a bunch of parallel D3-branes whose
worldvolume is the nullbrane, i.e.  $\RR^{1,3}/\ZZ$, four dimensional
Minkowski spacetime modded out by the simultaneous discrete action of
a null rotation in $\RR^{1,2}$ and a spacelike translation along
$\RR$, which was first introduced in \cite{fofs}.

The full anti-de Sitter action, including the deformation parameters,
 integrates to the following $\RR$-action on $\RR^{2,p}$:
\begin{equation}
  \begin{pmatrix}
    x^1\\x^2\\x^3\\x^4\\x^5\\x^6\\x^{2i+5}\\x^{2i+6}\\
  \end{pmatrix}
  \mapsto
  \begin{pmatrix}
    x^1 - t x^3 + \half t^2 (x^1-x^4)\\
    x^2 - t x^5 + \half t^2 (x^2-x^6)\\
    x^3 + t (x^4 - x^1) \\
    x^4 - t x^3 + \half t^2 (x^1-x^4)\\
    x^5 + t (x^6 - x^2) \\
    x^6 - t x^5 + \half t^2 (x^2-x^6)\\
    x^{2i+5}\cos\varphi_i t - x^{2i+6}\sin\varphi_i t\\
    x^{2i+6}\cos\varphi_i t + x^{2i+5}\sin\varphi_i t
  \end{pmatrix}~, \quad \forall\,\text{i}
\end{equation}
which is manifestly free of fixed points for any value of the rotation
parameters.

\subsubsection{Pure AdS}
\label{sec:dnrads}

Let us first consider the pure double null rotation in AdS$_5$.  This
was analysed in \cite{simon}.  We will extend this analysis by
discussing the isometries preserved by the quotient, constructing
suitable adapted coordinate systems, and examining the action on the
boundary of AdS. In the process, we will uncover interesting relations
to compactified plane waves.

The Killing vector that we quotient along is
\begin{equation}
\xi_{\AdS} = \be_{13} - \be_{34} + \be_{25} - \be_{56}.
\end{equation}
Its norm is $||\xi_{\AdS}||^2 = (x_1 + x_4)^2 + (x_2 + x_6)^2$.  This
is clearly positive semidefinite, and the quadric
$-(x_1+x_4)(x_1-x_4) - (x_2+x_6)(x_2-x_6) + x_3^2+x_5^2 = -1$ defining
the AdS embedding constrains the coordinates so that it is positive
definite.  There are four linearly independent commuting isometries in
$\fso(2,4)$:
\begin{equation}
\begin{aligned}[m]
\xi_1 &= \be_{13} - \be_{34} - \be_{25} + \be_{56}, \\
\xi_2 &= \be_{15} + \be_{23} - \be_{36} + \be_{45}, \\
\xi_3 &= \be_{12} - \be_{24} + \be_{16} + \be_{46}, \\
\xi_4 &= \be_{35} - \be_{12} + \be_{46}.
\end{aligned}
\end{equation}
These Killing vectors have the non-trivial commutation relations
\begin{equation}
[\xi_1, \xi_2] = -2 \xi_3, \quad [\xi_1, \xi_4] = 2 \xi_2, \quad
[\xi_2, \xi_4] = -2 \xi_1.
\end{equation}
They therefore form a Heisenberg algebra on which $\xi_4$ acts as an
outer automorphism.  The symmetry algebra of the quotient is hence
$(\mathfrak{h}(1) \rtimes \mathbb{R}) \oplus \mathbb{R}$.  The norms of
the Killing vectors are $||\xi_1||^2 = ||\xi_2||^2 = ||\xi_{AdS}||^2$,
$||\xi_3||^2 = 0$, $||\xi_4||^2 = -1$.

We want to construct adapted coordinates to describe this quotient; it
is convenient for studying causality to adapt them to $\xi_{\AdS}$,
$\xi_3$ and $\xi_4$.  Let us therefore seek to choose coordinates
$(t,u,\phi,\rho,\gamma)$ so that $\xi_3 = \partial_v$, $\xi_4 =
-\partial_t$, and $\xi_{\AdS} = \partial_\phi$.  This requires
\begin{equation}
  \begin{aligned}[m]
    \frac{\partial (x^4 - x^1)}{\partial \phi} &= 0, \quad \frac{\partial (x^4+
    x^1)}{\partial \phi} = -2 x^3, \\
    \frac{\partial (x^6 - x^2)}{\partial \phi} &= 0, \quad \frac{\partial (x^6
    + x^2)}{\partial \phi} = -2 x^5, \\
    \frac{\partial x^3}{\partial \phi} &= x^4 - x^1, \quad \frac{\partial x^5}
    {\partial \phi} = x^6 - +x^2, \\
    \frac{\partial (x^4 - x^1)}{\partial v} &= 0, \quad \frac{\partial (x^4 +
    x^1)}{\partial v} = -2 (x^6 - x^2), \\
    \frac{\partial (x^6 - x^2)}{\partial v} &= 0, \quad \frac{\partial (x^6 +
    x^2)}{\partial v} = -2 (x^4 - x^1), \\
    \frac{\partial x^3}{\partial v} &= 0, \quad \frac{\partial x^5}{\partial v} = 0,
    \\
    \frac{\partial (x^4 - x^1)}{\partial t} &= (x^6 - x^2), \quad
    \frac{\partial (x^4 + x^1)}{\partial t} = (x^6 + x^2), \\
    \frac{\partial (x^6 - x^2)}{\partial t} &= -(x^4 - x^1), \quad
    \frac{\partial (x^6 - x^2)}{\partial t} = -(x^4 + x^1), \\
    \frac{\partial x^3}{\partial t} &= x^5, \quad
    \frac{\partial x^5}{\partial t} = -x^3~.
  \end{aligned}
\end{equation}
There are two quantities independent of $\{t,v,\phi\}$: $(x^4-x^1)^2 +
(x^6-x^2)^2$ and $x^3\cdot (x^6 - x^2) - x^5\cdot (x^4 - x^1)$.  We will choose
coordinates $\{\rho,\psi\}$ so that
\begin{equation}
  \begin{aligned}[m]
    (x^4 - x^1)^2 + (x^6 - x^2)^2 &= e^{2\rho} \\
    x^3\cdot (x^6 - x^2) - x^5\cdot (x^4 - x^1) &= e^\rho \psi~;
  \end{aligned}
\end{equation}
we must take $-\infty < \rho < \infty$ and $-\infty < \psi < \infty$ to
obtain coordinates that cover the whole spacetime.  A coordinate
system satisfying all these conditions is
\begin{equation}
  \begin{aligned}[m]
    x^4 - x^1 &= e^{\rho} \sin t, \\
    x^4 + x^1 &= - e^{\rho} (2\phi \psi + 2v) \cos t -
    (e^{-\rho}  + (\psi^2+\phi^2) e^{\rho}) \sin t, \\
    x^6 - x^2 &= e^\rho  \cos t, \\
    x^6 + x^2 &=  e^{\rho} (2\phi \psi + 2v)\sin t -
    (e^{-\rho} + (\psi^2+\phi^2) e^{\rho} )\cos t,\\
    x^3 &= e^{\rho} (\psi \cos t + \phi \sin t),\\
    x^5 &= e^{\rho} (-\psi \sin t + \phi \cos t)~.
  \end{aligned}
\end{equation}
The $\AdS_5$ metric in these coordinates is
\begin{equation}
 \label{globaldnr}
  g_{dnr} = -dt^2 + d\rho^2 + e^{2\rho} ( d\psi^2 + d\phi^2 - 2 dt dv -
  4\psi dt d\phi), 
\end{equation}
and the other two Killing vectors are
\begin{equation}
  \begin{aligned}[m]
    \xi_1 &= -\cos 2t\, (\partial_\phi - 2 \psi
    \partial_v) + \sin 2t\, \partial_\psi,\\
    \xi_2 &= \sin 2t \,(\partial_\phi - 2 \psi
    \partial_v) + \cos 2t\, \partial_\psi~.
 \end{aligned}
\end{equation}
Even though we will not give the explicit details, it is easy to check
by working out the inverse coordinate transformation that this
coordinate system covers the whole of AdS.  Before any identification,
the range of all adapted coordinates is non-compact. The double null
rotation quotient is simply described by making the coordinate $\phi$
compact.

We would also like to understand the conformal boundary of this
quotient. First, we should note that even though the quotient is free
of fixed points in the bulk, its boundary has a continuous line of
them. The action generated by $B^{(1,2)}\oplus B^{(1,2)}$ integrates
to the real line, so the only possible fixed points are the ones for
which $\xi_{\AdS}$ vanishes.  These points are given by
\begin{equation*}
  x^4 - x^1 = x^6 - x^1 = x^3 = x^5 =0~.
\end{equation*}
The above does not belong to $\AdS_5$, since they do not satisfy the
quadric equation \eqref{eq:ads}.  This is indeed true for the bulk of
AdS (finite non-compact spacelike direction in global AdS), but there
is a continuous curve of fixed points on an infinite cylinder of axis,
global time $\tau$, and a maximal circle base. To see this, consider
the standard global description of $\AdS_5$,
\begin{equation*}
  \begin{aligned}[m]
    x^1 &= \cosh\chi\cos\tau ,\\
    x^2 &= \cosh\chi\sin\tau ,\\
    x^i & = \sinh\chi\,\hat x^i \,\,\, i=3,\dots ,6~,
  \end{aligned}
\end{equation*}
where $\{\hat x^i\}$ parametrise a 3-sphere of unit radius.  It is
easy to see that any solution to the fixed point conditions requires
$\chi\to\infty$, from which we already learn such points belong to the
boundary of $\AdS_5$.  It is also clear that $\hat x^3 = \hat x^5 =
0$.  Thus, such fixed points belong to a maximal circle in the
$x^4-x^6$ plane.  If the angular variable describing such a maximal
circle is $\varphi$ $(0\leq \varphi < 2\pi)$, the continuous line of
fixed points is determined by
\begin{equation*}
  \tau = \varphi \quad (\text{mod}\,2\pi)~.
\end{equation*}

Thus, the action of the quotient is well-defined on the global
boundary of AdS (i.e., the Einstein static universe) with a single
null line deleted. However, we know that the Einstein static universe
with a null line deleted is conformal to a symmetric plane
wave~\cite{bn}. This suggests that the boundary of (\ref{globaldnr})
should be described in terms of a plane wave. 

Inspired by this, and the analysis of the $B^{(2,2)}_\pm$ case in
section \ref{sec:notsl}, let us now make a coordinate transformation
$Z = e^{-\rho}$ in (\ref{globaldnr}). The metric then becomes
\begin{equation} \label{zdnr}
g_{dnr} = \frac{1}{Z^2} (- 2 dt dv - Z^2 dt^2 + dZ^2 +  d\psi^2 + d\phi^2 -
  4\psi dt d\phi), 
\end{equation}
where $0< Z < \infty$ covers the whole of $\AdS_{5}$. By rescaling the
metric by a factor of $Z^2$, we can conformally map global $\AdS_5$
into the space with metric
\begin{equation} \label{cembdnr}
\bar g = - 2 dt dv - Z^2 dt^2 + dZ^2 +  d\psi^2 + d\phi^2 -
  4\psi dt d\phi, 
\end{equation}
with the conformal boundary lying at $Z=0$. Since $\xi_{\AdS} =
\partial_\phi$ annihilates the conformal factor, this embedding
commutes with the quotient; we can regard the double null rotation as
conformally embedded in (\ref{cembdnr}) with $\phi$
compactified.\footnote{Note that this conformal embedding does not
  provide a true compactification of the spacetime, since
  (\ref{cembdnr}) is itself not compact. As noted above, this
  represents the necessary exclusion of the fixed points of the
  quotient in the Einstein static universe.}

Now, the space (\ref{cembdnr}) is simply a symmetric plane wave. This
can be made obvious by making the further coordinate
transformation\footnote{It is worth noting that there is a simple
  relation between these and the embedding coordinates for $\AdS_5$: $x^4
  -x^1 = (\sin U) /Z$, $x^4+x^1 = - (V \cos U + (X^2+Y^2+Z^2) \sin
  U)/Z$, $x^6-x^2 = (\cos U)/Z$, $x^6+x^2 = (V \sin U - (X^2+Y^2+Z^2)
  \cos U)/Z$, $x^3 = X/Z$, $x^5 = Y/Z$.}
\begin{equation} \label{symct}
 \begin{aligned}[m]
 V &= v + \psi \phi, \\
U &= t, \\
X  &= \psi \cos t + \phi \sin t, \\
Y &= -\psi \sin t + \phi \cos t,
 \end{aligned} 
\end{equation}
under which the metric becomes
\begin{equation} \label{sympw}
\bar g = - 2 dU dV - (X^2+Y^2+Z^2) dU^2 + dX^2 + dY^2 + dZ^2. 
\end{equation}

This provides an interesting alternative description of the double
null rotation, of interest independent of the question of the
conformal boundary.  As in section~\ref{sec:notsl}, this relation
between the symmetric plane wave and AdS is anticipated by previous
work, since they are both conformally flat spaces and hence
conformally embedded in the Einstein static universe. We see also that
AdS covers the half of the plane wave at $Z>0$, as we would expect,
since it covers half the Einstein static universe. What is remarkable
is that the isometry we want to quotient along commutes with the
conformal rescaling, as noted above.  In fact, not only does it do so;
all the unbroken symmetries of the double null rotation also do so, since
they do not involve $\partial_\rho$.  Thus, they are all symmetries of
the conformally related plane wave metric (\ref{sympw}). If we
introduce the usual basis for the Killing vectors of the plane wave,
\begin{equation}
\begin{aligned}[m]
\xi_{e_i} &= -\cos U \partial_{X^i} + X^i \sin U \partial_V, \\
\xi_{e_i^*} &= -\sin U \partial_{X^i} - X^i \cos U \partial_V, \\
\xi_{e_V} &= \partial_V, \\
\xi_{e_U} &= -\partial_U,
\end{aligned}
\end{equation}
we can identify the isometries of the double null rotation quotient as
\begin{equation}
  \begin{aligned}[m]
    \xi_{\AdS} &= - \xi_{e_1^*} - \xi_{e_2}, \\
    \xi_1 &= -\xi_{e_1^*} + \xi_{e_2},\\
    \xi_2 &= -\xi_{e_1} - \xi_{e_2^*}, \\
    \xi_3 &= \xi_{e_V}, \\
    \xi_4 &= \xi_{e_U} - \xi_{M_{12}}. 
  \end{aligned}
\end{equation}
Thus, the double null rotation is conformally related to a
compactification of the plane wave of the type considered
in~\cite{Michelson}. 

To return to the question of the conformal boundary
of the double null rotation, we see that it is given by the surface at
$Z=0$ in (\ref{cembdnr}), with metric
\begin{equation} \label{cpwbdy}
  g_\partial = - 2 dt dv +  d\psi^2 + d\phi^2 - 4\psi dt d\phi. 
\end{equation}
This is itself a compactified plane wave, as can be seen by the
application of the coordinate transformation (\ref{symct}). One might
be puzzled by this result, as one would have expected to find the
nullbrane as the conformal boundary of the double null rotation. We
demonstrate in appendix~\ref{sec:nullbrane} that the nullbrane is in
fact related to (\ref{cpwbdy}) by a further conformal
transformation. Thus, (\ref{cpwbdy}) and the nullbrane describe the
same conformal structure on the boundary. The description in terms of
the compactified plane wave (\ref{cpwbdy}) is preferable to the
nullbrane for two reasons: First, the nullbrane only covers a part of
the boundary [it corresponds to the region $-\pi/2 < t < \pi/2$ in
(\ref{cpwbdy})], so the former description is more global. Second, the
further conformal transformation to the nullbrane does not commute
with the symmetry $\xi_4$ of the double null rotation. If we work with
(\ref{cpwbdy}), all the unbroken symmetries of the bulk spacetime
after we perform the quotient are realised as symmetries of the
boundary (rather than conformal isometries). This should be a helpful
simplification in studying the holographic relation for this
spacetime.

The connection to plane waves also makes it easy to identify a time
function for the double null rotation. Writing the double null
rotation metric (\ref{zdnr}) in the form suitable for Kaluza-Klein
reduction along $\phi$,
\begin{equation}
  g = \frac{1}{Z^2} [-2dv dt -(Z^2 + 4 \psi^2) dt^2 + d\psi^2 + ( d\phi
  -2\psi dt)^2], 
\end{equation}
we see that the lower-dimensional spacetime would again be a plane
wave (up to conformal factor).  Hence, applying the
results of \cite{causal}, where time functions were found for general
plane waves, we can deduce that a suitable time function for the
nullbrane is
\begin{equation}
  \tau = t  + \frac{1}{2} \tan^{-1} \left( \frac{4 v}{1+Z^2+ 4\psi^2} \right).
\end{equation}
It is easy to check that 
\begin{equation}
  \nabla_\mu \tau \nabla^\mu \tau = - \frac{4 Z^2}{[(1+Z^2+4\psi^2)^2
    + 16v^2]}.
\end{equation}
Thus, $\tau$ is a good time function on AdS.  Since
$\mathcal{L}_{\xi_{\AdS}} \tau = 0$, its existence shows that the
double null rotation quotient of AdS preserves the property of stable
causality by the general argument of \cite{causal}.

As recently discussed in \cite{fofs2} \footnote{In \cite{simon}, it
  was claimed that the amount of supersymmetry preserved by the double
  null rotation quotient was $\nu=\frac{1}{4}$, but as shown in
  \cite{fofs2}, the latter is actually enhanced to
  $\nu=\frac{1}{2}$.}, the supersymmetry preserved by this double null
rotation quotient in $\AdS_5$, and actually in any higher dimensional
AdS spacetime embedded in a supergravity theory, is $\nu=\frac{1}{2}$.
That is, this configuration has sixteen supercharges.  It is
interesting to comment on the relation with the single null rotation
quotient.  In that case, we argued that the standard enhancement of
supersymmetry when taking the near horizon geometry was lost after the
identification.  This may suggest that the same phenomenon is taking
place in the double null rotation, since the action generated by the
latter is the combination of two commuting null rotations.  However,
the general solution to the eigenvalue problem
\begin{equation*}
  N\,\varepsilon = N_1\cdot N_2\,\varepsilon = 0,
\end{equation*}
where $N$ stands for the full double null rotation generator in the
spinorial representation, and $N_i$ $i=1,2$ stand for nilpotent
operators, is not given in terms of the intersection of kernels of the 
nilpotent operators associated with each of the null rotations, which
would give rise to $\nu=\frac{1}{4}$, but there exist non-trivial 
solutions \cite{fofs2} that enhance supersymmetry to one-half.  Thus, 
in this case, the double null rotation quotient preserves the same 
amount of supersymmetry as the corresponding asymptotically flat 
analogue in terms of parallel and coincident D3-branes in the 
nullbrane vacuum.

{\bf Deformation by $B^{(0,2)}$.} In order to turn on any deformation
parameter, we must consider higher dimensional AdS spacetimes.  In
particular, it is natural to consider $\AdS_7$, since this is very
naturally obtained in M-theory from the near horizon limit of
M5-branes.  If we denote by $\alpha$ the deformation parameter, the
deformed seven dimensional quotient can be written as
\begin{equation}
  g_{\AdS_7/\Gamma} = \cosh^2\chi\,g_{dnr} + d\chi^2 +
  \sinh^2\chi\,\left(d\varphi_1
  + \alpha\,d\phi\right)^2~.
\end{equation}
where $g_{dnr}$ stands for \eqref{globaldnr}.

Since we only turned on a single deformation parameter, $\alpha$,
the corresponding seven dimensional quotient, when embedded in string
theory, will break supersymmetry.  It is certainly possible to
construct supersymmetric versions of the latter by deforming the
orbifold action with a non-trivial action on $\Sph^4$.

\subsubsection{Sphere deformations}
\label{sec:dnrsphere}

Let us start our discussion on sphere deformations of the double null
rotation quotient by focusing on $\AdS_5\times S^5$. The family of
deformations is described by \eqref{eq:sphere5}, that is, by three
real parameters. As discussed in \cite{fofs2}, the only supersymmetric
loci in the fundamental region defined by the action of the Weyl group
is, besides the origin, given either by $\theta_1=\theta_2$ and $\theta_3=0$,
preserving $\nu=\frac{1}{4}$, or by $\theta_1 - \theta_2 \pm \theta_3
= 0$, preserving $\nu=\frac{1}{8}$.

The discussion for $\AdS_7\times S^4$ is analogous. In this case,
there exists a two parameter family of deformations, given by
\eqref{eq:sphere4}. The only supersymmetric loci in the fundamental
region defined by the action of the Weyl group is either the origin,
corresponding to the double null rotation quotient itself, or the line
$\theta_1=\theta_2$, which preserves $\nu=\frac{1}{4}$.

As an explicit example of a sphere deformation of the double null rotation
quotient, we shall focus on a supersymmetric deformation on
$\AdS_5\times \Sph^5$.  We will focus on the same sphere action considered
in section~\ref{sec:sdorbsphere}.  As before, we apply the general
formalism developed in \eqref{eq:dressing} for the full Killing vector
$\xi = \xi_{\AdS} + \xi_S$.  If we introduce adapted coordinates so
that $\xi = \partial_\phi$ by defining $\psi'= \psi - \gamma \phi$,
the full ten-dimensional metric on the quotient space will be
\begin{multline}
g = g_{dnr} + d\theta_1^2
  + \sin^2\theta_1\,d\theta_2^2 + \cos^2\theta_1\,\left(d\varphi_1 +
  2(d\psi' + \gamma d\phi)\right)^2 \\
  + \sin^2\theta_1\,\left((d\psi'+ \gamma d\phi)^2 + d\varphi^2 + 2\cos
  2\theta_2\,
  (d\psi'+ \gamma d\phi)\cdot \varphi\right)~.
\end{multline}
where $g_{dnr}$ denotes the metric on the quotient of AdS$_5$ given in
(\ref{globaldnr}). 

We could consider quotients involving both two forms $\oplus_i
B^{(0,2)}(\varphi_i)$ acting on AdS and sphere deformations. The
techniques required to deal with them are exactly the same as those
used above. The reader can find an analysis of their supersymmetry in \cite{fofs2}.

\section{Black holes as quotients}
\label{sec:bh}

In the previous section, we discussed causally regular quotients,
which arise in some cases where the Killing vector defining the
AdS orbifold is nowhere timelike.  One might think that these are the
cases of primary interest, since any other quotient will have at least
a region of closed timelike curves.  However, as is well known, certain
causally ill-behaved quotients can be given an interpretation as an
analogue of black holes \cite{btz1,btz2}.

The idea is that one can excise regions where closed timelike curves
will arise from the original spacetime, and consider the quotient just
of the remaining portion of AdS$_{p+1}$.  The resulting geometry will
be causally regular by construction, but will clearly not be
geodesically complete, having a `singularity' corresponding to the
boundary of the excised region.  This singularity is not a curvature
singularity in the classical geometry, but extending the spacetime
beyond it would introduce causal pathologies; it is therefore expected
on the basis of the chronology protection conjecture that quantum
corrections will lead to a true singularity at this location.  The
interesting question is whether this singularity is naked---that is,
visible from infinity---or concealed by an event horizon.  If it is
behind an event horizon, we view the quotient geometry as a black
hole, generalising the BTZ solution \cite{btz1,btz2}.

In this section, we will study which quotients can lead to
black holes of this type.  Unlike in the previous section, where
deformation on the sphere introduced qualitatively new possibilities,
we find that the  quotients with a black hole interpretation are
the BTZ quotients in AdS$_3$, and the higher-dimensional
generalisation of the non-rotating BTZ quotients, coupled with some
action on the sphere.

First, we need to establish what region of the spacetime we remove.
In \cite{hp}, where quotients acting just on the AdS factor were
considered, it was argued that we should remove the region where the
Killing vector $\xi_{\AdS}$ fails to be spacelike.  Clearly, the
quotient will contain closed timelike curves in this region. However,
it is not in general true that all closed timelike curves will pass
inside this region. In particular, for cases with $B^{(0,2)}(\varphi_i)$
components, this does not remove all the closed timelike curves. 

Closed timelike curves in the region where $\xi_{\AdS}$ is spacelike
can be constructed by an argument very similar
to that used in section~\ref{sec:cause}. As discussed at the beginning
of section~\ref{sec:nonsingular}, for any of our quotients, we can
construct a natural coordinate system \eqref{eq:adsslicing} on the AdS
part, in which we decompose AdS$_{p+1}$ in terms of an AdS$_{n+1}$ and a
$\Sph^{p-n-1}$ factors, where the Killing vector generating the quotient
is $\xi_{\AdS} = \xi_{\AdS_{n+1}} + \xi_r$, with $\xi_{\AdS_{n+1}}$
acting only on the AdS$_{n+1}$ part of the metric (\ref{eq:adsk}) and
containing the non-trivial block or blocks, while the $\xi_r$ is a
combination of rotations (the $B^{(0,2)}(\varphi_i)$ blocks) acting on
the unit sphere $\Sph^{p-n-1}$. Now consider an orbit where $\xi_{\AdS}$
is spacelike, but $\xi_{\AdS_{n+1}}$ is timelike. As in
section~\ref{sec:cause}, we can construct a closed curve which follows
the orbit of $\xi_{\AdS_{n+1}}$ on the AdS$_{n+1}$ factor and a
length-minimising geodesic on the $\Sph^{p-n-1}$ factor. There are
identified points which are separated by an arbitrarily large timelike
distance in the AdS$_{n+1}$ factor; since the separation on $\Sph^{p-n-1}$
is bounded, this closed curve will be timelike for sufficiently large
separation on the AdS$_{n+1}$ factor. Obviously, a similar argument
applies when we consider the deformation on the transverse sphere;
there will be closed timelike curves wherever the norm of the
non-trivial blocks taken on their own is timelike.

Thus, it would seem that a natural region to excise is the region
where $\xi_{\AdS_{n+1}}$ is timelike. That is, the region to excise is
determined by the norm of the non-trivial blocks, omitting all the
rotations (both $B^{(0,2)}(\varphi_i)$ and the rotations on transverse
spheres). Note however that this is still not sufficient to eliminate
the closed timelike curves in all cases. That is, the resulting
quotient is not guaranteed to be causally regular. However, this is
the only possibility we will consider here. It represents the natural
generalisation of the construction of black hole solutions
of~\cite{btz1,btz2} to higher dimensions. We will focus on seeing what
black analogues can be constructed by removing this portion of the
quotient. We will see that the resulting spacetime in the black hole
examples are in fact free of closed causal curves.

The singularity surface we consider is then where $\xi_{\AdS_{n+1}}
\cdot \xi_{\AdS_{n+1}} = 0$ in $\AdS_{p+1} \times \Sph^q$. Our main
concern for the rest of this section is to establish in which cases
this singularity surface is naked, and in which cases it is concealed
by an event horizon. Since $\xi_{\AdS_{n+1}}$ is a Killing field,
\begin{equation}
  \nabla_{\xi_{\AdS_{n+1}}} (\|\xi_{\AdS_{n+1}}\|^2) = 2\,i_{\xi_{\AdS_{n+1}}}
   \left(\nabla_{\xi_{\AdS_{n+1}}}\xi_{\AdS_{n+1}}\right) = 0,
\end{equation}
so $\xi_{\AdS_{n+1}}$ is always tangent to surfaces defined by
$\|\xi_{\AdS_{n+1}}\|^2 =$ constant.  Hence, the `singularity' defined by
$\|\xi_{\AdS_{n+1}}\|^2 = 0$ has a null tangent, and must be a timelike
or null surface.  We think of such a quotient as an analogue of a black
hole if there is a non-trivial event horizon $\dot{J}^-({\mathcal
  I}^+)$ in the quotient.  Since the singularity surface is timelike or
null, this can only happen if the singularity surface divides the
future null infinity ${\mathcal I}^+$ of the $\AdS_{p+1}$ spacetime
into disconnected regions.  The behaviour of the Killing vector on the
asymptotic boundary of the AdS spacetime is therefore essential in
determining if a given case is a black hole or not.

\subsection{AdS$_3$ black holes}

For the $\AdS_3$ case, the addition of a deformation on the sphere
does not significantly modify the analysis of \cite{btz2}: the only
quotients which lead to black holes are the ones whose AdS Killing
vector field is associated with the two-forms $B^{(1,1)} (\beta_1)
\oplus B^{(1,1)} (\beta_2)$, for $|\beta_1| \neq |\beta_2|$, and
$B^{(2,2)} (\beta)$ for $\beta \neq 0$, corresponding to non-extremal
and extremal black holes, respectively.  These AdS Killing vectors
correspond to type $I_b$ and type $II_a$ in the notation
of \cite{btz2} \footnote{Note that the $M=J=0$ black hole solutions
  of \cite{btz2}, obtained by quotienting by $B^{(1,2)}$, do not have
  a generalisation to include rotation on the sphere, as the
  associated AdS Killing vectors are nowhere timelike, so these give
  causally regular quotients once a non-trivial $\xi_{\Sph^3}$ is
  included, as described in the previous section.}.  When embedding
these black holes in string theory, it is certainly natural to embed
them in type IIB, in terms of $\AdS_3\times \Sph^3\times\TT^4$, coming
from the near horizon of the D1-D5 system.  Thus, the most general
Killing vector field giving rise to black holes is given by
\begin{equation}
  \xi = \xi_{BTZ} + \theta_1\,R_{12} + \theta_2\,R_{34}~,
\end{equation}
where we are using the notation introduced in
Section~\ref{sec:sphere}.

The metric on these solutions is easily constructed.  For simplicity,
we shall focus again on the deformation for which
$\theta_1=\theta_2=\gamma$.  Let us adopt BTZ coordinates on the AdS
space, so that $\xi_{\AdS_3} = \partial_\phi$, and adapted
coordinates on the sphere, so that $\xi_S = \partial_\psi$.  Then the
metric is
\begin{multline}
  g = -\frac{(r^2 - r_+^2)(r^2 - r_-^2)}{r^2} dt^2 + \frac{r^2 dr^2}{
  (r^2 - r_+^2) (r^2 - r_-^2)} + r^2 [ d\phi - \frac{r_- r_+}{r^2}
  dt]^2 \\ + d\theta^2 + d\chi^2 + d\psi^2 + 2 \cos 2\theta
  d\chi d\psi,
\end{multline}
and the quotient introduces the periodic identifications $\phi \sim
\phi + 2\pi m, \psi \sim \psi + 2\pi \gamma m$, $m \in \ZZ$.  If we
introduce a new coordinate $\tilde \psi = \psi - \gamma \phi$, then
$\xi = \partial_\phi$ and the metric in fully adapted coordinates is
\begin{multline}
  g = -\frac{(r^2 - r_+^2)(r^2 - r_-^2)}{r^2} dt^2 + \frac{r^2 dr^2}{
  (r^2 - r_+^2) (r^2 - r_-^2)} + r^2 [ d\phi - \frac{r_- r_+}{r^2}
  dt]^2 \\+ \gamma^2 d\phi^2 + 2 \gamma d\phi (d\tilde \psi + \cos
  2\theta d\chi) + d\theta^2 + d\chi^2 + d\tilde \psi^2 + 2 \cos
  2\theta d\chi d\tilde \psi.
\end{multline}
Note that the deformation on the sphere does not affect the leading
$r^2$ part of the metric at large distances, so the structure of the
asymptotic boundary of the black hole is not changed.  From the point
of view of Kaluza-Klein reduction over the sphere, this geometry is
described as the rotating BTZ black hole with a flat $SU(2)_L \subset
SO(4)$ gauge connection $A^3_{\phi} = \gamma$ turned on, in analogy
with previous discussions of conical defects \cite{bbkr}.  Since the
gauge field has zero stress-energy, it does not modify the
three-dimensional metric.  Its presence does however modify the
supersymmetry conditions \cite{bbkr}.  Unlike in the conical defect
case, we cannot make non-supersymmetric black hole solutions
supersymmetric by adding a deformation on the sphere, as we cannot
balance the hyperbolic black hole holonomy by a holonomy in $SU(2)$.

\subsection{Higher-dimensional black holes}

Let us now investigate what happens in higher dimensions.  For the
excision we are studying, the singularity is determined by the
non-trivial part of the AdS action, $\xi_{\AdS_{n+1}}$, and the
presence of horizons is determined by considering the intersection of
this singularity surface with the AdS boundary. We therefore focus on
the AdS part of the story, and only add in the sphere at the end.

We want to know if there is an event horizon in the quotient. Since
the location of the singularity is determined by $\xi_{\AdS_{n+1}}$,
it is natural to study this using the decomposition
(\ref{eq:adsslicing}). This considerably simplifies the task of
studying the higher-dimensional cases, by relating it to the
lower-dimensional classification. It would require considerable work to
determine directly from the form of the Killing vectors whether or not
event horizons exist.  By relating this question to the existence of
horizons in lower dimensions, we can avoid most of this work and also
gain some valuable insight into the differences between the AdS$_3$
case and higher dimensions.

For a Killing vector which does not contain a $B^{(2,3)}$ block, a
$B^{(2,4)}_\pm(\varphi)$ block, two $B^{(1,2)}$ blocks, or a
$B^{(1,2)}$ and a $B^{(1,1)}$ block, we can adapt the coordinate
system of (\ref{eq:adsk}) with $n=2$; that is, we can decompose
AdS$_{p+1}$ in terms of AdS$_3$ and $\Sph^{p-3}$ factors.  The Killing
vector then decomposes as $\xi_{\AdS} = \xi_{\AdS_3} + \xi_r$,
where $\xi_{\AdS_3}$ acts only on the AdS$_3$ part of the metric
(\ref{eq:adsk}) and contains the non-trivial block or blocks, while
the $\xi_r$ is a combination of rotations (the $B^{(0,2)}(\varphi_i)$
blocks) acting on the unit sphere $\Sph^{p-3}$.  Furthermore,
$\xi_{\AdS_3}$ is precisely the Killing vector associated to the
same type of quotient in the analysis of \cite{btz2}.

We would exploit this decomposition to simplify the problem of finding
horizons.  We will show that there is a simple condition on the action
in AdS$_3$ which will imply that the singularity is naked in
AdS$_{p+1}$.  The existence of a non-trivial event horizon in the
quotient spacetime implies that there are points in the singularity
surface $||\xi_{\AdS_3}||^2 =0$ which cannot be connected to the same
asymptotic region in both the past and the future.  Conversely, if a
point in AdS with $||\xi_{\AdS_3}||^2 =0$ lies on some timelike curve
which lies entirely in the region where $||\xi_{\AdS_3}||^2 \geq 0$ in
the bulk and starts and ends in some connected component of the region
of the boundary where $||\xi_{\AdS_3}||^2 >0$, this point on the
singularity will be naked in the quotient.  Thus, the existence of such
a curve implies the nakedness of the singularity.

Now, in the coordinates (\ref{eq:adsk}), we can consider the
restriction to the AdS$_3$ factor at some fixed point on the sphere
factor that $\xi_r$ acts on, and ask if there is such a curve which in
addition stays in this submanifold.  This will supply a sufficient
condition for nakedness of the singularity which can be expressed in
AdS$_3$ terms.  We therefore want to look for a timelike curve in
AdS$_3$ which connects points in the same connected component of the
region of the boundary where $||\xi_{\AdS_3}||^2 > 0$ through the
region where $||\xi_{\AdS_3}||^2 \geq 0$ in the bulk, and passing
through a point at $||\xi_{\AdS_3}||^2 = 0$. But this is the same
thing as the condition for a naked singularity in AdS$_3$: cases which
do not lead to black holes in AdS$_3$ do not lead to black holes in
higher dimensions either.  Horizons can only arise in the cases where
there is a horizon in the AdS$_3$ quotient.

Consider now the cases which give black holes in AdS$_3$; that is, the
$B^{(1,1)} (\beta_1) \oplus B^{(1,1)} (\beta_2)$, for $|\beta_1| \neq
|\beta_2|$, and $B^{(2,2)} (\beta)$ for $\beta \neq 0$. Consider first
the rotating black holes. We will see that there will be no horizons
in the higher-dimensional cases.  In the quotient of AdS$_3$, we
obtained a solution with an inner horizon and a timelike singularity,
so any point on the singularity surface was connected to the boundary
to both the past and future, but it was connected to different
components of the boundary, so this did not imply the absence of a
horizon.  In higher dimensions, however, we can describe the
asymptotic boundary in terms of an AdS$_3 \times \Sph^{p-3}$ metric,
\begin{equation}
g_{\partial} = g_{\AdS_3} + g_{\Sph^{p-3}}.
\end{equation}
Since the portion of the bulk of AdS$_3$ where $\xi_{\AdS_3}$ is
spacelike is connected, the portion of the boundary of AdS$_{p+1}$
where $\xi_{\AdS_3}$ is spacelike will be connected, and hence the
curves which link a point on the singularity to the boundary have
their endpoints in a single connected component of the region of the
boundary where $||\xi_{\AdS_3}||^2 >0$.  Thus, they imply that the
singularity is naked in the higher-dimensional quotients, as noted for
the case $p=3$ in \cite{hp}.

This leaves only the cases where we quotient by a Killing vector with
a single $B^{(1,1)}$ factor, which would correspond to a non-rotating
black hole in AdS$_3$. We will see shortly that this case does have
non-trivial event horizon for AdS$_{p+1}$, $p \geq 2$.  This is thus
the only case involving AdS$_3$ blocks with an event horizon in higher
dimensions.\footnote{We are again excluding the case of $B^{(1,2)}$,
  corresponding to an $M=0$ black hole, on the grounds that once we
  include rotation on the sphere, this will become a causally regular
  quotient.}

It remains to consider the Killing vectors containing blocks
$B^{(2,3)}$ and $B^{(2,4)}_\pm(\varphi)$, and the cases containing two
$B^{(1,2)}$ blocks or a $B^{(1,2)}$ block and a $B^{(1,1)}$ block.
However, these do not lead to any more examples with horizons.  For
two $B^{(1,2)}$ blocks, this is obvious, as the Killing vector is
nowhere timelike.  For the $B^{(2,3)}$ block, we can observe that it
was shown in \cite{hp} (where this case is called type $V$) that there
is no horizon in this case in AdS$_4$; this can easily be extended to
show that there is no horizon in higher dimensions by the arguments
used above. For a $B^{(1,2)}$ block and a $B^{(1,1)}$ block, we can
similarly appeal to the analysis of \cite{hp}.

For the $B^{(2,4)}_\pm(\varphi)$ blocks, we analyse the situation in
AdS$_5$, and appeal to the argument set forth above to extend the
conclusion to general dimensions.  In AdS$_5$, the Killing vector is
\begin{equation}
  \xi_{\AdS} = \be_{15} - \be_{35} \pm \be_{26} - \be_{46} + \varphi (\mp
  \be_{12} + \be_{34} + \be_{56})~.
\end{equation}
The norm of this Killing vector is
\begin{equation}
  \|\xi_{\AdS}\|^2 = -\varphi^2 + 4\varphi (x_6 (x_3- x_1) - x_5 (x_4
 \mp x_2)) + (x_3 - x_1)^2 + (x_4 \mp x_2)^2,
\end{equation}
where $\{x_1,\ldots,x_6\}$ are the $\RR^{2,4}$ embedding
coordinates.  Adapting a global coordinate system on $\AdS_5$,
\begin{equation}
  \begin{aligned}[m]
    x_1 & = \cosh \rho \cos t, \quad x_2 = \cosh \rho \sin t~,\\
    x_3 & = \sinh \rho \cos \theta \cos \phi, \quad x_4 = \sinh \rho \cos
    \theta \sin \phi, \\
    x_5 &= \sinh \rho \sin \theta \cos \psi, \quad x_6 = \sinh \rho
    \sin\theta \sin \psi~,
  \end{aligned}
\end{equation}
the norm becomes
\begin{multline*}
  \|\xi_{\AdS}\|^2 = -\varphi^2 + 4 \varphi \sinh \rho \sin \theta \left[
  - \cosh \rho \sin (\psi \pm t) + \sinh \rho \cos \theta \sin(\psi -
  \phi)\right] \\
  + \cosh^2 \rho + \sinh^2 \rho \cos^2 \theta - 2 \cosh \rho
  \sinh \rho \cos \theta \cos(\phi \pm t)~.
\end{multline*}
Thus, we see that the global time dependence of the norm is simply a
simultaneous rotation in the two angles $\phi,\psi$ on the $\Sph^3$ in
AdS$_5$.  Thus, the region of the boundary where the norm of the
Killing vector is spacelike is clearly connected, and
this case does not give rise to a black hole in any dimension.

Thus, the only quotient with a black hole interpretation for $p>2$ is
the quotient by an AdS Killing vector $B^{(1,1)}(\beta) \oplus_i
B^{(0,2)}(\varphi_i)$. The resulting quotient is the
higher-dimensional generalisation of the non-rotating BTZ black hole.
Special cases of this solution for $p=3,4$ have been discussed before
in \cite{hp,ban1,ban2}.\footnote{Note that in \cite{hp}, it was
  claimed that this does not lead to a black hole for $\varphi_i \neq
  0$. This is because \cite{hp} took the singularity surface to be $||
  \xi_{\AdS} ||^2 = 0$, which does not eliminate all closed timelike
  curves in this case. We take the singularity surface to be
  $||\xi_{\AdS_3}||^2=0$, cutting out more of the global AdS
  spacetime; this gives a causally regular spacetime which can be
  interpreted as a black hole.} As above, the natural coordinate
system on these quotients in general is the one given by the
decomposition \eqref{eq:adsslicing}. If we adopt adapted coordinates
for the $B^{(1,1)}(\beta)$ action on the $\AdS_3$ factor, this is
\begin{equation}
  g = \cosh^2 \chi \left( - (r^2 -1) dt^2 + \frac{dr^2}{r^2-1} + r^2
  d\phi^2 \right) + d\chi^2 + \sinh^2 \chi d\Omega_{p-3}~.
 \label{btzbh}
\end{equation}
where we have re-absorbed the length scale $r_+$ associated with the
black hole by rescaling coordinates, so the period of the angular
coordinate $\phi$ depends on $r_+$. The quotient makes identifications
in $\phi$ with some twist on the $\Sph^{p-3}$ determined by the
$\varphi_i$. We note that although these are deformations of the
higher-dimensional BTZ quotient by rotations, they do not look like
rotating black holes in the usual sense: $\partial_t$ is still
hypersurface-orthogonal, and there is a single horizon.

The special case where we consider a simple boost, so $\varphi_i = 0$,
was considered in detail in \cite{hp,ban1,ban2}. In this case the
quotient preserves, in addition to the symmetry associated with $\xi$,
an $\SO(1,p-1)$ symmetry in the orthogonal subspace. Various
coordinate systems were defined on the quotient which are adapted to
make some or all of this symmetry manifest in \cite{ban1,ban2}.  We
would like to briefly connect to that work by showing how our
preferred coordinate system above which makes the AdS$_3$ structure
manifest is connected to one of those coordinate systems.  

In \cite{ban2}, ``spherical'' coordinates were defined, in which the
metric takes the form
\begin{equation}
  g = (\rho^2 - 1)\left[ -\sin^2 \theta dt^2 +d\theta^2 + \cos^2 \theta
  d\Omega_{p-3}\right] + \frac{d\rho^2}{(\rho^2 -1)} + \rho^2 d\phi^2.
 \label{banbh}
\end{equation}
These coordinates are one example
of coordinates adapted to the $\SO(1,p-1) \times \SO(1,1)$ symmetry of
this spacetime.  They are related to \eqref{btzbh} by the coordinate
transformation
\begin{equation}
  \cos \theta = \frac{\sinh \chi}{\sqrt{\rho^2-1}}, \quad \rho = r \cosh\chi~.
\end{equation}
It is interesting to note that this shows that the $\SO(1,1)$ manifest
in (\ref{banbh}) is precisely the time translation of the BTZ black
hole.  Note that the spherical coordinates of (\ref{banbh}) cover more
of the spacetime than the BTZ coordinates of (\ref{btzbh}).  This
illustrates that while the coordinates we have constructed adapted to
the decomposition of the Killing vector in terms of lower-dimensional
quotients are useful, they are not the best coordinate system for
every purpose.

Another interesting coordinate system on this quotient is the `de
Sitter' coordinates of \cite{ban2}, which were used in
\cite{bubble,birm}, where this locally AdS$_{p+1}$ black hole arises
as the asymptotic behaviour of the bubble of nothing solution.  In
that context, it is convenient to adopt a coordinate system in which
the metric is
\begin{equation}
  g = (1+R^2) d\phi^2 + \frac{dR^2}{1+R^2} + R^2 \left[-d\tau^2 + \cosh^2
  \tau(d \tilde\theta^2 + \sin^2 \tilde\theta d\Omega_{p-3})\right]~.
 \label{bubblemet}
\end{equation}
These coordinates are adapted to the same $\SO(1,p-1) \times \SO(1,1)$
symmetry as in \eqref{banbh}.  The coordinate transformation relating
\eqref{bubblemet} to \eqref{banbh} is
\begin{equation}
  \rho^2 = 1+R^2, \quad \cos \theta = \cosh \tau \sin \tilde \theta~,
  \quad \tanh t = \frac{\tanh \tau}{\cos \tilde \theta}~.
\end{equation}
These `de Sitter' coordinates have the advantage that they cover the
whole exterior region of the black hole.  They demonstrate that the
black hole is not a static solution in higher dimensions; there is no
Killing vector which is timelike everywhere outside the black hole
event horizon.

As in the three-dimensional case, when we consider the quotient of
$\AdS_{p+1} \times \Sph^q$, we can write the AdS and sphere factors in
adapted coordinates separately, so that $\xi_{\AdS} = \partial_\phi$,
and $\xi_{\Sph} = \partial_\psi$.  Fully adapted coordinates are then
obtained by setting $\tilde \psi = \psi - \gamma \phi$, which
introduces $O(1)$ cross terms between AdS and sphere coordinates.
Again, from the Kaluza-Klein reduced point of view, what we are doing
is introducing a flat $\SO(q+1)$ gauge connection $A_\phi^a = \gamma$
on the black hole solution above, without modifying the metric.

One other issue deserves remarking on on the subject of black holes:
in \cite{ban2}, it was claimed that a rotating black hole solution
could be constructed by taking a quotient of AdS$_5$.  We want to
point out that this is not the same as the deformation by
$B^{(0,2)}(\varphi_i)$ discussed above; in fact, this quotient is not
a black hole.  The solution of \cite{ban2} was given by considering
$\AdS_5$ in the coordinates
\begin{equation}
  g = \sinh^2 \rho\left[ -\cos^2\theta d\tilde t^2 + d\theta^2 + \sin^2 \theta
  d\psi^2\right] + d\rho^2 + \cosh^2 \rho d\tilde \phi^2,
 \label{bancs}
\end{equation}
and making identifications along $\phi = \tilde \phi$ at fixed
$t =\frac{r_+ \tilde t - r_- \tilde \phi}{r_+^2-r_-^2}$.  This gives a
`black hole' metric of the form
\begin{multline}
  g = \cos^2 \theta \left[ - \frac{(r^2 - r_+^2)\cdot (r^2 - r_-^2)}{r^2}
  dt^2 + r^2 \left( d\phi - \frac{r_-}{r_+\cdot r^2}\, (r^2 -r_+^2) dt
  \right)^2 \right] \\
  + \frac{r^2 dr^2}{(r^2- r_+^2)\cdot (r^2 -r_-^2)}
  + \frac{(r^2 - r_+^2)}{(r_+^2 - r_-^2)}\, (d\theta^2 + \sin^2 \theta
  d\psi^2) + \frac{r_+^2\, (r^2 - r_-^2)}{(r_+^2 - r_-^2)} \sin^2 \theta
  d\phi^2,
 \label{banrot}
\end{multline}
where $r^2 = r_+^2 \cosh^2 \rho - r_-^2 \sinh^2 \rho$.  Since the
coordinates $\tilde t$ and $\tilde \phi$ in (\ref{bancs}) both
parametrise $\SO(1,1)$ symmetries (while $\chi$ parametrises an $\SO(2)$
symmetry), we can easily see that this quotient corresponds to the
rotating BTZ black hole type of quotient: that is, to a quotient by a
Killing vector formed from $B^{(1,1)} (\beta_1) \oplus B^{(1,1)}
(\beta_2)$, with $\beta_1 \beta_2 \neq 0$.  This can be seen explicitly
by noting that defining the new coordinates $\chi, \bar{r}$ by
\begin{equation}
  \begin{aligned}[m]
    \sinh^2 \chi & = \frac{(r^2 - r_+^2)}{(r_+^2 - r_-^2)} \sin^2 \theta~, \\
    \bar{r}^2 - r_-^2 & = \frac{r^2 - r_-^2}{\cosh^2 \chi}~,
  \end{aligned}
\end{equation}
we can rewrite \eqref{banrot} as
\begin{multline}
  g = \cosh^2 \chi \left[ - \frac{(\bar r^2 - r_+^2)\cdot (\bar r^2 - r_-^2)}
  {\bar r^2} dt^2 + \bar r^2 \left( d\phi + \frac{r_-}{r_+\, \bar r^2}\,
  (\bar r^2 - r_+^2) dt \right)^2 \right.\\
  + \left. \frac{\bar r^2\, d\bar r^2}{(\bar r^2- r_+^2)\cdot (\bar
  r^2 -r_-^2)} 
  \right] + d\chi^2 + \sinh^2 \chi d\psi^2~,
\end{multline}
showing that the quotient space has a rotating BTZ black hole factor
and a circle factor, as expected for this type of quotient.  Now, we
have argued above that the presence of a rotating BTZ black hole
factor implies that the region of the boundary of AdS$_5$ where the
Killing vector we are quotienting along is spacelike is connected.
Thus, this quotient cannot lead to an event horizon.  The apparent
presence of an event horizon in the coordinates \eqref{banrot} is
attributable to those coordinates not covering the whole of infinity.

\section{On Penrose limits of discrete quotients}
\label{sec:disc}

In Section~\ref{sec:cause}, we determined the subset of quotients of
$\AdS_{p+1}\times \Sph^q$ spacetimes having closed timelike curves.
In the main body of this work, we focused on the quotients which are
free of closed causal curves, or on those having them, but allowing a
black hole interpretation.  We would like to finish our work with some
short discussion regarding the relation of a subset of quotients of
$\AdS_{p+1}\times\Sph^q$ having closed timelike curves and not falling
in the black hole category and {\it compactified} plane waves and
Gödel-type universes, both having closed timelike curves.  The
relation between closed timelike curves in quotients of
$\AdS_{p+1}\times\Sph^q$ and compactified plane waves was already briefly
commented on in \cite{causal}.

That such a relation should exist is very intuitive, given the
existing relation between Penrose limits of $\AdS_{p+1}\times \Sph^q$
and plane waves \cite{Gueven,BFOHP,BFOP}, and the T-duality relation
between the latter and Gödel-type universes
\cite{BGHV,HarTak,MaozSimon}\footnote{The interplay between Penrose
  limits and quotients of AdS was also considered in \cite{AST},
  although their physical motivation was not related to closed
  timelike curves.}.  One possible motivation to make this connection
more precise could be the fact that AdS/CFT \cite{mald} could shed
some light on the issue of physics in the presence of closed timelike
curves.

In general, the operation consisting on taking the Penrose
limit of a given configuration $\eM$ does not commute with the
operation of considering a discrete quotient in $\eM$.  Even though we
do not have a general statement, it turns out that for abelian discrete
quotients whose generator belongs to the maximal compact subgroup of AdS,
that is for two-forms $B^{(2,0)}(\varphi)\oplus_i B^{(0,2)}(\varphi_i)$, the
following diagram commutes.
\begin{equation}
  \begin{CD}
    \Big(\text{AdS}_p\times\text{S}^q\Big)/\Gamma @>\text{T-duality}>> \eN \\
       @VV{\operatorname{\text{Penrose limit}}}V
       @VV{\operatorname{\text{Penrose limit}}}V \\
    \Big(\text{plane wave}\Big)/\Gamma     @>\text{T-duality}>>
       \text{G\"odel-type} 
  \end{CD}
\end{equation}

Let us make the connection more explicit.  Even though we could develop
the discussion in general, we shall focus on $\AdS_3\times\Sph^3$ for
algebraic simplicity.  Consider the quotient generated by
\begin{equation}
  \xi_c = \beta\,A_-\,\partial_\tau + \beta\,A_+\,\partial_\psi +
   \beta\left(\partial_\varphi + \partial_\chi\right) ,
 \label{eq:compact}
\end{equation}
where $\{\tau,\rho,\varphi\}$ are global coordinates in $\AdS_3$ and
$\{\theta,\psi,\chi\}$ are global coordinates in $\Sph^3$, whereas
$\beta$ is any non-vanishing real number and $A_\pm$ are defined as
\begin{equation}
  A_\pm = \left(1 \pm \frac{1}{4\beta^2 R^2}\right)~.
\end{equation}
The norm of such Killing vector field is given by
\begin{equation}
  \|\xi_c\|^2 = \frac{1}{16\beta^2 R^2}\left\{\cosh^2\rho (8\beta^2 R^2 - 1)
  + \cos^2\theta (8\beta^2 R^2 + 1)\right\}~.
 \label{eq:normc}
\end{equation}
Thus, $\|\xi_c\|^2>0$ whenever $8\beta^2\,R^2>0$ $\forall$ $\rho,\theta$.
Even if this property is satisfied, we know the corresponding discrete
quotient will have closed timelike curves, as proved in
section~\ref{sec:cause}.

It is convenient for our purposes to make the change of variables
\begin{equation}
  \begin{aligned}[m]
    \tau = \beta\,A_-\,u + x^- \quad &,\quad \varphi=\hat\varphi +
    \beta\,u~, \\
    \psi = \beta\,A_+\,u - x^- \quad &, \quad \chi = \hat\chi + \beta\,u~,
  \end{aligned}
\end{equation}
in which $\xi_c = \partial_u$.  The global metric describing the above
quotient of
$\AdS_3\times\Sph^3$ consists in rewriting the metric in the new
adapted coordinate
system and making $u$ compact.  The result is
\begin{multline}
  g = -R^2\,\left(\cosh^2\rho -\cos^2\theta\right)\,(dx^-)^2 +
  R^2\,\left(d\rho^2 +
  \sinh^2\rho\,d\hat\varphi^2 + d\theta^2 +
  \sin^2\theta\,d\hat\chi^2\right) \\
  + 2\beta\,R^2\,du\left(\sin^2\theta\,d\hat\chi +
  \sinh^2\rho\,d\hat\varphi - (A_-\,\cosh^2\rho + A_+\,\cos^2\theta )\,dx^-
  \right) \\
  + \|\xi_c\|^2\,du^2 ~.
 \label{eq:adssphecomp}
\end{multline}
The full type IIB configuration certainly includes a transverse $\TT^4$,
and some fluxes.  It will not be necessary for our purposes to write these
explicitly, but we shall keep in mind that we are working with a vacuum
in which no NS-NS three-form field strength is turned on.

We shall first show that the Penrose limit of \eqref{eq:adssphecomp} is
indeed a quotient of a plane wave.  The procedure is by now standard.  Thus,
we shall just state that one needs to rescale $x^-=R^{-2}\,v$, take the
limit $R\to\infty$ while focusing on the lightlike geodesic sitting at
$\rho=\theta=0$.  Thus, we also need the rescalings $\rho = \frac{r}{R}$
and $\theta=\frac{y}{R}$.  Following this prescription, and having in mind
that $u$ is compact, we can afterwards apply a T-dual transformation giving
rise to a Gödel-type spacetime, in particular, to one dual version of
$\eG_5$, following the conventions introduced in \cite{HarTak}.  Of course,
the dual configuration will have a non-vanishing NS-NS two-form potential,
by construction, due to the crossed-terms in the metric
\eqref{eq:adssphecomp}.

We would be interested in determining the spacetime that we get after
applying the upper horizontal transformation in the diagram above.
This corresponds to applying a T-duality transformation along
the orbits of $\partial_u$.  The T-dual metric that we get in this way
is given by
\begin{multline}
  g^\prime = -R^2\,\left(\cosh^2\rho -\cos^2\theta\right)\,(dx^-)^2
  + R^2\,\left(d\rho^2 +
  \sinh^2\rho\,d\hat\varphi^2 + d\theta^2 +
  \sin^2\theta\,d\hat\chi^2\right) \\
  - \frac{\beta^2\,R^4}{\|\xi_c\|^2}\,\left(
  \sin^2\theta\,d\hat\chi + \sinh^2\rho\,d\hat\varphi -
  \left(A_+\,\cos^2\theta
  + A_-\,\cosh^2\rho\right)\,dx^-\right)^2 \\
  + \|\xi_c\|^{-2}\,du^2~.
\end{multline}
It is a straightforward exercise to check that the Penrose limit of the
above metric gives rise to $\eG_5$.  The corresponding fluxes can also
be matched.  Note that $\|\xi_c\|^2\to 1$ in the Penrose limit, which
matches the construction given, for instance, in \cite{BGHV}.

Thus, indeed, it is possible to understand the physics of Gödel-type
universes as describing the physics of certain sectors of the dual field
theory associated with the discrete quotient of the original $\AdS_3\times
\Sph^3$, following \cite{BMN}.  However, we also see that the dual
field theory is living in a space with closed timelike curves.
One easy way to realise this fact is to note that the action of the Killing
vector field $\xi_c$ acts in the same way at any value of the non-compact
spacelike coordinate $\rho$ in $\AdS_3$, in particular at its conformal
boundary.  Actually, the argument applies to any $\AdS_{p+1}$ spacetime.
We thus learn that if $\AdS_{p+1}/\Gamma$ is the geometry of the
bulk, where $\Gamma$ stands for the discrete group associated with the
discrete quotient generated by $B^{(2,0)}(\varphi)\oplus_i
B^{(0,2)}(\varphi_i)$, its conformal boundary is given by
$\Big(\RR\times \Sph^{p-1}\Big)/\hat\Gamma$, where $\hat\Gamma$ stands
for the restriction of $\Gamma$ on the boundary.  The conformal boundary
quotient would possibly include a non-trivial action on the fields
coming from the R--symmetry group.  Thus, whenever $\hat\Gamma$
acts non-trivially on the real timelike $\RR$ axis, the boundary
theory will be defined in a base space having closed timelike curves,
and as such, it will be non-globally hyperbolic.  Therefore, any
holographic description for these scenarios involves an understanding
of field theory in non-globally hyperbolic spaces, which we are
generically missing.

\section*{Acknowledgements}

This work is part of a project conceived while two of the authors (JF
and JS) were participating in the programme \emph{Mathematical Aspects
  of String Theory} which took place in the Fall of 2001 at the Erwin
Schrödinger Institute in Vienna, and it is again our pleasure to
thank them for support and for providing such a stimulating
environment in which to do research.  JMF's participation in this
programme was made possible in part by a travel grant from the UK
PPARC.  JMF would like to acknowledge the support and hospitality of
the following institutions visited during the time that it took to
complete and write up this work: the IHÉS in the Summer of 2002, the
Weizmann Institute in May 2003 and CERN in August 2003.  In particular
he would like to thank Micha Berkooz for the invitation to visit the
Weizmann Institute.  JMF is a member of EDGE, Research Training
Network HPRN-CT-2000-00101, supported by The European Human Potential
Programme, and his research is partially supported by the UK EPSRC
grant GR/R62694/01.  JS would like to thank the hospitality of the
following institutions during the different stages of this project:
the School of Mathematics of the University of Edinburgh, the
Perimeter Institute in Waterloo and the Institute for Theoretical
Physics in the University of Amsterdam.  JS and SFR would like to thank
the Aspen Center for Physics for hospitality.  During the time it took
to complete this work, JS was supported by a
Marie Curie Fellowship of the European Community programme ``Improving
the Human Research Potential and the Socio-Economic Knowledge Base''
under the contract number HPMF-CT-2000-00480, by the Phil Zacharia
fellowship, by grants from the United States--Israel Binational
Science Foundation (BSF), the European RTN network HPRN-CT-2000-00122,
Minerva, by the United States Department of Energy under the grant
number DE-FG02-95ER40893 and the National Science Foundation under Grant
No.~PHY99-07949. OM is supported by a PPARC studentship.  
SFR is supported by the EPSRC.

\appendix

\section{Global vs Poincaré patch in AdS}
\label{sec:patch}

In this appendix, we will review the global coordinate and
Poincaré patch descriptions of $\AdS_{p+1}$.  We wish to remind
the reader of the expressions for the Killing vector fields generating
the isometries in these two coordinate systems.  For the Poincaré
patch, this will be useful for understanding the relation between
certain global AdS quotients and the near horizon limit of the
corresponding discrete quotients of brane geometries in
supergravity.  For global coordinates, this will be useful for
understanding the action of the Killing vectors on the ESU boundary of
AdS.

Considering first the Poincaré coordinates, let us define
$\{y^\mu,z\}$ $\mu=2,\dots ,p+1$ in terms of the flat embedding
coordinates in $\RR^{2,p}$ introduced in \eqref{eq:ads} by
\begin{equation}
  \begin{aligned}[m]
    x^\mu &= \frac{1}{z}\,y^\mu \quad \mu=2,\dots ,p+1 \\
    x^1 &= \frac{1}{2z}\left[z^2+
    \left(1+\eta_{\mu\nu}\,y^\mu y^\nu\right)\right] \\
    x^{p+2} &= \frac{1}{2z}\left[z^2-
    \left(1-\eta_{\mu\nu}\,y^\mu y^\nu \right)\right]~.
  \end{aligned}
 \label{eq:patchmap}
\end{equation}
In these coordinates, the AdS$_{p+1}$ metric is
\begin{equation}
  g = \frac{1}{z^2} ( \eta_{\mu\nu} dy^\mu dy^\nu + dz^2).
\end{equation}
The explicit symmetries in this form of the metric are the
Poincaré symmetries acting on the slices of constant $z$.
Using the identities
\begin{equation*}
  \frac{\partial x^\mu}{\partial y^\nu} = \frac{1}{z}\delta^\mu_\nu
  \quad , \quad \frac{\partial x^{p+2}}{\partial y^\nu} =
  \frac{\partial x^1}{\partial y^\nu}=\eta_{\nu\mu}\,x^\mu
  \quad , \quad x^1-x^{p+2}= \frac{1}{z} ~,
\end{equation*}
we see that these
are related to the usual $\fso(2,p)$ basis by
\begin{equation}
  \begin{aligned}[m]
    P_\mu &= \partial_{y^\mu} \,\rightarrow\, -
    \left(\be_{1\mu} - \be_{\mu p+2}\right) \\
    L_{\mu\nu} &= y_\mu\partial_{y^\nu} - y_\nu\partial_{y^\mu}
     \,\rightarrow\, \be_{\mu\nu}~.
  \end{aligned}
 \label{eq:match}
\end{equation}
Therefore, timelike translations in the Poincaré patch correspond
to a null rotation with two timelike directions in global AdS,
which is mapped to the two-form $B^{(2,1)}$.  On the other hand,
spacelike translations in the Poincaré patch correspond to a
standard null rotation with two spacelike directions, or equivalently,
to $B^{(1,2)}$.  Finally, Lorentz transformations in the Poincaré
patch are mapped to Lorentz transformations in $\RR^{2,p}$.

The other symmetries in $\fso(2,p)$ are realised as conformal
symmetries acting on the slices of constant $z$ together with a
suitable $\partial_z$ component:
\begin{equation}
\be_{1\mu} + \be_{\mu p+2} = - \eta_{\sigma\nu} y^\sigma y^\nu
\partial_{y^\mu} + 2 y_\mu
y^\nu \partial_{y^\nu} + 2 z y_\mu \partial_z,
\end{equation}
\begin{equation}
\be_{1 p+2} = - y^\mu \partial_{y^\mu} - z \partial_z.
\end{equation}

A convenient global coordinate system on AdS$_{p+1}$ is defined in
terms of the embedding coordinates by
\begin{equation}
\begin{aligned}[m]
x_1 &= \cosh \chi \sin \tau, \\
x_2 &= \cosh \chi \cos \tau, \\
x_m &= \sinh \chi \,\hat x_m, \quad  m = 3,\ldots,p+2,
\end{aligned}
\end{equation}
where the $\hat x_m$ are embedding coordinates for an $\Sph^{p-1}$,
$\sum_m \hat x_m^2 = 1$.  The metric in this coordinate system is
\begin{equation}
g = -\cosh^2 \chi d\tau^2 + d\chi^2 + \sinh^2 \chi d\Omega_{p-1}.
\end{equation}
The explicit symmetries of this form of the metric are the
time-translation
\begin{equation}
\be_{12} = \partial_\tau,
\end{equation}
and the $\fso(p)$ symmetries of the sphere,
\begin{equation}
\be_{mn} = \hat x_m \partial_{\hat x_n} - \hat x_n \partial_{\hat x_m},
\quad m,n = 3,\ldots,p+2.
\end{equation}
The other Killing vectors are
\begin{equation}
\begin{aligned}[m]
\be_{1m} &= \cos \tau \tanh \chi \hat x_m \partial_\tau + \sin \tau \hat
x_m \partial_\chi + \sin \tau \coth \chi (\delta_{mn} - \hat x_m \hat
x_n) \partial_{\hat x_n}, \\
\be_{2m} &= - \sin \tau \tanh \chi \hat x_m \partial_\tau + \cos \tau \hat
x_m \partial_\chi + \cos \tau \coth \chi (\delta_{mn} - \hat x_m \hat
x_n) \partial_{\hat x_n}, 
\end{aligned}
\end{equation}
where $m,n = 3,\ldots,p+2$.

\section{Symmetry-adapted coordinates for nullbranes}
\label{sec:nullbrane}

As a byproduct of our investigations of the quotients of Anti-de
Sitter space in this paper---most particularly, the studies of the
double null rotations in section~\ref{sec:dnrdef}---we were led to
realise that there is a rich structure of symmetries in the nullbrane
quotients of flat space which has not been fully exploited in previous
work on these solutions.

The nullbrane is a quotient of flat $\mathbb{R}^{1,3}$ by a
combination of a null rotation and a translation \cite{fofs},
\begin{equation}
  \xi = \partial_4 -\be_{12} + \be_{23} = \partial_4 + (x^1-x^3)\partial_2
  + x^2(\partial_1 + \partial_3 ),
\end{equation}
where $x^1$ is the timelike coordinate and $\{x_2,x_3,x_4\}$ are
spacelike ones.  The norm of this Killing vector is $\|\xi\|^2 = (x_1
- x_3)^2 +1 $, so it is spacelike everywhere.  This quotient was shown
to be free of closed causal curves in \cite{fofs}.  There are three
Killing vectors in the $\fso(1,3) \ltimes \mathbb{R}^4$ Poincaré
algebra on $\mathbb{R}^{1,3}$ which commute with this $\xi$,
\begin{equation}
  \begin{aligned}[m]
    \xi_1 &= -\partial_4 -\be_{12} + \be_{23}, \\
    \xi_2 &= \partial_2 -(\be_{14} + \be_{34}) , \\
    \xi_3 &= \partial_1 + \partial_3~.
  \end{aligned}
\end{equation}
These have norms $\|\xi_1\|^2 = \|\xi_2\|^2 = \|\xi\|^2$ and
$\|\xi_3\|^2 = 0$.  The only non-trivial commutation relation is
$[\xi_1, \xi_2] = -2\xi_3$.  The coordinates defined on the nullbrane
in \cite{fofs} do not make any of these additional symmetries
manifest.  We will now construct an adapted coordinate system which
makes the $\xi_2$ and $\xi_3$ symmetries manifest: that is, we want
$\xi = \partial_{\bar\phi}$, $\xi_2 = \partial_{\bar\psi}$ and $\xi_3
= \partial_{\bar v}$.  This requires
\begin{equation}
  \begin{aligned}[m]
    \frac{\partial x^1}{\partial \bar\phi} = \frac{\partial
      x^3}{\partial\bar\phi} =x^2, & \frac{\partial x_2}{\partial\bar
      \phi} = x^1 - x^3,
    \frac{\partial x^4}{\partial \bar\phi} = 1, \\
    \frac{\partial x^1}{\partial \bar\psi} = \frac{\partial
      x^3}{\partial\bar\psi} =x^2, & \frac{\partial
      x_2}{\partial\bar\psi} = 1, \frac{\partial
      x^4}{\partial \bar\psi} = x^1-x^3, \\
    \frac{\partial x^1}{\partial \bar v} &= \frac{\partial
      x^3}{\partial\bar v} = 1~.
  \end{aligned}
\end{equation}
Since $x^1-x^3$ is independent of $ \bar\phi,\bar\psi,\bar v$, we will
choose to define 
coordinates so that $x^1 - x^3 = \bar u$.  A suitable coordinate system is
\begin{equation} \label{nbct1}
  \begin{aligned}[m]
    x^1 + x^3 &= 2 \bar \phi \bar\psi + \bar u (\bar\phi^2 +
    \bar\psi^2) + 2\bar v,   \\
    x^1 - x^3 &= \bar u,\\
    x^2 &= \bar\psi + \bar u \bar\phi, \\
    x^4 &= \bar\phi + \bar u \bar\psi~.
  \end{aligned}
\end{equation}
In these coordinates, the flat metric is
\begin{equation} \label{symnbr}
  g = -2d\bar u d\bar v + (1+\bar u^2) (d\bar \psi^2 + d\bar \phi^2) +
  4\bar u d\bar \phi d\bar \psi~.
\end{equation}
The nullbrane is constructed by compactifying the $\bar \phi$ coordinate.
The determinant of the metric is $-\det g = (1-\bar u^2)^2$, so this
coordinate system breaks down at $\bar u= \pm 1$, where the expressions for
$x^2$ and $x^4$ lose their linear independence. Thus, although these
are symmetry-adapted coordinates, they do not provide global
coordinates for the spacetime. 

It is interesting to note that in these coordinates, the solution
resembles a plane wave written in Rosen coordinates.  For the
uncompactified solution, this is not unexpected; flat space is a
trivial plane wave.  The interesting observation is that the
compactification of $\phi$ preserves this structure. By a slight
change in the coordinate system, we can make a more direct relation to
a non-trivial plane wave, and at the same time obtain global
coordinates. Instead of (\ref{nbct1}), we set 
\begin{equation} \label{nbct2}
  \begin{aligned}[m]
    x^1 + x^3 &= 2 \phi \psi + u (\phi^2 + \psi^2) + 2v,   \\
    x^1 - x^3 &= u,\\
    x^2 &= \psi + u \phi, \\
    x^4 &= \phi - u \psi~.
  \end{aligned}
\end{equation}
The flat metric is now 
\begin{equation}  \label{glnbr}
  g = -2du dv + (1+u^2) (d\psi^2 + d\phi^2) - 4\psi d\phi du~.
\end{equation}
The determinant of the metric is $-\det g = (1+u^2)^2$, so this is now
a global coordinate system. 

The price we pay is that the symmetry $\xi_2$ is no longer manifest;
on the other hand, this form treats the two Killing vectors $\xi_1$
and $\xi_2$ more symmetrically.  In these coordinates, $\xi =
\partial_\phi$, $\xi_3 = \partial_v$, while the other two Killing
vectors are
\begin{equation}
  \begin{aligned}[m]
  \xi_1 &= -\frac{1-u^2}{1+u^2} \partial_\phi + \frac{2u}{1+u^2}
  \partial_\psi + 2\psi \frac{1-u^2}{1+u^2} \partial_v~, \\
  \xi_2 &= \frac{2u}{1+u^2} \partial_\phi + \frac{1-u^2}{1+u^2}
  \partial_\psi - 2\psi \frac{2u}{1+u^2} \partial_v~.  
  \end{aligned}
\end{equation}
The inverse coordinate transformation is
\begin{equation}
  \begin{aligned}[m]
    u &= x^1 - x^3, \\
    \phi &= \frac{x^4 + (x^1 - x^3) x^2}{[1+ (x^1 - x^3)^2]},  \\
    \psi &= \frac{x^2 - (x^1 - x^3) x^4}{[1+ (x^1 - x^3)^2]},   \\
    2v &= (x^1 + x^3) - \frac{(x^1-x^3)}{1 + (x^1 - x^3)^2} 
    ((x^2)^2 + (x^4)^2) \\
 &- \frac{2}{[1 + (x^1 - x^3)^2]^2}
    (x^4+(x^1-x^3)x^2)(x^2-(x^1-x^3) x^4).  
  \end{aligned}
 \label{invtransnbr}
\end{equation}

The advertised relation to the plane wave can be seen if we now set $u
= \tan U$. Then
\begin{equation}  \label{glnbr2}
  g = \frac{1}{\cos^2 U} [-2dU dv + d\psi^2 + d\phi^2 - 4\psi d\phi
  dU]~.
\end{equation}
The metric in square brackets is a conformally flat plane wave.
Furthermore, the symmetry $\xi = \partial_\phi$ that we quotient along
annihilates the conformal factor, so we can think of the nullbrane as
conformally related to a compactified plane wave. The plane wave
nature of this solution can be instantly recognised after the further
coordinate transformation
\begin{equation}
 \begin{aligned}[m]
 V&= v + \psi \phi, \\
X  &= \psi \cos U + \phi \sin U, \\
Y &= -\psi \sin U + \phi \cos U,
 \end{aligned} 
\end{equation}
which brings the metric to the form
\begin{equation} \label{cpw}
g = \frac{1}{\cos^2 U} [ -2dU dV - (X^2+Y^2) dU^2 + dX^2 + dY^2]. 
\end{equation}
This form makes little of the symmetry explicit. The Killing vector we
are quotienting along is 
\begin{equation}
\xi = \sin U \partial_X + \cos U \partial_Y + (X \cos U - Y \sin U)
\partial_V, 
\end{equation}
and the other symmetries of the quotient are 
\begin{equation}
\begin{aligned}[m]
\xi_1 &= \sin U \partial_X - \cos U \partial_Y + (X \cos U + Y \sin U)
\partial_V,\\
\xi_2 &= \cos U \partial_X + \sin U \partial_Y + (-X \sin U + Y \cos U)
\partial_V,\\
\xi_3 &= \partial_V.
\end{aligned} 
\end{equation}
Note that not only does $\xi$ annihilate the conformal factor; so do
the other isometries. Thus, all the isometries of the nullbrane are
related to isometries of the conformally related compactified plane
wave. We can recognise them as 
\begin{equation}
\begin{aligned}[m]
\xi &= - \xi_{e_1^*} - \xi_{e_2}, \\
\xi_1 &= -\xi_{e_1^*} + \xi_{e_2},\\
\xi_2 &= -\xi_{e_1} - \xi_{e_2^*}, \\
\xi_3 &= \xi_{e_V},
\end{aligned}
\end{equation}
where we write the isometries of the plane wave in the usual basis
\begin{equation}
\begin{aligned}[m]
\xi_{e_i} &= -\cos U \partial_{X^i} + X^i \sin U \partial_V, \\
\xi_{e_i^*} &= -\sin U \partial_{X^i} - X^i \cos U \partial_V, \\
\xi_{e_V} &= \partial_V, \\
\xi_{e_U} &= -\partial_U. 
\end{aligned}
\end{equation}
Thus, the quotient of the plane wave that is conformally related to
the nullbrane is of the type considered in~\cite{Michelson}. The
additional symmetry $\xi_{e_U}$ that would be present in the plane
wave is broken by the conformal factor. As we saw in
section~\ref{sec:dnrdef}, this is precisely the additional symmetry
that appears in the double null rotation. 

As in section~\ref{sec:dnrdef}, in addition to exposing this relation
to the plane waves, the global coordinates (\ref{glnbr2}) allow us to
easily find a global time function for the nullbrane, hence
demonstrating that it is a stably causal solution. We first rewrite
the nullbrane metric in a form suitable for Kaluza-Klein reduction
along $\phi$,
\begin{equation}
 g = \frac{1}{\cos^2 U} [-2dU dv -4 \psi^2 dU^2 + d\psi^2 + ( d\phi
 -2\psi dU)^2]. 
\end{equation}
We see that Kaluza-Klein reduction will give a plane wave metric in
one dimension lower (up to conformal factor).  Hence, applying the
results of \cite{causal}, a suitable time function for the
nullbrane is
\begin{equation}
\tau = U  + \frac{1}{2} \tan^{-1} \left( \frac{4 v}{1+4\psi^2} \right).
\end{equation}
It is easy to check that 
\begin{equation}
\nabla_\mu \tau \nabla^\mu \tau = - \frac{4 \cos^2 U}{[(1+4\psi^2)^2 +
  16v^2]} = - \frac{4 (1+u^2)}{[(1+4\psi^2)^2 +
  16v^2]}.
\end{equation}
Thus, $\tau$ is a good time function on flat space, and since
$\mathcal{L}_\xi \tau = 0$, the nullbrane is stably causal by the
general argument of \cite{causal}.

\providecommand{\href}[2]{#2}\begingroup\raggedright\endgroup


\begin{thebibliography}{10}

\bibitem{melvin}
M.~A. Melvin, ``Pure magnetic and electric geons,'' Phys. Lett. {\bf 8} (1964)
65--70.

\bibitem{GibbWilt}
G.~W. Gibbons and D.~L. Wiltshire, ``Space-time as a membrane in higher
  dimensions,'' Nucl. Phys. {\bf B287} (1987) 717,
\href{http://xxx.lanl.gov/abs/hep-th/0109093}{{\tt hep-th/0109093}}.

\bibitem{gibmaed}
G.~W. Gibbons and K.-I. Maeda, ``Black holes and membranes in higher
  dimensional theories with dilaton fields,'' Nucl. Phys. {\bf B298} (1988)
741.

\bibitem{KOST}
J.~Khoury, B.~A. Ovrut, P.~J. Steinhardt, and N.~Turok, ``The ekpyrotic
  universe: Colliding branes and the origin of the hot big bang,'' Phys. Rev. D
  {\bf 64} (2001) 123522,
\href{http://xxx.lanl.gov/abs/hep-th/0103239}{{\tt hep-th/0103239}}.

\bibitem{KOSST}
J.~Khoury, B.~A. Ovrut, N.~Seiberg, P.~J. Steinhardt, and N.~Turok, ``From big
  crunch to big bang,'' Phys. Rev. D {\bf 65} (2002) 086007,
\href{http://xxx.lanl.gov/abs/hep-th/0108187}{{\tt hep-th/0108187}}.

\bibitem{BHKN}
V.~Balasubramanian, S.~F. Hassan, E.~Keski-Vakkuri, and A.~Naqvi, ``A
  space-time orbifold: A toy model for a cosmological singularity,'' Phys. Rev.
  D {\bf 67} (2003) 026003,
\href{http://xxx.lanl.gov/abs/hep-th/0202187}{{\tt hep-th/0202187}}.

\bibitem{CorCos1}
L.~Cornalba and M.~S. Costa, ``A new cosmological scenario in string theory,''
  Phys. Rev. D {\bf 66} (2002) 066001,
\href{http://xxx.lanl.gov/abs/hep-th/0203031}{{\tt hep-th/0203031}}.

\bibitem{joan}
J.~Sim\'on, ``The geometry of null rotation identifications,'' JHEP {\bf 06}
  (2002) 001,
\href{http://xxx.lanl.gov/abs/hep-th/0203201}{{\tt hep-th/0203201}}.

\bibitem{lms1}
H.~Liu, G.~Moore, and N.~Seiberg, ``Strings in a time-dependent orbifold,''
  JHEP {\bf 06} (2002) 045,
\href{http://xxx.lanl.gov/abs/hep-th/0204168}{{\tt hep-th/0204168}}.

\bibitem{lawrence}
A.~Lawrence, ``On the instability of {3D} null singularities,'' JHEP {\bf 11}
  (2002) 019,
\href{http://xxx.lanl.gov/abs/hep-th/0205288}{{\tt hep-th/0205288}}.

\bibitem{fabmcg}
M.~Fabinger and J.~McGreevy, ``On smooth time-dependent orbifolds and null
  singularities,'' JHEP {\bf 06} (2003) 042,
\href{http://xxx.lanl.gov/abs/hep-th/0206196}{{\tt hep-th/0206196}}.

\bibitem{lms2}
H.~Liu, G.~Moore, and N.~Seiberg, ``Strings in time-dependent orbifolds,'' JHEP
  {\bf 10} (2002) 031,
\href{http://xxx.lanl.gov/abs/hep-th/0206182}{{\tt hep-th/0206182}}.

\bibitem{horpol}
G.~T. Horowitz and J.~Polchinski, ``Instability of spacelike and null orbifold
  singularities,'' Phys. Rev. D {\bf 66} (2002) 103512,
\href{http://xxx.lanl.gov/abs/hep-th/0206228}{{\tt hep-th/0206228}}.

\bibitem{CKR}
B.~Craps, D.~Kutasov, and G.~Rajesh, ``String propagation in the presence of
  cosmological singularities,'' JHEP {\bf 06} (2002) 053,
\href{http://xxx.lanl.gov/abs/hep-th/0205101}{{\tt hep-th/0205101}}.

\bibitem{BCKR}
M.~Berkooz, B.~Craps, D.~Kutasov, and G.~Rajesh, ``Comments on cosmological
  singularities in string theory,'' JHEP {\bf 03} (2003) 031,
\href{http://xxx.lanl.gov/abs/hep-th/0212215}{{\tt hep-th/0212215}}.

\bibitem{EGKR}
S.~Elitzur, A.~Giveon, D.~Kutasov, and E.~Rabinovici, ``From big bang to big
  crunch and beyond,'' JHEP {\bf 06} (2002) 017,
\href{http://xxx.lanl.gov/abs/hep-th/0204189}{{\tt hep-th/0204189}}.

\bibitem{EGR}
S.~Elitzur, A.~Giveon, and E.~Rabinovici, ``Removing singularities,'' JHEP {\bf
  01} (2003) 017,
\href{http://xxx.lanl.gov/abs/hep-th/0212242}{{\tt hep-th/0212242}}.

\bibitem{PioBer}
B.~Pioline and M.~Berkooz, ``Strings in an electric field, and the {M}ilne
  universe,'' JCAP {\bf 0311} (2003) 007,
\href{http://xxx.lanl.gov/abs/hep-th/0307280}{{\tt hep-th/0307280}}.

\bibitem{CorCosrev}
L.~Cornalba and M.~S. Costa, ``Time-dependent orbifolds and string cosmology,''
  Fortsch. Phys. {\bf 52} (2004) 145--199,
\href{http://xxx.lanl.gov/abs/hep-th/0310099}{{\tt hep-th/0310099}}.

\bibitem{fofs}
J.~Figueroa-O'Farrill and J.~Sim\'on, ``Generalized supersymmetric
  fluxbranes,'' JHEP {\bf 12} (2001) 011,
\href{http://xxx.lanl.gov/abs/hep-th/0110170}{{\tt hep-th/0110170}}.

\bibitem{DGGH0}
F.~Dowker, J.~P. Gauntlett, S.~B. Giddings, and G.~T. Horowitz, ``On pair
  creation of extremal black holes and {Kaluza-Klein} monopoles,'' Phys. Rev. D
  {\bf 50} (1994) 2662--2679,
\href{http://xxx.lanl.gov/abs/hep-th/9312172}{{\tt hep-th/9312172}}.

\bibitem{DGGH1}
F.~Dowker, J.~P. Gauntlett, G.~W. Gibbons, and G.~T. Horowitz, ``The decay of
  magnetic fields in {Kaluza-Klein} theory,'' Phys. Rev. D {\bf 52} (1995)
  6929--6940,
\href{http://xxx.lanl.gov/abs/hep-th/9507143}{{\tt hep-th/9507143}}.

\bibitem{DGGH2}
F.~Dowker, J.~P. Gauntlett, G.~W. Gibbons, and G.~T. Horowitz, ``Nucleation of
  $p$-branes and fundamental strings,'' Phys. Rev. D {\bf 53} (1996)
  7115--7128,
\href{http://xxx.lanl.gov/abs/hep-th/9512154}{{\tt hep-th/9512154}}.

\bibitem{GSflux}
M.~Gutperle and A.~Strominger, ``Fluxbranes in string theory,'' JHEP {\bf 06}
  (2001) 035,
\href{http://xxx.lanl.gov/abs/hep-th/0104136}{{\tt hep-th/0104136}}.

\bibitem{arkady}
J.~G. Russo and A.~A. Tseytlin, ``Supersymmetric fluxbrane intersections and
  closed string tachyons,'' JHEP {\bf 11} (2001) 065,
\href{http://xxx.lanl.gov/abs/hep-th/0110107}{{\tt hep-th/0110107}}.

\bibitem{CGS}
C.-M. Chen, D.~V. Gal'tsov, and S.~A. Sharakin, ``Intersecting
  {M}-fluxbranes,'' Grav. Cosmol. {\bf 5} (1999) 45,
\href{http://xxx.lanl.gov/abs/hep-th/9908132}{{\tt hep-th/9908132}}.

\bibitem{CG}
M.~S. Costa and M.~Gutperle, ``The {Kaluza-Klein Melvin solution in
  M}-theory,'' JHEP {\bf 03} (2001) 027,
\href{http://xxx.lanl.gov/abs/hep-th/0012072}{{\tt hep-th/0012072}}.

\bibitem{Saffin}
P.~M. Saffin, ``Gravitating fluxbranes,'' Phys. Rev. D {\bf 64} (2001) 024014,
\href{http://xxx.lanl.gov/abs/gr-qc/0104014}{{\tt gr-qc/0104014}}.

\bibitem{CHC}
M.~S. Costa, C.~A.~R. Herdeiro, and L.~Cornalba, ``Flux-branes and the
  dielectric effect in string theory,'' Nucl. Phys. {\bf B619} (2001) 155--190,
\href{http://xxx.lanl.gov/abs/hep-th/0105023}{{\tt hep-th/0105023}}.

\bibitem{EmparanFlux}
R.~Emparan, ``Tubular branes in fluxbranes,'' Nucl. Phys. {\bf B610} (2001)
  169--189,
\href{http://xxx.lanl.gov/abs/hep-th/0105062}{{\tt hep-th/0105062}}.

\bibitem{BrecherSaffin}
D.~Brecher and P.~M. Saffin, ``A note on the supergravity description of
  dielectric branes,'' Nucl. Phys. {\bf B613} (2001) 218--236,
\href{http://xxx.lanl.gov/abs/hep-th/0106206}{{\tt hep-th/0106206}}.

\bibitem{Brechersaffin2}
D.~Brecher and P.~M. Saffin, ``Decay modes of intersecting fluxbranes,'' Phys.
  Rev. D {\bf 67} (2003) 125013,
\href{http://xxx.lanl.gov/abs/hep-th/0302206}{{\tt hep-th/0302206}}.

\bibitem{Uranga}
A.~M. Uranga, ``Wrapped fluxbranes,''
\href{http://xxx.lanl.gov/abs/hep-th/0108196}{{\tt hep-th/0108196}}.

\bibitem{EmpGut}
R.~Emparan and M.~Gutperle, ``From p-branes to fluxbranes and back,'' JHEP {\bf
  12} (2001) 023,
\href{http://xxx.lanl.gov/abs/hep-th/0111177}{{\tt hep-th/0111177}}.

\bibitem{steif}
G.~T. Horowitz and A.~R. Steif, ``Singular string solutions with nonsingular
  initial data,'' Phys. Lett. {\bf B258} (1991)
91--96.

\bibitem{btz1}
M.~Ba\~nados, C.~Teitelboim, and J.~Zanelli, ``The black hole in
  three-dimensional space-time,'' Phys. Rev. Lett. {\bf 69} (1992) 1849--1851,
\href{http://xxx.lanl.gov/abs/hep-th/9204099}{{\tt hep-th/9204099}}.

\bibitem{btz2}
M.~Ba\~nados, M.~Henneaux, C.~Teitelboim, and J.~Zanelli, ``Geometry of the
  (2+1) black hole,'' Phys. Rev. D {\bf 48} (1993) 1506--1525,
\href{http://xxx.lanl.gov/abs/gr-qc/9302012}{{\tt gr-qc/9302012}}.

\bibitem{hp}
S.~Holst and P.~Peldan, ``Black holes and causal structure in anti-de {S}itter
  isometric spacetimes,'' Class. Quant. Grav. {\bf 14} (1997) 3433--3452,
\href{http://xxx.lanl.gov/abs/gr-qc/9705067}{{\tt gr-qc/9705067}}.

\bibitem{ban1}
M.~Ba\~nados, ``Constant curvature black holes,'' Phys. Rev. D {\bf 57} (1998)
  1068--1072,
\href{http://xxx.lanl.gov/abs/gr-qc/9703040}{{\tt gr-qc/9703040}}.

\bibitem{ban2}
M.~Ba\~nados, A.~Gomberoff, and C.~Martinez, ``Anti-de {S}itter space and black
  holes,'' Class. Quant. Grav. {\bf 15} (1998) 3575--3598,
\href{http://xxx.lanl.gov/abs/hep-th/9805087}{{\tt hep-th/9805087}}.

\bibitem{cousshenn}
O.~Coussaert and M.~Henneaux, ``Self-dual solutions of 2+1 {E}instein gravity
  with a negative cosmological constant,''
\href{http://xxx.lanl.gov/abs/hep-th/9407181}{{\tt hep-th/9407181}}.

\bibitem{simon}
J.~Sim\'on, ``Null orbifolds in {AdS}, time dependence and holography,'' JHEP
  {\bf 10} (2002) 036,
\href{http://xxx.lanl.gov/abs/hep-th/0208165}{{\tt hep-th/0208165}}.

\bibitem{bns}
V.~Balasubramanian, A.~Naqvi, and J.~Sim\'on, ``A multi-boundary {AdS} orbifold
  and {DLCQ} holography: A universal holographic description of extremal black
  hole horizons,''
\href{http://xxx.lanl.gov/abs/hep-th/0311237}{{\tt hep-th/0311237}}.

\bibitem{HorMar}
G.~T. Horowitz and D.~Marolf, ``A new approach to string cosmology,'' JHEP {\bf
  07} (1998) 014,
\href{http://xxx.lanl.gov/abs/hep-th/9805207}{{\tt hep-th/9805207}}.

\bibitem{BehLus}
K.~Behrndt and D.~Lust, ``Branes, waves and {AdS} orbifolds,'' JHEP {\bf 07}
  (1999) 019,
\href{http://xxx.lanl.gov/abs/hep-th/9905180}{{\tt hep-th/9905180}}.

\bibitem{GhoMuk}
B.~Ghosh and S.~Mukhi, ``Killing spinors and supersymmetric {AdS} orbifolds,''
  JHEP {\bf 10} (1999) 021,
\href{http://xxx.lanl.gov/abs/hep-th/9908192}{{\tt hep-th/9908192}}.

\bibitem{Cai}
R.-G. Cai, ``Constant curvature black hole and dual field theory,'' Phys. Lett.
  {\bf B544} (2002) 176--182,
\href{http://xxx.lanl.gov/abs/hep-th/0206223}{{\tt hep-th/0206223}}.

\bibitem{BDHRS}
P.~Bieliavsky, S.~Detournay, M.~Herquet, M.~Rooman, and P.~Spindel, ``Global
  geometry of the 2+1 rotating black hole,'' Phys. Lett. {\bf B570} (2003)
  231--236,
\href{http://xxx.lanl.gov/abs/hep-th/0306293}{{\tt hep-th/0306293}}.

\bibitem{BRSbtz}
P.~Bieliavsky, M.~Rooman, and P.~Spindel, ``Regular poisson structures on
  massive non-rotating btz black holes,'' Nucl. Phys. {\bf B645} (2002)
  349--364,
\href{http://xxx.lanl.gov/abs/hep-th/0206189}{{\tt hep-th/0206189}}.

\bibitem{AST}
M.~Alishahiha, M.~M. Sheikh-Jabbari, and R.~Tatar, ``Spacetime quotients,
  penrose limits and conformal symmetry restoration,'' JHEP {\bf 01} (2003)
  028,
\href{http://xxx.lanl.gov/abs/hep-th/0211285}{{\tt hep-th/0211285}}.

\bibitem{mcinnes2003}
B.~McInnes, ``Orbifold physics and de sitter spacetime,''
\href{http://xxx.lanl.gov/abs/hep-th/0311055}{{\tt hep-th/0311055}}.

\bibitem{FHLT}
B.~Fiol, C.~Hofman, and E.~Lozano-Tellechea, ``Causal structure of d = 5 vacua
  and axisymmetric spacetimes,'' JHEP {\bf 0402} (2004) 034,
\href{http://xxx.lanl.gov/abs/hep-th/0312209}{{\tt hep-th/0312209}}.

\bibitem{mcinnes2004}
B.~McInnes, ``String theory and the shape of the universe,''
\href{http://xxx.lanl.gov/abs/hep-th/0401035}{{\tt hep-th/0401035}}.

\bibitem{fofs2}
J.~Figueroa-O'Farrill and J.~Simon, ``Supersymmetric {Kaluza-Klein} reductions
  of {AdS} backgrounds,''
\href{http://xxx.lanl.gov/abs/hep-th/0401206}{{\tt hep-th/0401206}}.

\bibitem{mr}
O.~Madden and S.~F. Ross, ``Quotients of anti-de {Sitter} space,''
\href{http://xxx.lanl.gov/abs/hep-th/0401205}{{\tt hep-th/0401205}}.

\bibitem{godel}
K.~{G\"odel}, ``An example of a new type of cosmological solutions of
  {E}instein's field equations of graviation,'' Rev. Mod. Phys. {\bf 21} (1949)
447--450.

\bibitem{GGHPR}
J.~P. Gauntlett, J.~B. Gutowski, C.~M. Hull, S.~Pakis, and H.~S. Reall, ``All
  supersymmetric solutions of minimal supergravity in five dimensions,'' Class.
  Quant. Grav. {\bf 20} (2003) 4587--4634,
\href{http://xxx.lanl.gov/abs/hep-th/0209114}{{\tt hep-th/0209114}}.

\bibitem{BGHV}
E.~K. Boyda, S.~Ganguli, P.~Horava, and U.~Varadarajan, ``Holographic
  protection of chronology in universes of the {G\"odel} type,'' Phys. Rev. D
  {\bf 67} (2003) 106003,
\href{http://xxx.lanl.gov/abs/hep-th/0212087}{{\tt hep-th/0212087}}.

\bibitem{HarTak}
T.~Harmark and T.~Takayanagi, ``Supersymmetric {G\"odel} universes in string
  theory,'' Nucl. Phys. {\bf B662} (2003) 3--39,
\href{http://xxx.lanl.gov/abs/hep-th/0301206}{{\tt hep-th/0301206}}.

\bibitem{penrose}
R.~Penrose, ``Any space-time has a plane wave as a limit,'' in {\em
  Differential geometry and relativity}, M. Cahen and M. Flato, eds.,
pp.~271--275. 
\newblock Reidel, Dordrecht-Holland, 1976.

\bibitem{Gueven}
R.~Gueven, ``Plane wave limits and {T}-duality,'' Phys. Lett. {\bf B482} (2000)
  255--263,
\href{http://xxx.lanl.gov/abs/hep-th/0005061}{{\tt hep-th/0005061}}.

\bibitem{BFOHP}
M.~Blau, J.~Figueroa-O'Farrill, C.~Hull, and G.~Papadopoulos, ``Penrose limits
  and maximal supersymmetry,'' Class. Quant. Grav. {\bf 19} (2002) L87--L95,
\href{http://xxx.lanl.gov/abs/hep-th/0201081}{{\tt hep-th/0201081}}.

\bibitem{BFOP}
M.~Blau, J.~Figueroa-O'Farrill, and G.~Papadopoulos, ``Penrose limits,
  supergravity and brane dynamics,'' Class. Quant. Grav. {\bf 19} (2002) 4753,
\href{http://xxx.lanl.gov/abs/hep-th/0202111}{{\tt hep-th/0202111}}.

\bibitem{FigSimBranes}
J.~Figueroa-O'Farrill and J.~Sim\'on, ``Supersymmetric {Kaluza-Klein reductions
  of M2 and M5} branes,'' Adv. Theor. Math. Phys. {\bf 6} (2003) 703--793,
\href{http://xxx.lanl.gov/abs/hep-th/0208107}{{\tt hep-th/0208107}}.

\bibitem{FigSimGrav}
J.~Figueroa-O'Farrill and J.~Sim\'on, ``Supersymmetric {Kaluza-Klein reductions
  of M-waves and MKK- monopoles},'' Class. Quant. Grav. {\bf 19} (2002)
  6147--6174, \href{http://xxx.lanl.gov/abs/hep-th/0208108}{{\tt
  hep-th/0208108}}.
Erratum: ibid. \textbf{21} (2004) 337.

\bibitem{DLP}
M.~J. Duff, H.~Lü, and C.~N. Pope, ``Supersymmetry without supersymmetry,''
  Phys. Lett. {\bf B409} (1997) 136--144,
  \href{http://xxx.lanl.gov/abs/hep-th/9704186}{{\tt hep-th/9704186}}.

\bibitem{DuffLuPopeUnt}
M.~J. Duff, H.~Lü, and C.~N. Pope, ``$\textrm{AdS}_5 \times {S}^5$ untwisted,''
  Nucl. Phys. {\bf B532} (1998) 181--209,
  \href{http://xxx.lanl.gov/abs/hep-th/9803061}{{\tt hep-th/9803061}}.

\bibitem{PSS}
C.~N. Pope, A.~Sadrzadeh, and S.~R. Scuro, ``Timelike {H}opf duality and type
  {IIA*} string solutions,'' Class. Quant. Grav. {\bf 17} (2000) 623--641,
  \href{http://xxx.lanl.gov/abs/hep-th/9905161}{{\tt hep-th/9905161}}.

\bibitem{HullHolonomy}
C.~M. Hull, ``Holonomy and symmetry in {M}-theory,''
  \href{http://xxx.lanl.gov/abs/hep-th/0305039}{{\tt hep-th/0305039}}.

\bibitem{causal}
V.~E. Hubeny, M.~Rangamani, and S.~F. Ross, ``Causal inheritance in plane wave
  quotients,'' Phys. Rev. D {\bf 69} (2004) 024007, 
\href{http://xxx.lanl.gov/abs/hep-th/0307257}{{\tt hep-th/0307257}}.

\bibitem{MaozSimon}
L.~Maoz and J.~Sim\'on, ``Killing spectroscopy of closed timelike curves,''
JHEP {\bf 0401} (2004) 051,
\href{http://xxx.lanl.gov/abs/hep-th/0310255}{{\tt hep-th/0310255}}.

\bibitem{doCarmo}
M.~do~Carmo, {\em Riemannian Geometry}.
\newblock Birkhäuser, 1992.

\bibitem{bbkr}
V.~Balasubramanian, J.~de~Boer, E.~Keski-Vakkuri, and S.~F. Ross,
  ``Supersymmetric conical defects: Towards a string theoretic description of
  black hole formation,'' Phys. Rev. D {\bf 64} (2001) 064011,
\href{http://xxx.lanl.gov/abs/hep-th/0011217}{{\tt hep-th/0011217}}.

\bibitem{mart:orb1}
E.~J. Martinec and W.~McElgin, ``String theory on {AdS} orbifolds,'' JHEP {\bf
  04} (2002) 029,
\href{http://xxx.lanl.gov/abs/hep-th/0106171}{{\tt hep-th/0106171}}.

\bibitem{mart:orb2}
E.~J. Martinec and W.~McElgin, ``Exciting {AdS} orbifolds,'' JHEP {\bf 10}
  (2002) 050,
\href{http://xxx.lanl.gov/abs/hep-th/0206175}{{\tt hep-th/0206175}}.

\bibitem{bn}
D.~Berenstein and H.~Nastase, ``On lightcone string field theory from super
  {Yang-Mills} and holography,''
\href{http://xxx.lanl.gov/abs/hep-th/0205048}{{\tt hep-th/0205048}}.

\bibitem{strom:ads2}
A.~Strominger, ``{AdS}$_2$ quantum gravity and string theory,'' JHEP {\bf 01}
  (1999) 007,
\href{http://xxx.lanl.gov/abs/hep-th/9809027}{{\tt hep-th/9809027}}.

\bibitem{cvetic}
M.~Cvetic, H.~Lu, and C.~N. Pope, ``Spacetimes of boosted p-branes, and {CFT}
  in infinite- momentum frame,'' Nucl. Phys. {\bf B545} (1999) 309--339,
\href{http://xxx.lanl.gov/abs/hep-th/9810123}{{\tt hep-th/9810123}}.

\bibitem{Michelson}
J.~Michelson, ``({T}wisted) toroidal compactification of pp-waves,'' Phys. Rev.
  D {\bf 66} (2002) 066002,
\href{http://xxx.lanl.gov/abs/hep-th/0203140}{{\tt hep-th/0203140}}.

\bibitem{bubble}
V.~Balasubramanian and S.~F. Ross, ``The dual of nothing,'' Phys. Rev. D {\bf
  66} (2002) 086002,
\href{http://xxx.lanl.gov/abs/hep-th/0205290}{{\tt hep-th/0205290}}.

\bibitem{birm}
D.~Birmingham and M.~Rinaldi, ``Bubbles in anti-de {S}itter space,'' Phys.
  Lett. {\bf B544} (2002) 316--320,
\href{http://xxx.lanl.gov/abs/hep-th/0205246}{{\tt hep-th/0205246}}.

\bibitem{mald}
J.~M. Maldacena, ``The large {N} limit of superconformal field theories and
  supergravity,'' Adv. Theor. Math. Phys. {\bf 2} (1998) 231--252,
\href{http://xxx.lanl.gov/abs/hep-th/9711200}{{\tt hep-th/9711200}}.

\bibitem{BMN}
D.~Berenstein, J.~M. Maldacena, and H.~Nastase, ``Strings in flat space and pp
  waves from {$N = 4$} super {Yang Mills},'' JHEP {\bf 04} (2002) 013,
\href{http://xxx.lanl.gov/abs/hep-th/0202021}{{\tt hep-th/0202021}}.

\end{thebibliography}
\end{document}